\documentclass[useAMS,usenatbib,floatfix,onecolumn]{mn2e}

\usepackage{times}
\usepackage{graphicx}
\usepackage{natbib}
\usepackage{color}

\def\pdeg{\ifmmode $\setbox0=\hbox{$^{\circ}$}\rlap{\hskip.11\wd0 .}$^{\circ}
          \else \setbox0=\hbox{$^{\circ}$}\rlap{\hskip.11\wd0 .}$^{\circ}$\fi}
\def\arcs{\ifmmode {^{\scriptscriptstyle\prime\prime}}
          \else $^{\scriptscriptstyle\prime\prime}$\fi}
\def\arcm{\ifmmode {^{\scriptscriptstyle\prime}}
          \else $^{\scriptscriptstyle\prime}$\fi}
\newdimen\sa  \newdimen\sb
\def\parcs{\sa=.07em \sb=.03em
     \ifmmode \hbox{\rlap{.}}^{\scriptscriptstyle\prime\kern -\sb\prime}\hbox{\kern -\sa}
     \else \rlap{.}$^{\scriptscriptstyle\prime\kern -\sb\prime}$\kern -\sa\fi}
\def\parcm{\sa=.08em \sb=.03em
     \ifmmode \hbox{\rlap{.}\kern\sa}^{\scriptscriptstyle\prime}\hbox{\kern-\sb}
     \else \rlap{.}\kern\sa$^{\scriptscriptstyle\prime}$\kern-\sb\fi}

\def\reff@jnl#1{{\rm#1\/}}
\def\apj{\reff@jnl{ApJ}}       
\def\apjs{\reff@jnl{ApJS}}     
\def\apjl{\reff@jnl{ApJL}}     
\def\aaps{\reff@jnl{A\&AS}}    
\def\mnras{\reff@jnl{MNRAS}}   
\def\prd{\reff@jnl{Phys.\ Rev.\ D}}    

\def\simpropto{\,{\lower3.0pt\hbox{$\propto$}\atop\raise1.0pt\hbox{$\sim$}}\,}

\def\P{{\mathbf P}}

\newcommand{\vW}{\mathbf{W}}
\newcommand{\vK}{\mathbf{K}}

\newcommand{\beq}{\begin{equation}}
\newcommand{\eeq}{\end{equation}}

\newcommand{\tC}{\widetilde{C}}
\newcommand{\tN}{\widetilde{N}}
\newcommand{\ta}{\tilde{a}}

\newcommand{\n}{{\bf{n}}}

\newcommand{\VEV}[1]{\langle#1\rangle}

\newcommand{\wjjj}[6]
{{
\left(
\begin{array}{lcr} #1 & #2 & #3 \\#4 & #5 & #6 \end{array}
\right)
}}

\newcommand{\be}{\begin{equation}}
\newcommand{\ee}{\end{equation}}

{\begin{enumerate}\setlength{\itemsep}{0mm}}%
{\end{enumerate}}
{\begin{enumerate}\setlength{\itemsep}{0mm}}%
{\end{enumerate}}

\begin{document}

\title[{\tt XFaster}  applied to Planck]{Application of {\tt XFaster}  power spectrum and likelihood estimator to Planck }
\author[G. Rocha, C. R. Contaldi, J. R. Bond \& K. M. G\'orski]{G. Rocha$^{1,2}$, C. R. Contaldi$^{3}$, J. R. Bond$^{4}$ \& K. M. G\'orski$^{1,5}$ \\
$^{1}$ Jet Propulsion Laboratory, California Institute of Technology, 4800 Oak Grove Drive, Pasadena CA 91109, U.~S.~A. \\
$^{2}$ Department of Physics, California Institute of Technology, Pasadena, 91125, U.~S.~A. \\
$^{3}$ Department of Physics, Imperial College, University of London, South Kensington Campus, London SW7 2AZ, U.~K. \\
$^{4}$ The Canadian Institute for Theoretical Astrophysics, CITA, University of Toronto, 60 St. George Street, Toronto, Ontario, M5S 3H8, Canada \\
$^{5}$ Warsaw University Observatory, Aleje Ujazdowskie 4, 00478 Warszawa, Poland}

\date{Accepted ---. Received ---; in original form \today}

\date{27 October 2010}
\pagerange{\pageref{firstpage}--\pageref{lastpage}} \pubyear{2009}
\maketitle
\label{firstpage}

\label{firstpage}

\maketitle

\begin{abstract}
We develop the {\tt XFaster} Cosmic Microwave Background
(CMB) temperature and polarization anisotropy power spectrum and
likelihood technique for the Planck CMB satellite mission. 
We give an overview of this estimator and its current
implementation and present the results of applying this algorithm to
simulated Planck data.
We show that it can accurately extract the power spectrum of Planck
data for the high-$\ell$ multipoles range.
We compare the {\tt XFaster} approximation for the  likelihood to other high-$\ell$ likelihood approximations such as Gaussian and Offset
Lognormal and a low-$\ell$ pixel-based likelihood.  We show that the
{\tt XFaster}  likelihood is not only accurate at high-$\ell$, but
also performs well at moderately low multipoles.  We also
present  results for cosmological parameter Markov Chain Monte Carlo
estimation with the {\tt XFaster} likelihood.
 As long as the low-$\ell$ polarization and temperature
power are properly accounted for, {\it e.g.},  by adding an adequate
low-$\ell$ likelihood ingredient, the input parameters are recovered
to a high level of accuracy.

\end{abstract}

\begin{keywords}
Cosmology: observations -- methods: data analysis -- cosmic microwave background
\end{keywords}

\section{Introduction}
\label{intro}

Power spectrum estimation plays a crucial role in CMB data analysis.
Primordial curvature fluctuations form a homogeneous, isotropic, and nearly Gaussian random field in most early universe scenarios, inflationary or otherwise. To the extent that fluctuations are Gaussian, the power spectrum describes their statistical properties fully. An immediate consequence is that the CMB temperature and polarization primary
anisotropies, linearly responding to the primordial fluctuations, form an isotropic nearly Gaussian random field, characterized by their own angular power spectra.

With the advent of large, high-quality data sets, especially that from the Planck mission  (\cite{bluebook06,planckmission09}), we can measure the angular power spectrum accurately over a wide range of angular scales.  Power spectrum estimation can be viewed as a significant data compression. For instance, 1~year of observations from a Planck High Frequency Instrument (HFI) channel produces roughly $N_{tod}\sim 2 \times 10^{11}$ data samples.  This reduces to $N_{map}\sim 5 \times 10^7$ pixels via mapmaking and finally to $N_{pse}\sim 3 \times 10^3$ values in the power spectra.

Planck is a full-sky experiment with beams ranging in size from 30\arcm\ to 5\arcm, and with high resolution maps encompassing tens of millions of pixels.  Direct extraction of science from the pixelized maps is computationally expensive and in fact unfeasible.
Accurate estimation of the angular power spectrum for Planck enables the extraction of science with minimal loss of information.

A number of approaches have been developed to estimate the angular power spectrum from CMB data (for a review see \cite{efstathiou04,varenna10}). Such estimators can be divided into three classes: codes accurate at large-angular scales (low multipoles $\ell$), including codes that evaluate or sample from the likelihood function directly; codes accurate at small angular scales (high-$\ell$) that characterize the statistics of an unbiased frequentist estimator for the power spectrum;  and hybrid codes that can be aplied to both low and high-$\ell$. The first class comprises Maximum Likelihood Estimators (MLE) either in Fourier space (\citet{gorski94a,gorski97,gorski94b,gorski96}) or in Real space (\citet{bunn95,hancock97}) such as MADspec (\cite{madspec99,madspec09}), Quadratic Maximum Likelihood (QML) estimators (\citet{wright94,hamilton97, tegmark97,DABJK00,knox99}) such as Bolpol (\cite{bolpol09}), Gibbs samplers such as Commander (\cite{commander1,commander2,commander3}), and Importance Samplers combined with a Copula-based approximation to the likelihood such as Teasing (\cite{teasing09}).  The second class comprises quadratic $Pseudo-C_{\ell}$ (PCL) or Master method codes (\cite{master02,master01}) such as Romaster, cRomaster (\cite{cromaster}), Xpol (\cite{xpol,archeops_cl2}), and Crosspect  (\cite{varenna10}), angular correlation function codes such as Spice (\cite{spice}) and Polspice (\cite{polspice}), and  Quadratic Maximum likelihood (QML) codes such as {\tt XFaster}  (\cite{XFaster-like1,faster-boom01,xfaster-boom03} and this paper, see Section~\ref{pse} for more details).
The third class consists of hybrid power spectrum estimators such as a QML estimator at low-$\ell$ combined with a PCL estimator at high-$\ell$ (\cite{efstathiou04}) and, to some extent, the {\tt XFaster} method alone.

{\tt XFaster} (\cite{XFaster-like1}) was first developed to give rapid and accurate
power spectra determinations from bolometer data for the Boomerang
long-duration balloon experiments, first for total anisotropy
(\cite{faster-boom01}) and then for polarization (\cite{xfaster-boom03}). 
{\tt XFaster} is a quadratic maximum likelihood estimator formulated in the isotropic, diagonal approximation of the Master method (\cite{master02}). The noise becomes a diagonalised Monte-Carlo-estimated bias and the signal is summed into bands to reduce  correlations induced by sky cuts.
In this sense {\tt XFaster} is an extension of the traditional Master estimators where the pseudo-$C_{\ell}$ quantity is replaced by the quadratic MLE expression and uncertainties are given by the Fisher matrix.

This method has been compared with other high-$\ell$ codes such as Polspice, Romaster, Xpol, CrossSpec within Planck working group $C_{\ell}$ Temperature and Polarization, CTP.  A detailed account of this comparison will be given in a paper by the Planck CTP working group (\cite{varenna10}).  A full and detailed account of {\tt XFaster}  as a standalone method will be given elsewhere (\cite{XFaster-like1}). Here we  give an overview of the method, but our main goal is to show its adequacy to extract the power spectrum from Planck data.

An interesting feature of the method is that it provides a natural expression for the likelihood based on the assumption that the cut-sky harmonic coefficients $a_{\ell m}$ follow the same distribution as those of the full-sky harmonics.  We compare our approximate likelihood to the exact full-sky likelihood (the inverse Wishart distribution) and to the pixel based likelihood (i.e., multivariate Gaussian of the pixel's $I$, $Q$, and $U$ Stokes parameters). We show that {\tt XFaster} agrees well with the exact likelihoods at moderate low-$\ell$ multipoles as well.

Using the {\tt XFaster} power spectrum and likelihood estimator, we show how to go straight from the map to parameters, bypassing the band power spectrum estimation step.  Alternatively, we show how to use the band power spectrum estimated with {\tt XFaster} in combination with any likelihood approximation to estimate parameters. In particular, in \cite{GRLike09} we compare parameters estimated with {\tt XFaster} and  the Offset Lognormal Bandpower likelihood to those obtained with the {\tt XFaster}  likelihood.
From our analysis we conclude that as long as the low-$\ell$ polarization is properly accounted for (by adding an adequate low-$\ell$ likelihood ingredient), we recover the input parameters accurately.

This paper is organized as follows: Section~\ref{planck-sims} describes the map and the Monte Carlo simulations for two phases of increasing complexity studied in the CTP working group; Section~\ref{pse-like} gives an overview of the {\tt XFaster} power spectrum and likelihood estimator, including the estimation of kernels, transfer or filter functions, and window functions;
Section~\ref{results} shows the results of applying {\tt XFaster} to Planck simulations in several different ways.  It includes the impact of beam asymmetries on power spectrum and cosmological parameter estimation, comparison of the {\tt XFaster} likelihood to other likelihood aproximations at high-$\ell$ and to pixel-based likelihood at low-$\ell$, and cosmological parameter estimation. Section~\ref{conc} gives conclusions.

\section{Simulations}
\label{planck-sims}

The Planck satellite (\cite{bluebook06, planckmission09}) is a full-sky experiment with beams ranging in size from 30\arcm\ to 5\arcm.
The Low Frequency Instrument (LFI) covers 30, 44, and 70\,GHz; the High Frequency Instrument (HFI) covers 100, 143, 216, 353, 545, and 857\,GHz.  From the second Lagrangian point of the Earth-Sun system ($L_{2}$) Planck scans nearly great circles on the sky, covering the full sky twice over the course of a year  (\cite{dupac05}).  Planck spins at 1\,rpm around an axis that is repointed roughly 30 times per day along a cycloidal path, with the spin axis moving in a 7\pdeg5 circle around the anti-Sun direction with a period of six months.  This ensures that all feeds cover the ecliptic pole regions fully.  We also include small perturbations to the pointing, with spin axis nutation and variations in the satellite spin rate.
For the analysis presented here we consider the 70\,GHz LFI channel.

The simulations used in this work include CMB and realistic detector noise only, and are  specified by the scanning strategy (as described above), telescope beams, and detector properties.   To mimic a more sensitive combination of channels, the white noise level was taken to be lower than that expected for the real 70\,GHz channel.  We used a  single  observed map containing CMB and noise as well as Monte Carlo simulations of signal and noise.  Technical details of the simulations are given in \cite{varenna10}.

We considered all twelve detectors of the 70\,GHz LFI channel.
The beams of the detectors have FWHMs of
13\arcm--14\arcm, so the maps were made with $N_{\mathrm{side}} = 1024$, corresponding
to a pixel size of 3\parcm4. Two sets of maps were provided, one 12-detector
map to be used in the auto-spectrum mode,  and three 4-detector maps to be used in the cross-spectrum mode.

\begin{figure}
\begin{center}
\includegraphics[width=9cm]{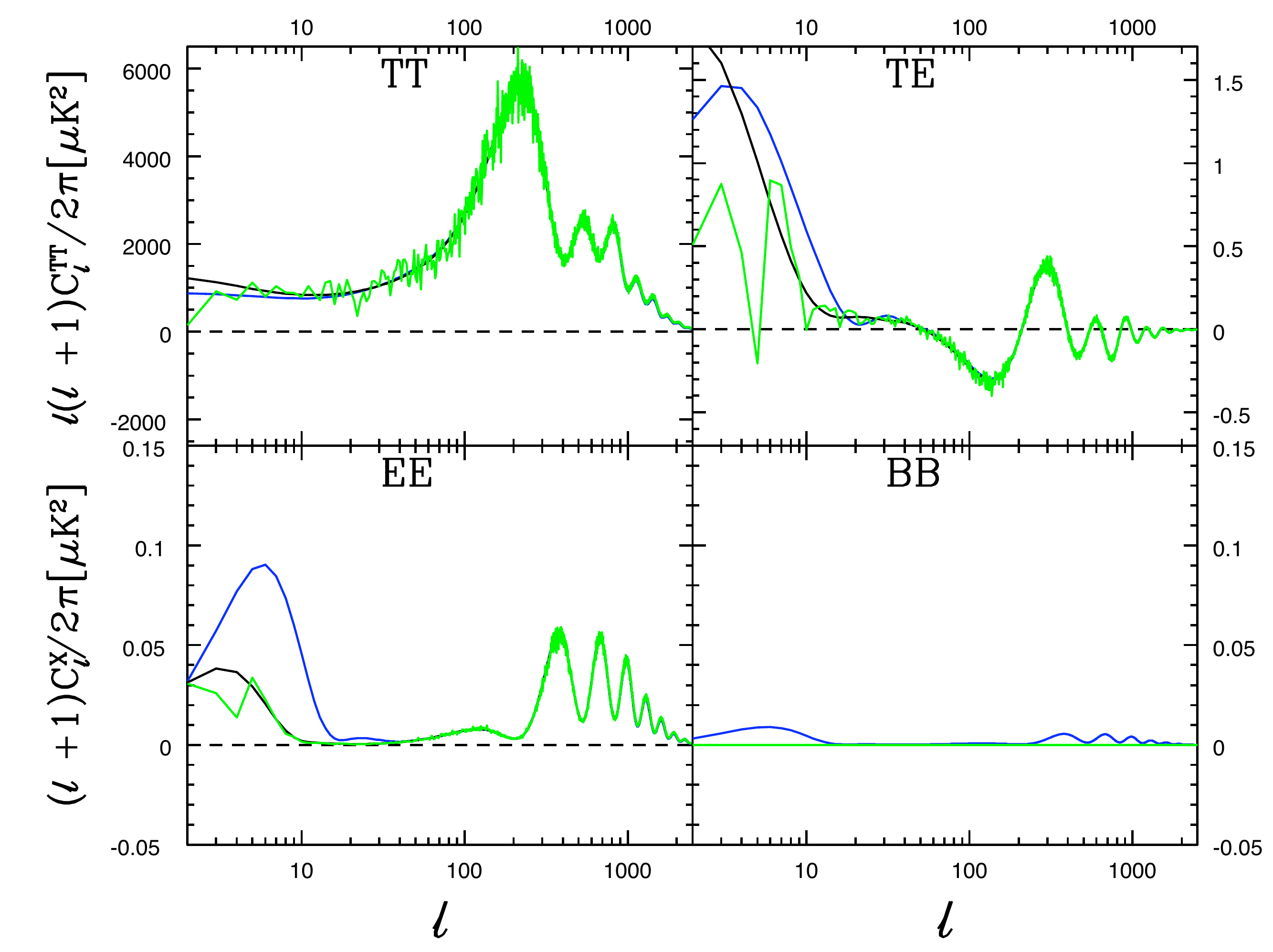}
\caption{Angular power spectrum, $C_{\ell}$, of the Planck CMB reference sky, obtained from the reference sky spherical harmonics, $a_{\ell m}$, used to generate the signal of the 70\,GHz observed map (green solid line); $C_{\ell}$ used to generate the Monte Carlo simulations for Phase~1 (blue solid line); $C_{\ell}$ used to generate the Monte Carlo simulations for Phase~2 (black solid line).}
\label{cl-models}
\end{center}
\end{figure}

The input sky signal used to generate the observed map was the CMB map
derived from the WMAP 1-year data used in a previous CTP map-making exercise (\cite{trieste,varenna10}).
It is derived from the Planck CMB reference sky available in
\footnote{http://www.sissa.it/~planck/reference-sky/CMB/alms/alm-cmb-reference-template-microKthermodynamic-nside2048.fits}.
Hence the large scale structure of the observed map  is a WMAP constrained realization.
The angular power spectrum of the $a_{lm}$ is plotted in Figure~\ref{cl-models}.

Simulations in four steps of increasing complexity were used.  For historical reasons we refer to these steps as Phases 1a, 1b, 2a, and 2b.

\smallskip\noindent
{\bf Phase 1a}---Data were simulated with symmetric beams and isotropic white noise.  The power spectrum was estimated from the full sky.

\smallskip\noindent
{\bf Phase 1b}---Data were simulated with symmetric beams and anisotropic white noise determined by the scan strategy.  A sky cut was applied in the calculation of the power spectrum to mimic the effects of removing the galactic plane from observed data.

\smallskip
\noindent
For Phases 1a and 1b the input CMB sky was convolved with a Gaussian beam of 14\arcm\ FWHM, and the noise realizations were generated assuming an RMS of 69.28\,$\mu$K per pixel in temperature and am RMS of 97.97\,$\mu$K per pixel in the $Q$ and $U$ polarization components.

The observed maps were generated in the pixel domain.
Monte Carlo signal simulations were generated from the WMAP 1-year best-fit $\Lambda$CDM power
spectrum\footnote{available in http://lambda.gsfc.nasa.gov/product/map/dr1/lcdm.cfm} plotted in Figure~\ref{cl-models} (blue solid line).
One hundred Monte Carlo realizations were generated directly in the pixel domain for both the signal and noise using {\tt HEALPix} tools (\cite{healpix}) and our own simulator.

\smallskip\noindent
{\bf Phase 2a}---Data were simulated with both correlated $1/f$ and anisotropic white noise.  Symmetric beams were assumed, all Gaussian with FWHM 14\arcm.  Noise effects induced by temperature fluctuations of the 20-K hydrogen sorption cooler were also included.

\smallskip\noindent
{\bf Phase 2b}---Data were simulated with both correlated $1/f$ and anisotropic white noise.  Asymmetric beams were used, specifically, elliptical Gaussians fit to the central parts of realistic beams calculated by a full diffraction code for the Planck optical system.  For the twelve beams, the geometric mean of the major and minor axis FWHMs ranged from 12\parcm43 to 13\parcm03.  Major axis to minor axis ratios varies from 1.22 to 1.26.

\smallskip
\noindent
For Phases 2a and 2b  the white noise per sample was 2025.8\,$\mu$K; the $1/f$ noise power spectrum had a knee frequency of 0.05\,Hz and a slope of $-1.7$.
\medskip

\begin{figure}
\begin{center}
\hbox{
\includegraphics[width=6cm]{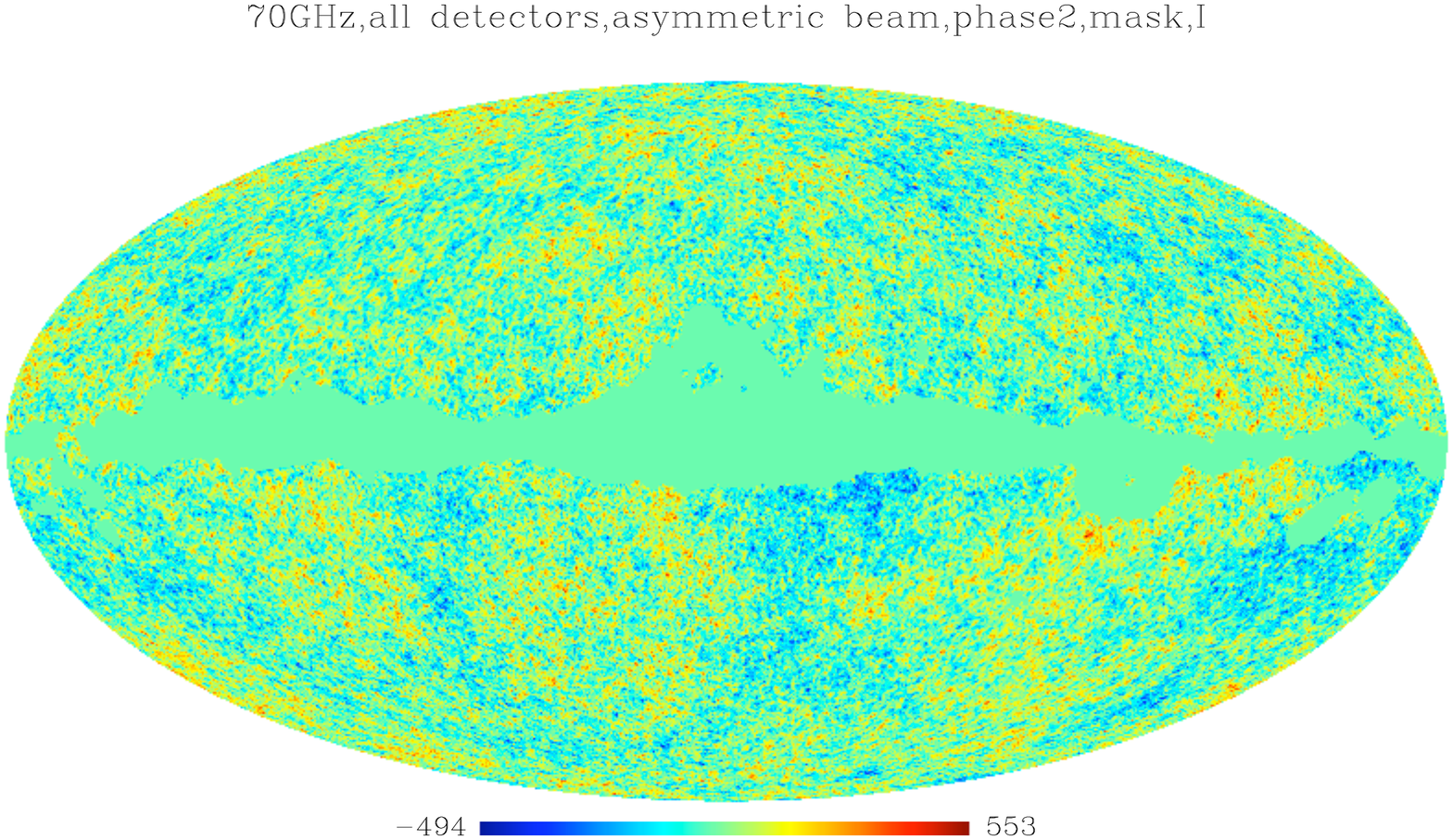}
\includegraphics[width=6cm]{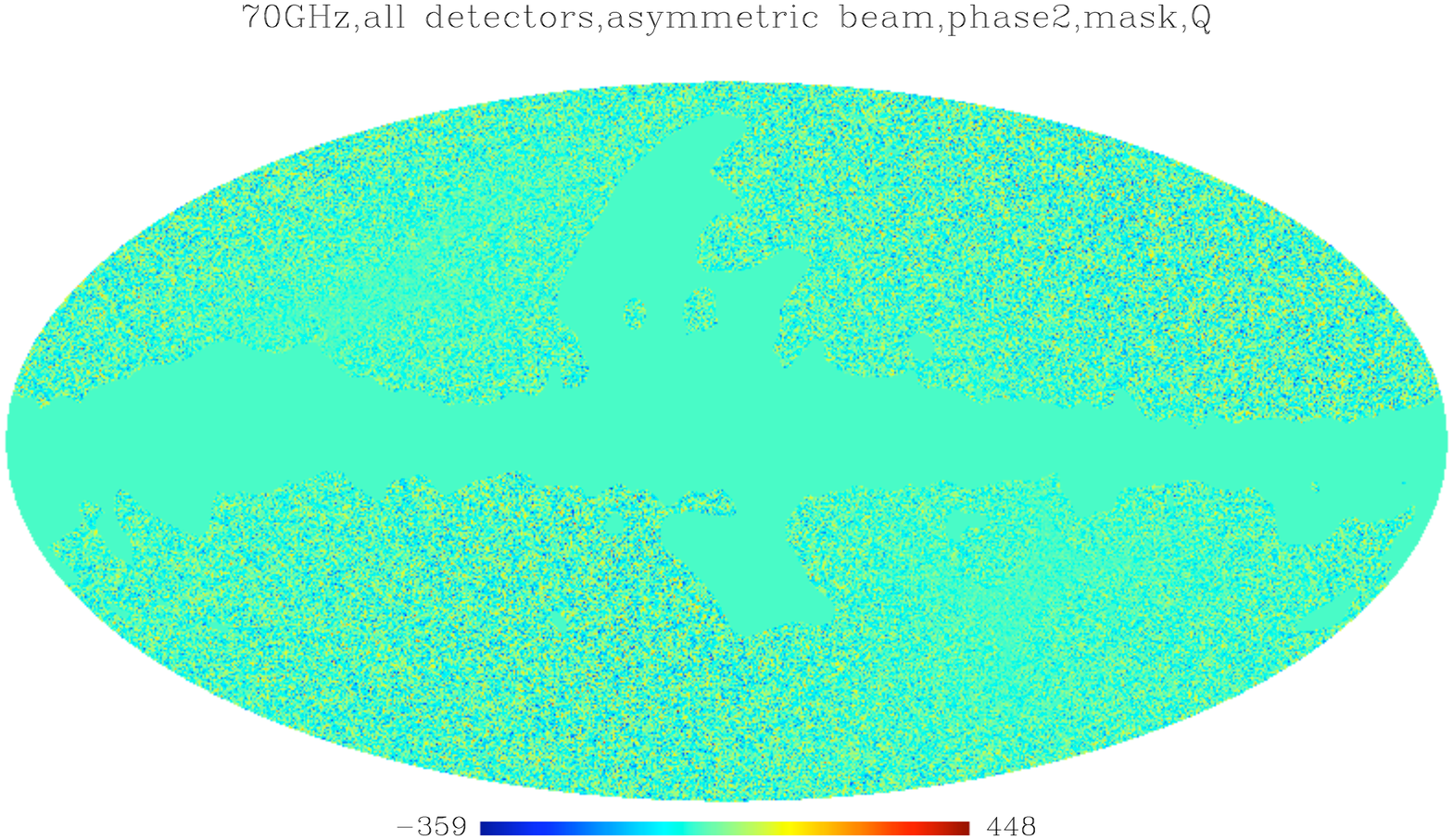}
\includegraphics[width=6cm]{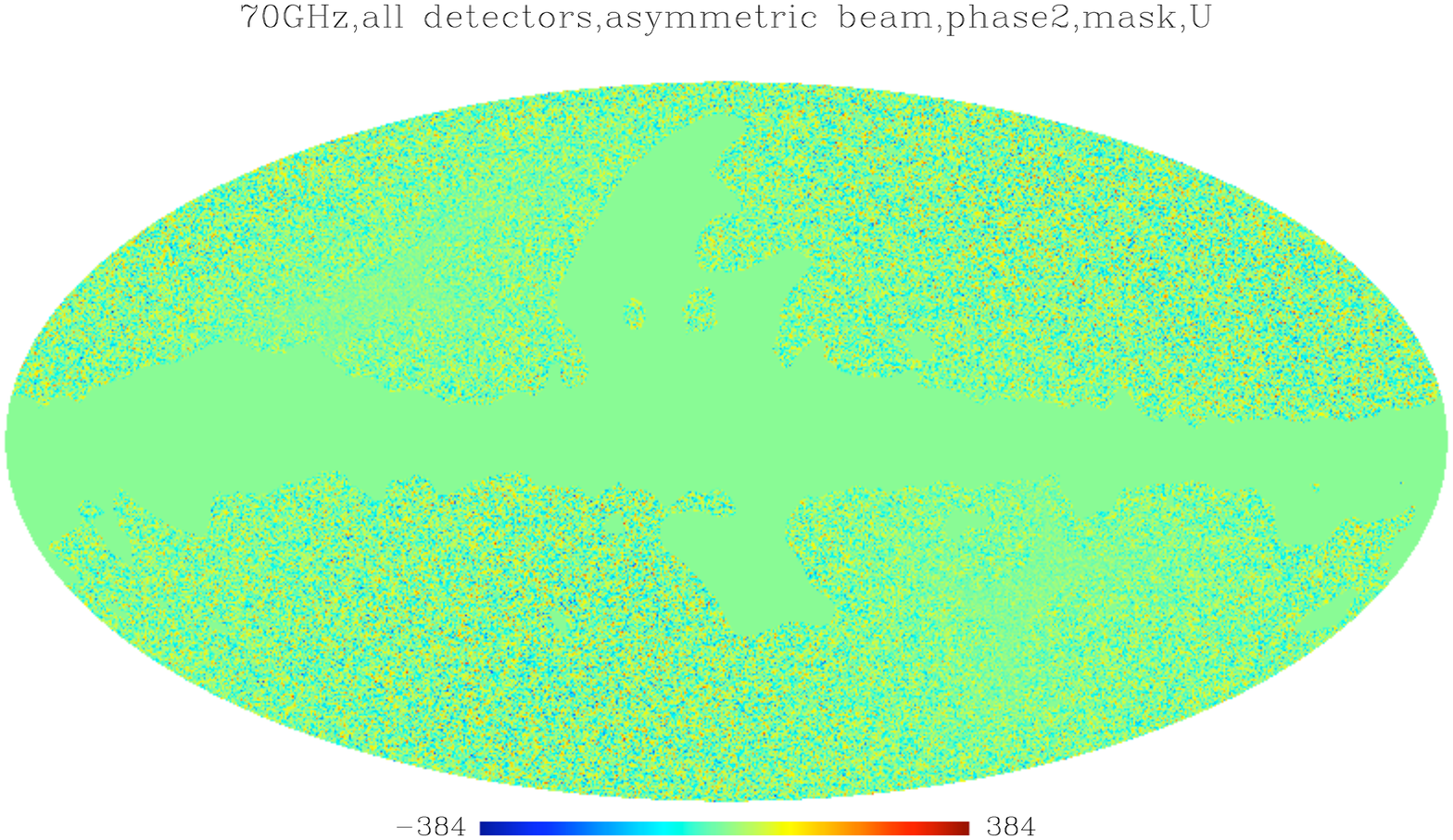}
}
\caption{70\,GHz ($I$, $Q$, $U$) Stokes parameters of the observed map (from left to right hand side) for Phase~2b, generated with all detectors, convolved with the asymmetric beam and with mask (for galaxy plus missing pixels) applied.}
\label{maps-masks}
\end{center}
\end{figure}

The observed maps were made from time-ordered data (TOD) using the destriping mapmaker {\tt Springtide} (\cite{cambridge,helsinki,paris,trieste,springtide}).  The TOD were generated using modules of the Planck simulator pipeline, {\tt LevelS} (\cite{levels}). Where a sky cut was applied in the analysis of the maps, the cut was made at the boundary where the total intensity of the diffuse foregrounds and point sources exceeded twice the CMB sigma.  Masks for missing pixels due to the scanning strategy, if any, were also considered.
Figure~\ref{maps-masks} shows the observed map for Phase2b using all twelve  detectors, with the mask for galaxy plus missing pixels applied.

As with Phases 1a and 1b, 100 Monte Carlo signal simulations were generated from the first year WMAP+CBI+ACBAR best fit $\Lambda$CDM power
spectrum$^{2}$ with BB mode power set to zero (see Figure~\ref{cl-models}).  
For the symmetric beam case only the noise TODS were generated, while the signal was simulated in the map domain.
For the asymmetric beam case both signal and noise simulations were generated in the time domain.

As mentioned above the large scale structure of the observed map is derived from real
observations, i.e., a WMAP constrained realization, hence it is not necessarily consistent with the best-fit spectrum at low multipoles.
This discrepancy will become evident later when comparing the power spectrum estimated from the observed map with the best-fit theoretical spectrum, as well as when comparing the cosmological parameters estimated with {\tt XFaster}  power spectrum and likelihood and the theoretical best fit parameters. As the Monte Carlo simulations are realizations of the WMAP 1-year best-fit $\Lambda$CDM power spectrum for Phases 1a and 1b, and the first year WMAP+CBI+ACBAR best fit $\Lambda$CDM power spectrum for Phases 2a and 2b, such discrepancy is no longer present.  Parameters estimated from these Monte Carlo simulations maps are now close to the WMAP best fit parameters.

The choice of 70\,GHz for the simulations was driven by practical matters of computational resources having to do with the size of the TOD, the number of pixels in the maps, and the number of multipoles that had to be calculated.  The HFI channels have higher angular resolution and sensitivity, and  will extend to smaller angular scales with reduced error bars.  Increases in computational capability over time make it possible now to generate thousands of Monte Carlo simulations at the higher frequencies as well.  Results will be presented in a future publication \cite{GRhfiXFaster10,varenna10}.

%

\section{{\tt XFaster}  Power spectrum and Likelihood estimator}
\label{pse-like}

\subsection{{\tt XFaster}  Power spectrum estimator \label{pse}}
\label{pse}

{\tt XFaster}  (\cite{XFaster-like1,faster-boom01,xfaster-boom03}) is an iterative, maximum likelihood, quadratic band power estimator
(\cite{DABJK00,knox99,hamilton97, tegmark97,tegmarkcosta01}) based on a diagonal approximation to the quadratic Fisher matrix estimator. It is a QML estimator formulated in the isotropic, diagonal approximation of the Master method (\cite{master02}).

It is common to expand the pixel temperature fluctuations (Stokes $I$), $T(\hat{\n})$, on the celestial sphere in terms of spherical harmonic functions, $Y_{\ell m}$,  as
\be
 T(\hat{\n}) = \sum_{\ell m} a_{\ell m} Y_{\ell m}(\hat{\n}),
\ee
with coefficients $a_{\ell m}$.

The maximum likelihood estimator (MLE) is based on a Gaussian assumption for the likelihood of the observed data (e.g.,  the pixel temperature, $T$, or its  spherical harmonic transform $a_{\ell m}$):
\begin{equation}
L(\mathbf{d} | \mathbf{p}) = \frac{1}{(2 \pi) ^{N/2} | \mathbf{C}| ^{1/2} } \exp \left( -\frac{1}{2}  \mathbf{d} \  \mathbf{C}^{-1}  \mathbf{d}^{t} \right)
\end{equation}
where $\mathbf{C}$ is the covariance of the data, and $\mathbf{p} $ is the set of model parameters.
$\mathbf{C} (\mathbf{p}) = \mathbf{S} (\mathbf{p}) + \mathbf{N} $, where $\mathbf{S}$ is the sky signal and $\mathbf{N}$ is the noise.
For single dish, full sky observations an isotropic signal is diagonal in the spherical harmonic space and can be described by an $m$-averaged power spectrum $C_{\ell}$ on each multipole, i.e., $S_{\ell m, \ell'm'} = \delta_{\ell \ell'} \delta_{mm'} C_{\ell}$.  The noise  is generally not diagonal.

All high-$\ell$ codes assume the data to be Gaussian distributed, however, they differ from {\tt XFaster} on the devised unbiased frequentist power spectrum estimator. All algorithms form quadratic functions of the data.  Pseudo-$C_{\ell}$ (PCL) codes estimate  $ C_{\ell} =\frac{1}{2 \ell+1} \sum_{m=-\ell}^{m=\ell} \frac{ | a_{\ell m}|^2} {4\pi}$ using fast spherical transforms.  Example of such algorithms are all Master type codes [\cite{master02,master01}] such as Romaster, cRomaster (\cite{cromaster}), Xpol (\cite{xpol,archeops_cl2}), and Crosspect  (\cite{varenna10}). Others calculate the angular correlation function $C(\theta)= \frac{1}{4 \pi} \sum_{\ell \ge 2} (2 \ell+1) C_{\ell} P_{\ell}(\cos{\theta})$ (where $P_{\ell}(\cos{\theta})$ is the Legendre polynomial and $\theta$ is an angular separation on the sky)  using fast  evaluation of the 2-point correlation function such as [Spice (\cite{spice}), Polspice (\cite{polspice}], though this is achieved using fast spherical transforms.  {\tt XFaster} (\cite{XFaster-like1,faster-boom01,xfaster-boom03}, instead uses the Quadratic Maximum likelihood (QML) expression as derived below.

It can be shown that the$(\ell,m)$ space maximum likelihood solution for the power spectrum is given by (\cite{DABJK00,hamilton97,tegmark97,tegmarkcosta01}):
\begin{equation}
C_{\ell}= \frac{1}{2}   \sum_{\ell'} {\cal{F}}_{\ell \ell'}^{-1} \>{\rm Tr}  \left[ \mathbf{C}^{-1}  \frac{\partial \mathbf{S}} {\partial C_{\ell'}} \  \mathbf{C}^{-1}  (\mathbf{C^{obs} } - \mathbf{N} )\right],
\label{MLCL}
\end{equation}
where $C_{\ell m, \ell' m'}^{obs} = a_{\ell m}^{obs} a_{\ell' m'} ^{obs *}$ is the quadratic in the coefficients of the expansion of the observed map and
${\bf{\cal{F}}}$ \footnote{It is traditional in the literature to use  ${\cal{F}}$ to represent the Fisher matrix and F  its ensemble average, however to avoid confusion with $F_{\ell}$ that denotes the filter or transfer function we resort to ${\cal{F}}$ for the ensemble average of the Fisher matrix.} is the  Fisher information matrix for the $C_{\ell}$ parameters (= curvature matrix in the ensemble average limit), given by:
\begin{equation}
{\cal{F}}_{\ell \ell'} = \frac{1}{2} \>{\rm Tr} \left[ \frac{\partial \mathbf{S}}{\partial C_{\ell}} \ \mathbf{C}^{-1}  \frac{\partial \mathbf{S}}{\partial C_{\ell'}} \ \mathbf{C}^{-1} \right].
\end{equation}
An iterative scheme can be employed to reach the maximum likelihood estimate for the $C_{\ell}$: start with an initial guess; compute $\mathbf{\cal{F}}$; evaluate~Eq~\ref{MLCL}. However, matrix operations become prohibitive for dimensions larger than a few thousand.

To circumvent this problem {\tt XFaster} recasts the estimator in the isotropic, diagonal approximations of the Master methods (\cite{master02,master01}) simplifying the calculations above. In this case the noise becomes a diagonalised Monte Carlo estimated bias and the signal is  summed into bands to average down the correlations induced by any reduced sky coverage. 
In this case for a single mode, say temperature alone, the covariance of the observed cut-sky modes is approximated by
\begin{equation}
\tilde{C}_{\ell m, \ell' m'} = \delta_{\ell \ell'} \delta_{m m'} ( \tilde{C}_{\ell} + \langle \tilde{N}_{\ell} \rangle ),
\label{Cll'}
\end{equation}
where $\tilde{C}_{\ell}$ is the cut-sky model power spectrum. 
 In our case the cut-sky power spectrum is parameterized through
a set of deviations $q_{\ell}$ from a template full-sky 'shape' spectrum $C_{\ell}^{(S)}$,
\begin{equation}
\tilde{C}_{\ell} = \sum_{\ell'} K_{\ell \ell'} B_{\ell'}^{2} F_{\ell'} C_{\ell'}^{(S)} q_{\ell'},
\label{Cl-cut}
\end{equation}
where $K_{\ell \ell'} $ is the coupling matrix due to the cut sky observations (see section~\ref{kernels}),
$F_{\ell}$ is a transfer or filter function accounting for the effect of pre-filtering the data both in time and spatial domain (see section~\ref{transfer}), and $ B_{\ell}$ expresses the effect of a finite beam.
For the case where the spectrum is parametrized in bands we consider band power deviations $q_{b}$
\begin{equation}
\tilde{C}_{\ell}= \sum_{b} q_{b} \tilde{C}_{b \ell}^{S} = \sum_{b} q_{b} \sum_{\ell'} K_{\ell \ell'} B_{\ell'}^{2} F_{\ell'} C_{\ell'}^{S} \chi_{b} (\ell'),
\label{Clcut}
\end{equation}
where
$\chi_{b} (\ell)$  is a binning function.
Assume for simplicity flat binning with $\chi_{b} (\ell)=1$ within the band and zero  outside.
 The ML solution for the $q_{b}$ is
 \begin{equation}
 q_{b} = \frac{1}{2} \sum_{b'} {\cal{F}}_{b b'}^{-1} \sum_{\ell} (2 \ell +1 ) g \frac{ \tilde{C}_{b' \ell}^{S}}{( \tilde{C}_{\ell} + \langle \tilde{N}_{\ell} \rangle )^{2} }(\tilde{C}_{\ell}^{obs} - \langle \tilde{N}_{\ell} \rangle),
 \label{qb}
 \end{equation}
where isotropy  reduces the trace as Tr $\rightarrow$ $\sum_{\ell} (2 \ell +1 ) g$, and
$g$  describes the effective degrees of freedom in the maps
 (which may be reduced by additional weighting of the modes such as filtering or pixel weighting), and is related to
the moments of the pixel weighting and the sky coverage, see \cite{master02}, and it can be further impacted by the binning of the power spectrum:
\begin{eqnarray}
g= f_{sky}\Delta\ell \frac{w_{2}^{2}}{w_{4}} & {\rm where} & Êf_{sky}w_{i}=\frac{1}{4 \pi} \int_{ 4 \pi} W^{i} (\hat{\n}) d \nÊ
\end{eqnarray}
where$\ W(\hat{\n})$ is the window or mask applied to the data, $f_{sky}w_{i}$ Êis the i-th moment of the arbitrary weighting
scheme, and $\Delta \ell$ is the width of the multipole bins.\\
The expression for the Fisher matrix is now given by
\begin{equation}
{\cal{F}}_{b b'} = \frac{1}{2} \sum_{\ell} (2 \ell +1) g \frac{\tilde{C}_{b\ell}^{S} \tilde{C}_{\ell b'}^{S}}{(\tilde{C}_{\ell} +  \langle \tilde{N}_{\ell} \rangle )^{2}},
\label{fisher}
\end{equation}

For polarization sensitive observations the data include the $I$, $Q$, and $U$ Stokes parameters. As mentioned before, the $I$ map is expanded in terms of spherical harmonics while the $Q$ and $U$ maps are expanded in spin-2 spherical harmonics, $_2 Y_{\ell m}$, to obtain E and B (grad-type or curl-type) polarization coefficients:

\beq
(Q \pm i U ) (\hat{\n}) = \sum_{\ell m} (a^{E}_{\ell m} \pm i a^{B}_{\ell m} ) _{\pm 2} Y_{\ell m}(\hat{\n})
\eeq

There are six spectra representing the six independent elements of the $3 \times 3$ covariance matrix of the
$(\ta_{\ell m}^{T}, \ta_{\ell m}^{E}, \ta_{\ell m}^{B})$ vector:

\begin{eqnarray}
\tC_{\ell}^{TT} &=& \sum_{b} q_{b}^{TT} \tC_{b \ell}^{(S) TT}+\tN_{\ell}^{TT}\\
\tC_{\ell}^{EE} &=&  \sum_{b}  (q_{b}^{EE}  {_{+}\tC}_{b \ell}^{(S) EE}+q_{b}^{BB}  {_{-}\tC}_{b \ell}^{(S)BB}) + \tN_{\ell}^{EE}\\
\tC_{\ell}^{BB} &=&  \sum_{b}  (q_{b}^{BB}  {_{+}\tC}_{b \ell}^{(S) BB}+q_{b}^{EE}  {_{-}\tC}_{b \ell}^{(S)EE}) + \tN_{\ell}^{BB}\\
\tC_{\ell}^{TE} &=& \sum_{b} q_{b}^{TE} \tC_{b \ell}^{(S) TE}+\tN_{\ell}^{TE}\\
\tC_{\ell}^{TB} &=& \sum_{b} q_{b}^{TB} \tC_{b \ell}^{(S) TB}+\tN_{\ell}^{TB}\\
\tC_{\ell}^{EB} &=& \sum_{b} q_{b}^{EB} \tC_{b \ell}^{(S) EB}+\tN_{\ell}^{EB}.
\end{eqnarray}

The template shape matrices are defined using the various coupling kernels for the different polarization types.  The transfer functions are distinct for each polarization type:

\begin{eqnarray}
\tC_{b \ell}^{(S) TT} &=& \sum_{\ell'} K_{\ell \ell'} F_{\ell'}^{TT} B_{\ell'}^{2} C_{\ell'}^{(S) TT} \chi_{b}(\ell),\\
\pm \tC_{b \ell}^{(S) EE} &=& \sum_{\ell'} {_{\pm}K}_{\ell \ell'} F_{\ell'}^{EE} B_{\ell'}^{2} C_{\ell'}^{(S) EE} \chi_{b}(\ell),\\
\pm \tC_{b \ell}^{(S) BB} &=& \sum_{\ell'} {_{\pm}K}_{\ell \ell'} F_{\ell'}^{BB} B_{\ell'}^{2} C_{\ell'}^{(S) BB} \chi_{b}(\ell),\\
\tC_{b \ell}^{(S) TE} &=& \sum_{\ell'} {_{\times}K}_{\ell \ell'} F_{\ell'}^{TE} B_{\ell'}^{2} C_{\ell'}^{(S) TE} \chi_{b}(\ell),\\
\tC_{b \ell}^{(S) TB} &=& \sum_{\ell'} {_{\times}K}_{\ell \ell'} F_{\ell'}^{TB} B_{\ell'}^{2} C_{\ell'}^{(S) TB} \chi_{b}(\ell),\\
\tC_{b \ell}^{(S) EB} &=& \sum_{\ell'} ({_{+}K}_{\ell \ell'} - {_{-}K}_{\ell \ell'}) F_{\ell'}^{EB} B_{\ell'}^{2} C_{\ell'}^{(S) TB} \chi_{b}(\ell).
\end{eqnarray}

For simplicity the beam $B_{\ell}$ is assumed independent of polarization.  (In principle, it could also be treated distinctly for each of the $T$, $E$, and $B$ modes.)
The mask coupling kernels, $K_{\ell \ell'}$, and the two additional polarization mask coupling kernels, $_{\pm}K_{\ell \ell'}$, $_{\times}K_{\ell \ell'}$ are defined in section~\ref{kernels}.

Extending the above formalism to polarization, the {\tt XFaster}  estimator takes a matricial form, implemented trivially since the matrix $C$ is now block diagonal:\\
$\tilde{\mathbf{C}} \rightarrow diag( \tilde{\mathbf{D}}_{\ell_{min}}, \tilde{\mathbf{D}}_{\ell_{min}+1},\dots, \tilde{\mathbf{D}}_{\ell_{max}}) $,
where each multipole's covariance is a $3 \times 3$ matrix:
\begin{equation}
\tilde{\mathbf{D}}_{\ell} =
\left(
\begin{array}{ccc}
\tilde{C}_{\ell}^{TT} & \tilde{C}_{\ell}^{TE} & \tilde{C}_{\ell}^{TB} \\
\tilde{C}_{\ell}^{TE} & \tilde{C}_{\ell}^{EE} & \tilde{C}_{\ell}^{EB}  \\
\tilde{C}_{\ell}^{TB} &   \tilde{C}_{\ell}^{EB} & \tilde{C}_{\ell}^{BB} \\
\end{array}
\right),
\end{equation}
Similarly its inverse is a block diagonal of the inverses of  $\tilde{\mathbf{D}}_{\ell}$ matrices and therefore simple to compute. The noise covariance matrix is also of this form, the $\tilde{N}_{\ell}^{X Y}$ in each block diagonal is obtained by noise only Monte Carlo simulations.
The band power deviations, $q_{b}$, take now the following form:
\begin{equation}
 q_{b} = \frac{1}{2} \sum_{b'} {\cal{F}}_{b b'}^{-1} \sum_{\ell} (2 \ell +1 ) g \>{\rm Tr}  \left [ { \tilde{\mathbf{D}}_{\ell}}^{-1}  \frac{\partial \mathbf{\tilde{S}}}{\partial q_{b'}} \  { \tilde{\mathbf{D}}_{\ell}}^{-1}  \left({\tilde{\mathbf{D}}_{\ell}}^{obs}  - \tilde{\mathbf{N}}_{\ell} \right) \right],
 \label{qb-p}
\end{equation}
and the Fisher matrix is now given by:
\begin{equation}
{\cal{F}}_{b b'} = \frac{1}{2} \sum_{\ell} (2 \ell+1) g \>{\rm Tr} \left[ \tilde{\mathbf{D}}_{\ell}^{-1}  \frac{\partial \mathbf{\tilde{S}}_{\ell}}{\partial q_{b}} \ \mathbf{\tilde{D}}_{\ell}^{-1}  \frac{\partial \mathbf{\tilde{S}}_{\ell}}{\partial q_{b'}} \right].
\end{equation}
where the band index, b, spans bands in all polarization types.
The derivatives of the signal matrices with respect to the deviations $q_{b}$ are given by:

\begin{eqnarray}
\frac{\partial \mathbf{\tilde{S}}}{\partial q_{b}} \vert_{b :: TT} &=& \left(
\begin{array}{ccc}
\tilde{C}_{b \ell}^{(S) TT} & 0 & 0 \\
0 & 0 & 0  \\
0 & 0 & 0 \\
\end{array}
\right),
\frac{\partial \mathbf{\tilde{S}}}{\partial q_{b}}|_{b :: TE} = \left(
\begin{array}{ccc}
0 & \tilde{C}_{b \ell}^{(S) TE}  & 0 \\
\tilde{C}_{b \ell}^{(S) TE}  & 0 & 0  \\
0 & 0 & 0 \\
\end{array}
\right)\\
\frac{\partial \mathbf{\tilde{S}}}{\partial q_{b}}|_{b :: EE} &=& \left(
\begin{array}{ccc}
0 & 0 & 0 \\
0 & {_{+}\tilde{C}}_{b \ell}^{(S) EE} & 0  \\
0 & 0 & {_{-}\tilde{C}}_{b \ell}^{(S) EE} \\
\end{array}
\right),
\frac{\partial \mathbf{\tilde{S}}}{\partial q_{b}}|_{b :: BB} = \left(
\begin{array}{ccc}
0 & 0  & 0 \\
0  &  {_{-}\tilde{C}}_{b \ell}^{(S) BB} & 0  \\
0 & 0 & {_{+}\tilde{C}}_{b \ell}^{(S) BB} \\
\end{array}
\right)\\
\frac{\partial \mathbf{\tilde{S}}}{\partial q_{b}}|_{b :: TB} &=& \left(
\begin{array}{ccc}
0 & 0 & \tilde{C}_{b \ell}^{(S) TB} \\
0 & 0 & 0  \\
\tilde{C}_{b \ell}^{(S) TB}& 0 & 0\\
\end{array}
\right),
\frac{\partial \mathbf{\tilde{S}}}{\partial q_{b}}|_{b :: EB} = \left(
\begin{array}{ccc}
0 & 0  & 0 \\
0  & 0 & \tilde{C}_{b \ell}^{(S) EB}  \\
0 & \tilde{C}_{b \ell}^{(S) EB} & 0 \\
\end{array}
\right).
\end{eqnarray}
These derivatives include contributions  from both $(+)$ and $(-)$ kernels in the case EE and BB due to the geometrical leakage.

\begin{figure}
\begin{center}
\includegraphics[width=13cm]{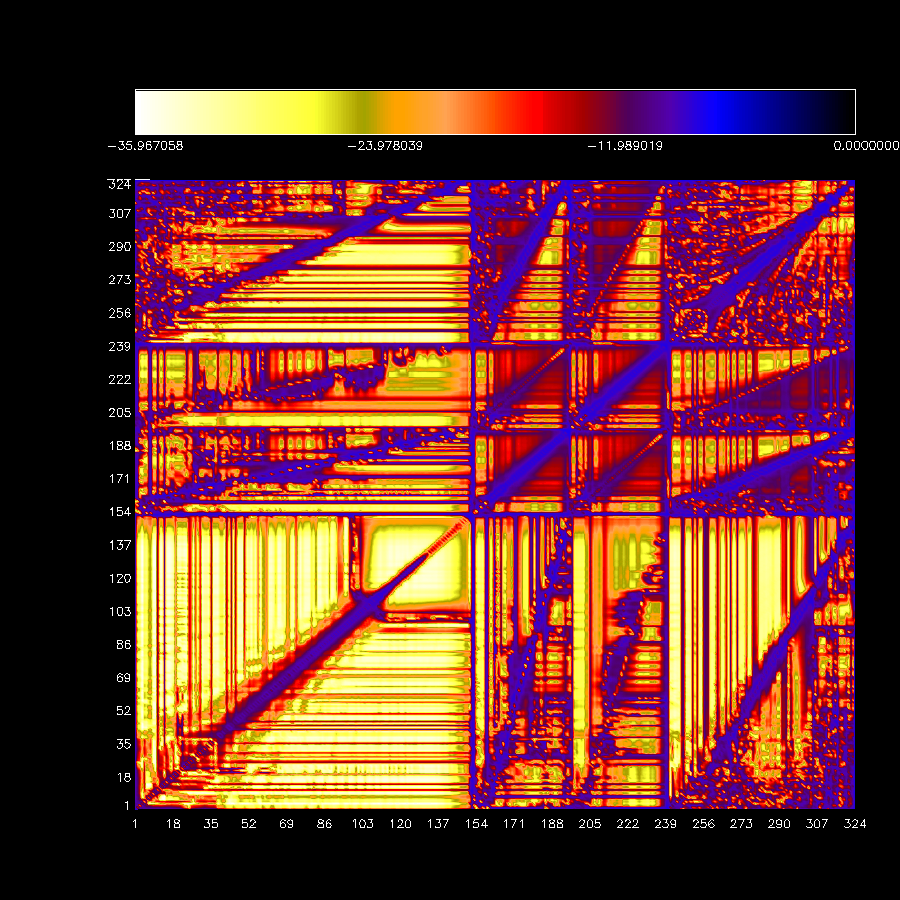}
\caption{ Logarithm of the absolute value of the `normalized' (ie set to 1 at the maximum value) inverse of the Fisher matrix (covariance matrix) for phase2, symmetric beam case.  TT, EE, BB and TE modes are displayed sequentially from bottom left-hand side corner to the upper right-hand side corner along the diagonal.}
\label{invfisher}
\end{center}
\end{figure}

This estimator makes use of the Monte Carlo pseudo-$C_{\ell}$ formalism of Master methods to estimate noise bias and linear filter functions.
It requires noise only Monte Carlo simulations to estimate the noise bias and signal only Monte Carlo simulations to estimate the filter function. Contrary to the conventional pseudo-$C_{\ell}$ based methods it does not requires signal+noise Monte Carlo simulations to estimate the uncertainty (variance) of the band powers, which is given by the Fisher matrix instead, a byproduct of the method.
The Fisher matrix is computed self-consistently and runs over all band powers and polarizations. It takes into account all the correlations in the approximation used (coupling kernel and diagonal noise bias) and can therefore be used to compute, self-consistently, all ancillary information required in the estimation process, correlations, window functions, etc. In Figure~\ref{invfisher} we plot the inverse of the fisher matrix for phase2, symmetric beam case. Note that the Fisher matrix is not diagonal.

Furthermore {\tt XFaster}  can estimate both auto-spectra and cross-spectra jointly, using the full covariance of the $a_{lm}$'s,  via a multiple-map analysis.

As mentioned above, the noise is generally not diagonal. For Planck, the noise is white to good approximation at small angular scales; however, at large angular scales instrumental characteristics such as $1/f$ noise and thermal fluctuations combine with the scan strategy to produce significant off-diagonal correlations in the noise. Therefore, the {\tt XFaster} approximation is not optimal at low-$\ell$ multipole range. We show in Sections~\ref{like-results} and \ref{par-results}  that {\tt XFaster} is a very good approximation for $ \ell > 30$.  We have no intention of using {\tt XFaster} for Planck at low-$\ell$. Instead, we will combine one of the codes adequate at low-$\ell$ (listed in Section~\ref{intro}) with {\tt XFaster} (or another high-$\ell$ estimator) into a hybrid estimator of the power spectrum that covers the entire multipole range.

Previous CMB experiments have customarily binned power spectra in multipole bands. The main reason for this is that it ``enhances'' the signal-to-noise ratio of the spectra.  It  also averages down correlations due to the reduced sky coverage at the same time as the coupling matrices due to the cut-sky  attempt to correct the cut-sky effect.
The {\tt XFaster} power spectrum can be computed multipole by multipole, i.e., for each $\ell$, or in multipole bands. The band power spectra are given in Section~\ref{pse-results}. To estimate cosmological parameters we can use band power spectra with any high-$\ell$ likelihood approximation, e.g., with the Offfset Lognormal Bandpower likelihood presented in \cite{GRLike09}.
 However, as the {\tt XFaster} likelihood is estimated multipole by multipole we can bypass the band power spectrum estimation step and estimate parameters directly from the maps (via its raw pseudo-$C_{\ell}$). Slices of the {\tt XFaster} likelihood and parameter constraints are given in Section~\ref{like-results} and Section~\ref{par-results}.

As kernels and transfer (filter) functions are an important ingredient in power spectrum estimation we describe next how they are computed within {\tt XFaster}  approach.


\subsubsection{Kernels \label{kernels}}
\label{kernels}

The effect of masking the sky is to produce a power spectrum that is a linear combination of the full sky power spectrum multipoles on the sky.
The coupling matrix due to the cut sky observations, $K_{\ell \ell'} $, encodes this effect, it only depends on the geometry of the mask or window and is easily computable.
\newline
Considering a  window function $W(\hat{\n})$, and ignoring the effects of beam convolution and filtering effects due to any pre-processing of the timelines, the ensemble averages of the cut-sky $\VEV{\tC_{\ell}} $ and the full-sky angular power  spectrum $ \VEV{C_{\ell}}$ can be related by:

  \beq
	\VEV{\tC_{\ell}} = \sum_{\ell'} K_{\ell\ell'} \VEV{C_{\ell'}},
	\label{eq:coupling_final}
\eeq
with coupling matrix, $K_{\ell\ell'}$ given by:
\beq
K_{\ell\ell'} = \frac{2 \ell'+1}{4\pi} \sum_{\ell"} J(\ell,\ell',\ell";0,0,0)^2  W_{\ell"}^{2}
\eeq
 where $J(l,l',l";0,0,0) = \wjjj{\ell}{\ell'}{\ell''}{0}{0}{0}$ is the 3j symbol, and $W_{\ell}^{2}$ is the power spectrum of the window function $W(\hat{\n})$, that is
 $W_{\ell}^{2} = (2 \ell + 1) {\cal{W}}_{\ell}$ with:

\begin{eqnarray}
 {\cal{W}}_{\ell} = \frac{1}{2 \ell + 1} \sum_{m} | W_{\ell m} | ^{2} & {\rm and} &  W_{\ell m} = \int d \n W(\hat{\n}) Y^{*}_{\ell, m} (\hat{\n}) \\
 {\cal{W}}_{0}=W_{0}^{2}= 4 \pi f_{sky}^{2} w_{1}^{2}  &  {\rm and }&  \sum_{\ell \ge 0} W_{\ell}^{2} = \sum_{\ell \ge 0}  (2 \ell + 1) {\cal{W}}_{\ell}=4 \pi f_{sky} w_{2}
\end{eqnarray}
Hence
\beq
K_{\ell\ell'} = \frac{2 \ell'+1}{4\pi} \sum_{\ell''} (2 \ell''+1) \mathcal{W}_{\ell''}  \wjjj{\ell}{\ell'}{\ell''}{0}{0}{0}^{2}
\eeq

Extending the above to polarized data we consider the additional polarization mask coupling kernels defined as follows:

\begin{eqnarray}
_{\pm} K_{\ell \ell'} &=& \frac{2 \ell' +1}{16 \pi} \sum_{L} (2L+1) \mathcal{W}_{L} \wjjj{\ell}{\ell'}{L}{2}{-2}{0}^{2} \left(1 \pm(-1)^{ \ell+\ell'+L} \right) \\
_{\times}  K_{\ell \ell'} &=& \frac{2 \ell' +1}{8 \pi} \sum_{L} (2L+1) \mathcal{W}_{L} \wjjj{\ell}{\ell'}{L}{2}{-2}{0} \wjjj{\ell}{\ell'}{L}{0}{0}{0} \left(1+ (-1)^{ \ell+ \ell'+L} \right)
\end{eqnarray}

These kernels account for the leakage of power between E and B modes induced by the usage of the full-sky $_{\pm 2} Y_{\ell m} (\hat{\n})$ basis on a cut-sky.

\begin{figure}
\begin{center}
\hspace*{+1cm}
\hbox{
\includegraphics[width=7cm]{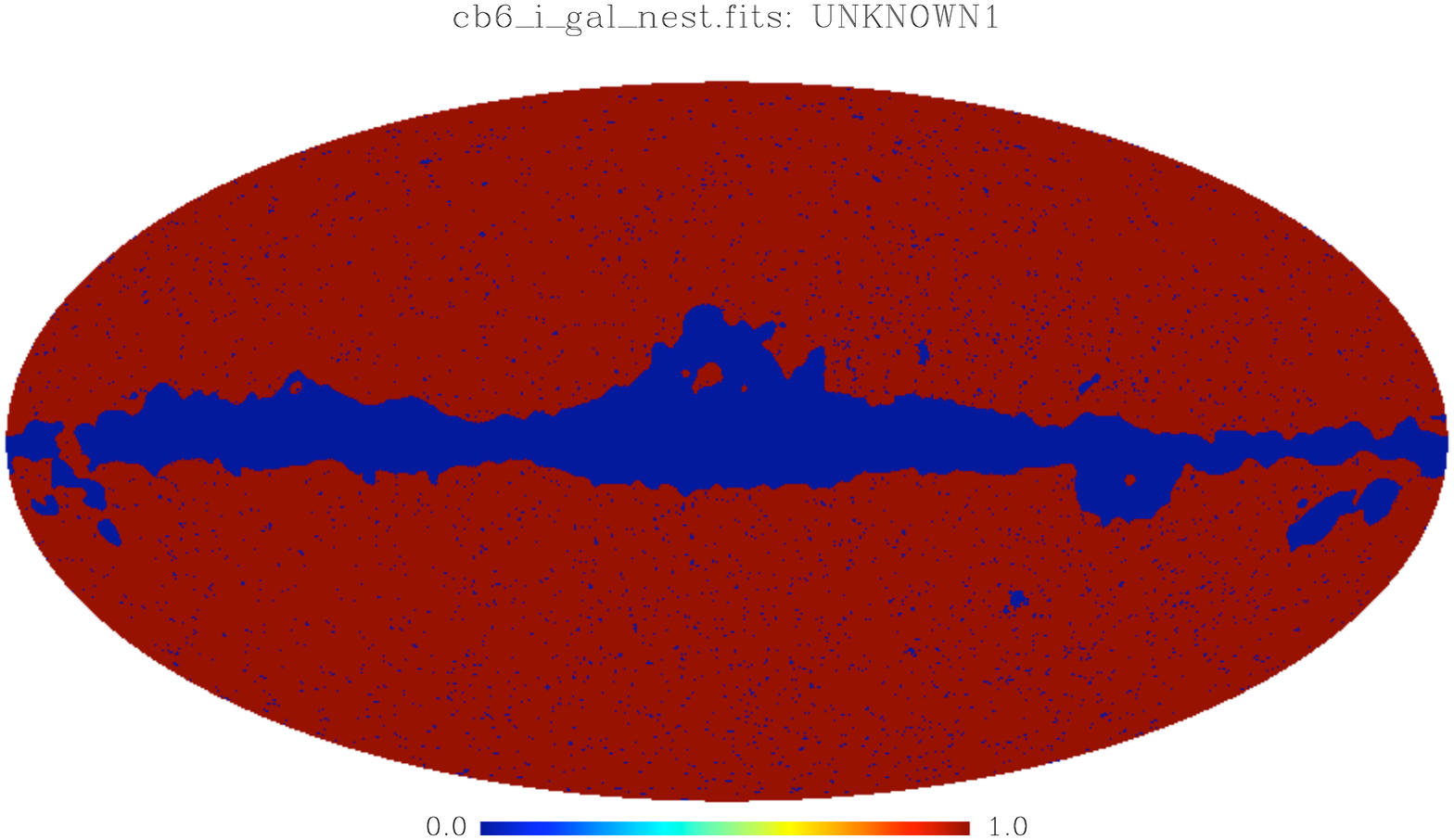}
\includegraphics[width=7cm]{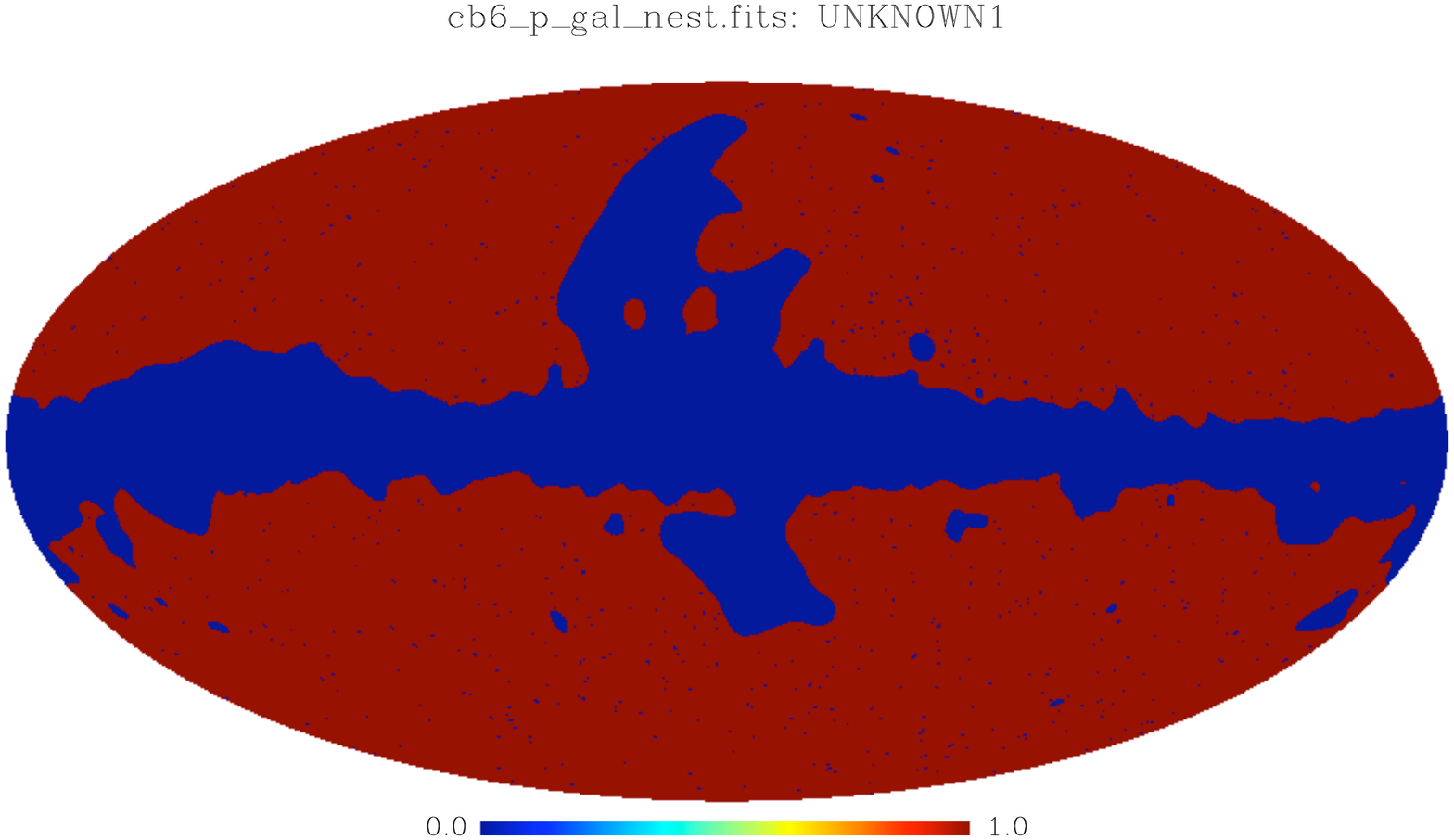}
}
\caption{Mask for Temperature (left hand side) and polarization (right hand side) used in Phase~2. The $f_{\rm sky} \simeq 0.85$ for Temperature and $f_{\rm sky} \simeq 0.73$ for Polarization.}
\label{mask-kernel}
\end{center}
\end{figure}

In Figure~\ref{mask-kernel} we plot the masks for temperature and polarization used in Phase~2 (see section~\ref{planck-sims}).

%


\subsubsection{Transfer or Filter functions \label{transfer}}
\label{transfer}

To compute $F_{\ell}$ we start with Eq.~\ref{Clcut} for the signal only Monte Carlo simulations, and  as  $F_{\ell}$ is assumed to be smooth over each multipole bin we move  $F_{\ell}$  out of the summation for each bin $b$, to get:
\begin{equation}
\tilde{C}_{\ell}= \sum_{b} q_{b} \tilde{C}_{b \ell}^{S} = \sum_{b} q_{b} F_{b}  \sum_{\ell'} K_{\ell \ell'} B_{\ell'}^{2} C_{\ell'}^{S} \chi_{b} (\ell')
\label{Fb}
\end{equation}
and proceed with the iterative scheme as one would to estimate $q_{b}$ but now  we estimate the transfer function, $F_{b}$ instead. This is achieved by fixing $q_{b} =1$, varying $F_{b}$ and considering the signal only Monte Carlo simulations.
 \begin{equation}
 F_{b} = \frac{1}{2} \sum_{b'} {\cal{F}}_{b b'}^{-1} \sum_{\ell} (2 \ell +1 ) g \frac{ \tilde{C}_{b' \ell}^{S}}{( \tilde{C}_{\ell} + \langle \tilde{N}_{\ell} \rangle )^{2} }\langle \tilde{S}_{\ell}  \rangle
 \label{qbFb-t}
 \end{equation}
where $\langle \tilde{S}_{\ell}  \rangle$ is the average of the signal only Monte Carlo simulations.

Extending to polarization we have:

\begin{equation}
 F_{b} = \frac{1}{2} \sum_{b'} {\cal{F}}_{b b'}^{-1} \sum_{\ell} (2 \ell +1 ) g Tr  \left [ { \tilde{\mathbf{D}}_{\ell}}^{-1}  \frac{\partial \mathbf{\tilde{S}}}{\partial q_{b'}} \  { \tilde{\mathbf{D}}_{\ell}}^{-1}   \VEV{\tilde{\mathbf{S}}_{\ell}}  \right],
 \label{qbFb-p}
\end{equation}


\subsection{{\tt XFaster} likelihood estimator \label{like}}
\label{like}

A very attractive feature of the {\tt XFaster}  power spectrum estimator is that is naturally provides a likelihood, i.e., the Probability of the observed cut sky data given the model.
In the {\tt XFaster}  approximation, considering only one mode (say Temperature alone), the likelihood takes the following form, up to a constant (where $\tilde{A}$ means $A$ estimated on the cut-sky):

\begin{equation}
\ln L= -\frac{1}{2} \sum_{\ell} g (2 \ell +1) \left( \frac{\tilde{C_{\ell}}^{obs}}{ \left( \tilde{C_{\ell}} + \left< \tilde{N_{\ell}}\right> \right)} + \ln \left( \tilde{C_{\ell}} + \left< \tilde{N_{\ell}}\right> \right) \right)
\label{XFaster-like}
\end{equation}
where $\tilde{C}_{\ell}$ is the cut-sky model power spectrum given by Eq.~\ref{Clcut} in Section~\ref{pse}, for the case where the spectrum is parametrized in bands we consider band power deviations $q_{b}$.

Extending to Temperature and Polarization we have:
\begin{equation}
\ln L= -\frac{1}{2} \sum_{\ell} g (2 \ell +1) \left(Tr \left ({\tilde{\mathbf{D}_{\ell}^{obs}}}{ \left( \tilde{\mathbf{D}_{\ell}} + \left< \tilde{\mathbf{N}_{\ell}}\right> \right)^{-1}} \right)+ \ln  \left| \tilde{\mathbf{D}_{\ell}} + \left< \tilde{\mathbf{N}_{\ell}}\right> \right| \right)
\label{XFaster-like}
\end{equation}
Where $\tilde{\mathbf{N}_{\ell}}$ and $\tilde{\mathbf{D}_{\ell}}$ are given in section~\ref{pse}.

An interesting point to note is that {\tt XFaster} likelihood follows intuitively from the usual full-sky ideal case exact likelihood (an inverse Gamma distribution for temperature alone and an inverse Wishart distribution for temperature + polarization (see e.g., \cite{GRLike09}).

Here we use one dimensional slices as an approximation to investigate the non-Gaussianity of the likelihood.  One samples in each deviation $q_{b}$ direction individually around the maximum likelihood solution $q_{b}^{*}$. This approximation is adequate if the band powers are not heavily correlated. Note that the likelihood slices are estimated along the bands and not along each $\ell$, and hence will be affected by the binning procedure.
To compare {\tt XFaster} likelihood to other approximations we make use of slices computed along the bandpower spectrum deviations, $q_{b}$.
Such likelihood slices for the 70\,GHz observed map are plotted in Section~\ref{like-results}.

When estimating parameters with {\tt XFaster} likelihood, by default estimated multipole by multipole, we do not make use of the bandpower spectra. It is in this sense that {\tt XFaster} likelihood allows to go straight from the maps to parameters bypassing the band power spectrum step. It only requires the raw pseudo-$C_{\ell}$ of the observations plus the kernel and transfer function to relate the cut-sky pseudo-$C_{\ell}$ to the full-sky $C_{\ell}$.

\subsubsection{Window functions \label{window}}
\label{window}

To compare the theoretical power spectrum to the observed power spectrum and to estimate parameters, we must construct an operator for obtaining theoretical bandpowers
from model power spectra $C_{\ell}^{T}$.  Following \cite{DABJK00} we define a logarithmic integral
\beq
\mathcal{I}[f_{\ell}] = \sum_{\ell}  \frac{\ell + \frac{1}{2}}{\ell (\ell + 1)} f_{\ell},
\eeq
which is used to calculate the expectation values for the deviations $q_{b}$(when a shape model, $C_{\ell}^{S}$ is considered), or bandpowers $C_{b}$ (when $C_{\ell}^{S}$ is assumed to be flat).
\beq
\langle q_{b} \rangle = \frac{\mathcal{I}\left[ W_{\ell}^{b} \mathcal{C}_{\ell} \right] }{\mathcal{I} \left[ W_{\ell}^{b} \mathcal{C}_{\ell}^{(S)} \right] } \qquad
\langle C_{b} \rangle = \frac{\mathcal{I} \left[ W_{\ell}^{b} \mathcal{C}_{\ell} \right] }{\mathcal{I} \left[ W_{\ell}^{b} \right] },
\eeq
where $W_{\ell}^{b}$ is the band power window function, and $\mathcal{C}^{(S)} = \ell(\ell + 1) C_{\ell}^{(S)} / 2 \pi$.

We define normalized window functions to be
\beq
\mathcal{I} \left [ W_{\ell}^{b} \mathcal{C}_{\ell}^{(S)} \right ] = 1.
\eeq
By taking the ensemble average limit of  Eq.~ (\ref{qb}) and using the fact that
\beq
\langle ( \tC_{\ell}^{obs} - \tN_{\ell} ) \rangle \rightarrow \tC_{\ell}
\eeq
we obtain
\beq
W_{\ell}^{b} = \frac{4 \pi}{(2 \ell + 1) } \sum_{b'} {\cal{F}}_{b b'}^{-1} \sum_{\ell'} g (2 \ell' +1 ) \frac{\tC_{b' \ell'}^{(S)}}{(\tC_{\ell'} + \langle \tN_{\ell'}\rangle ) ^{2}} K_{\ell \ell'} F_{\ell} B_{\ell}^{2}.
\eeq
Extending to polarization:
\be
W_{\ell}^{b} =  \frac{4 \pi}{(2 \ell + 1) } \sum_{b'} {\cal{F}}_{b b'}^{-1} \sum_{\ell'} g (2 \ell' +1 ) {\rm Tr}  \left[ \vW_{b' \ell'}  \vK_{\ell'} \right]
\ee
where $\vW_{b \ell}  = { \tilde{\mathbf{D}}_{\ell}}^{-1}  \frac{\partial \mathbf{\tilde{S}}}{\partial q_{b}} {\tilde{\mathbf{D}}_{\ell}}^{-1}$, and $\vK_{\ell}$ gives the cut-sky response to the individual full-sky multipoles:

\begin{equation}
\vK_{\ell}=
\left(
\begin{array}{ccc}
K_{\ell' \ell} F_{\ell}^{TT} B_{\ell}^2 &  _{\times}  K_{\ell' \ell} F_{\ell}^{TE} B_{\ell}^2 &  _{\times}  K_{\ell' \ell} F_{\ell}^{TB} B_{\ell}^2 \\
_{\times}  K_{\ell' \ell} F_{\ell}^{TE} B_{\ell}^2 &   _{+} K_{\ell' \ell} F_{\ell}^{EE} B_{\ell}^2 + _{-} K_{\ell' \ell} F_{\ell}^{BB} B_{\ell}^2 & \left( _{+} K_{\ell' \ell} - _{-} K_{\ell' \ell} \right) F_{\ell}^{EB} B_{\ell}^2 \\
_{\times}  K_{\ell' \ell} F_{\ell}^{TB} B_{\ell}^2 &   \left( _{+} K_{\ell' \ell} - _{-} K_{\ell' \ell} \right) F_{\ell}^{EB} B_{\ell}^2 &  _{+} K_{\ell' \ell} F_{\ell}^{BB} B_{\ell}^2 + _{-} K_{\ell' \ell} F_{\ell}^{EE} B_{\ell}^2 \\
\end{array}
\right).
\end{equation}
These window functions were used and compared to the top hat window functions in \cite{GRLike09} using the {\tt XFaster}  bandpower spectra and the Offset Lognormal Bandpower likelihood.  However as the {\tt XFaster} likelihood is estimated for each $\ell$, a comparison of observed to theoretical power spectrum does not make use of such windows. Instead the raw pseudo-$C_{\ell}$ of the observations, the kernels, and transfer or filter functions are all that is required for such comparison (see section~\ref{par-results}).

\subsection{The algorithm}
\label{pse-like-comp}

The power spectrum is estimated by the following procedure:

\begin{itemize}

\item Generate Monte Carlo simulations of time-ordered data (TOD) for both signal and noise.  The noise must have the same characteristics as the observed data, and in practice must be determined from the observed data.  The simulated signal, on the other hand, can be almost anything, as it is simply a tracer of the effects of time and spatial domain filtering in the process, and used to calculate the transfer function $F_\ell$.  In practice, it is convenient to use an approximate model of the CMB to generate the signal.

\item Make maps of the TOD using the same mapmaking code as used for the observations.

\item Estimate the pseudo-$C_{\ell}$ spectra and the spherical harmonic coefficients $a_{\ell m}$ from the signal only maps to get the transfer function $F_{\ell}$.

\item Estimate the pseudo-$C_{\ell}$ spectra from the noise only maps to compute the noise bias $ \langle \tilde{N}_{\ell} \rangle $.  The pseudo-spectra can be computed with anafast of {\tt HEALPix} package (\cite{healpix}) when the masks are the same for temperature and polarization, otherwise we use a specific code from the suite of {\tt XFaster} modules.

\item Iterate Equations \ref{qb} and \ref{fisher} to obtain an estimate of $q_{b}$.
In the  diagonal, isotropic approximation of {\tt XFaster}, the computational cost of the iterative estimator is very small compared with that of the TOD generation and map-making stages.

\item The iterative estimator yields the Fisher information matrix automatically. An estimate of the band power covariance is given by $\mathbf{\cal{F}}^{-1}$,  therefore we automatically get the uncertainty on  the estimator.

\end{itemize}

Large ensembles of signal+noise simulations are not required to estimate the band power covariance matrix as in the Master procedure, cutting the cost of Monte Carlo simulations by $1/3$. Furthermore the covariance is not biased by an assumed model (which at very least requires the Master procedure to be run twice to be close to unbiased errors).

As mentioned in Section~\ref{pse} {\tt XFaster}  can estimate both auto-spectra and cross-spectra jointly, using the full covariance of the $a_{lm}$,  via a multiple-map analysis.

\subsubsection{Computational scaling}

The overall scaling for {\tt XFaster}  without accounting for the signal and noise  Monte Carlo simulations should go as $\ell_{max} (n_{maps} \times n_{pol})^3$ for the internal Fisher calculation, where $n_{pol}$ is either 1 or 3, with a further scaling of $(n_{bins})^3$ for the outer iteration step.

Currently the code is not optimised for speed.  It could be sped up substantially by parallelizing the Fisher computation and would then scale linearly with number of processors.

Approximate times for a single CPU for Phase2, with CTP binning and using 30 Fisher iterations are as follows:

\begin{itemize}
\item $a_{lm}$ and $C_{\ell}$ from the 100 Monte Carlo simulated maps: $\simeq  8$ hours
($\simeq$ 5 minutes each)
\item kernel: 30 minutes
\item transfer function: 30 minutes (less if one relaxes the binning)
\item power spectrum: 1 hour
\item average mode to check for possible bias: 15--20 minutes
\end{itemize}

\section{Results}
\label{results}

\subsection{Results: Power spectrum \label{pse-results}}
\label{pse-results}

We estimated transfer (filter) functions, kernels, and the power spectrum for  the observed map described in Section~\ref{planck-sims}. We also computed the power spectrum for the  average of the signal+noise simulated maps. This average mode run  checks for possible biases of the power spectrum estimator itself. In principle, the estimator if unbiased should follow closely the input signal $C_{\ell}$ model used to generate the signal simulations.

\begin{figure}
\begin{center}
\vbox{
\hbox{
\includegraphics[width=9cm]{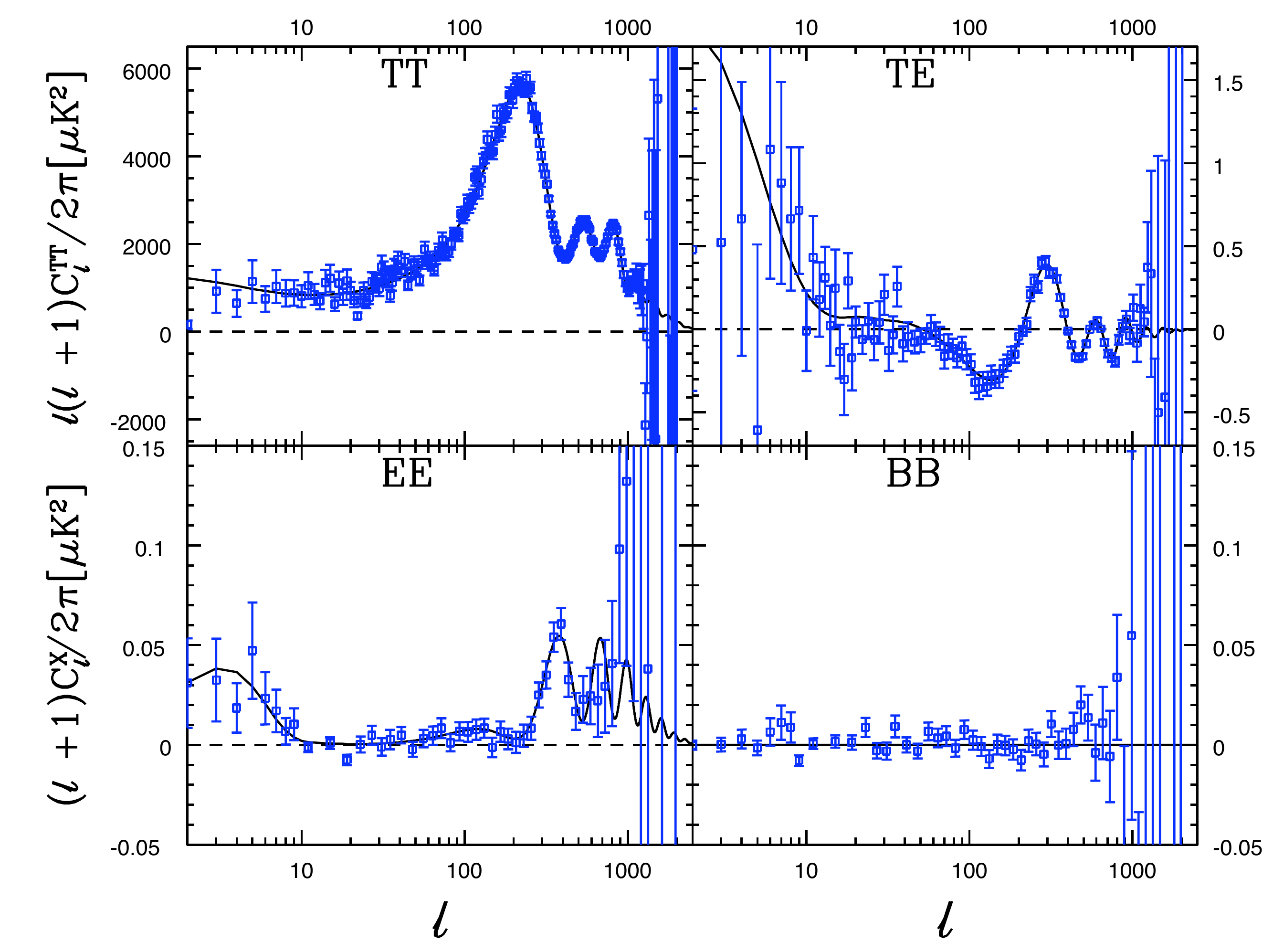}
\includegraphics[width=9cm]{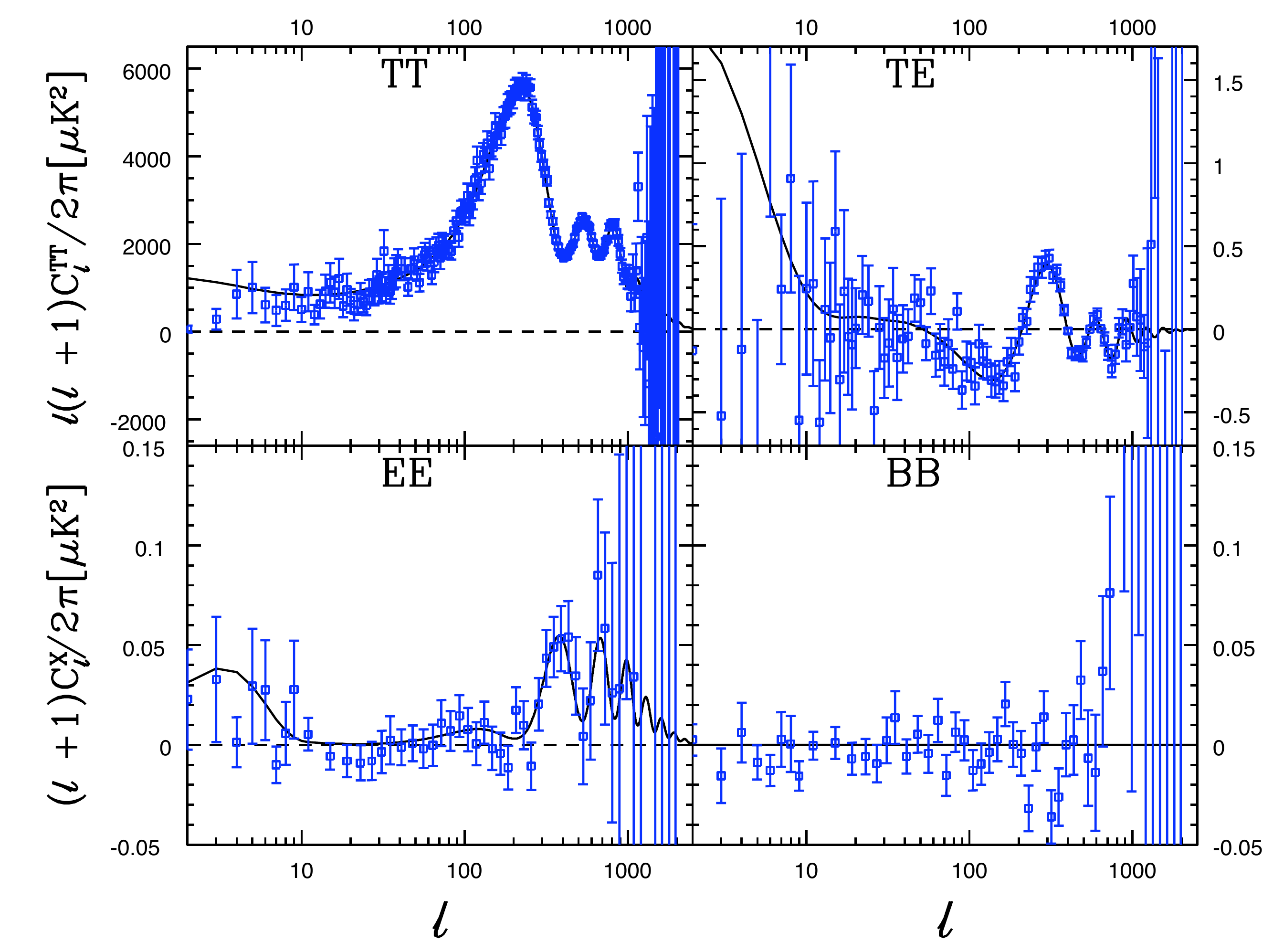}
}
\hbox{
\includegraphics[width=9cm]{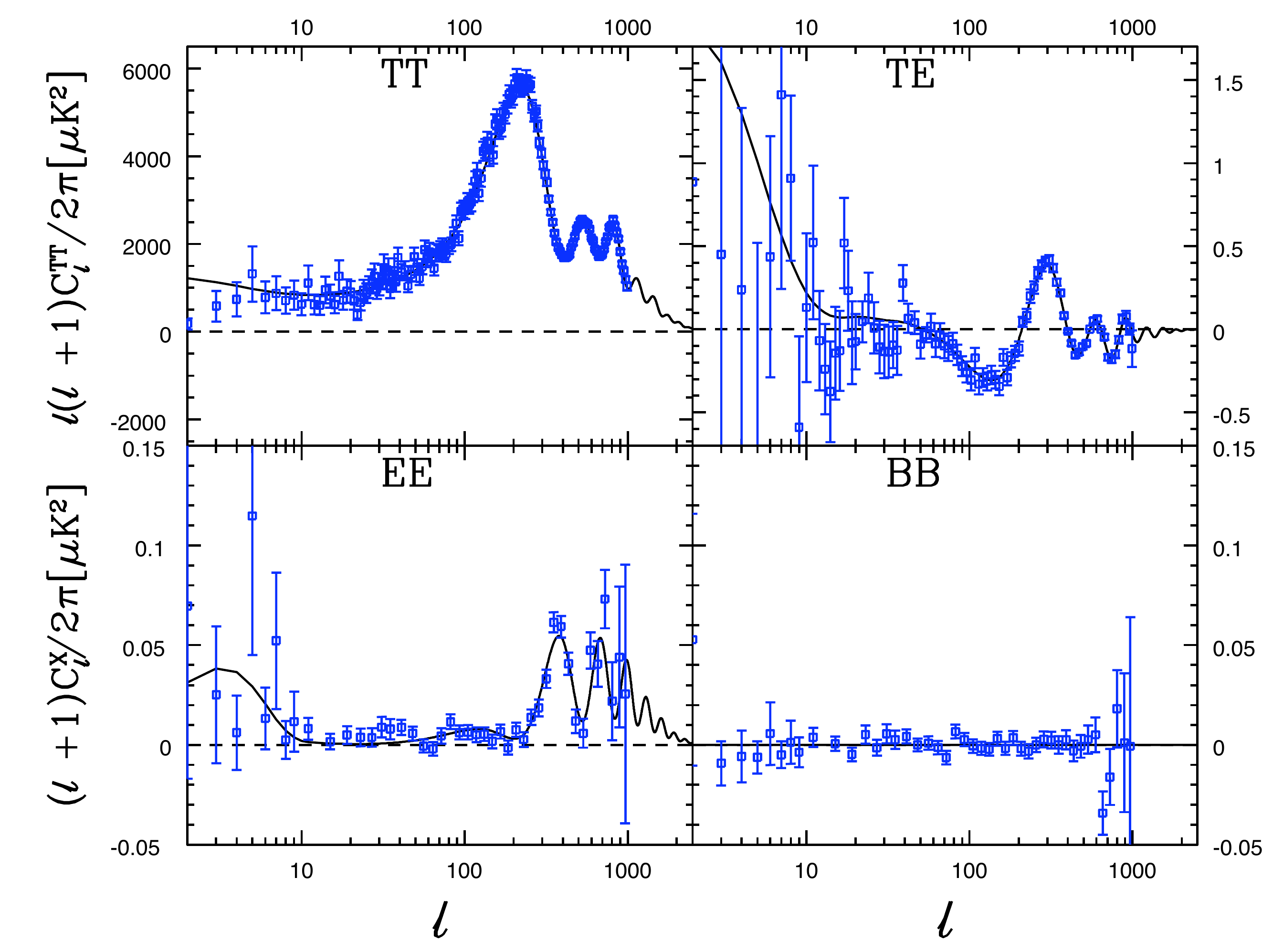}
\includegraphics[width=9cm]{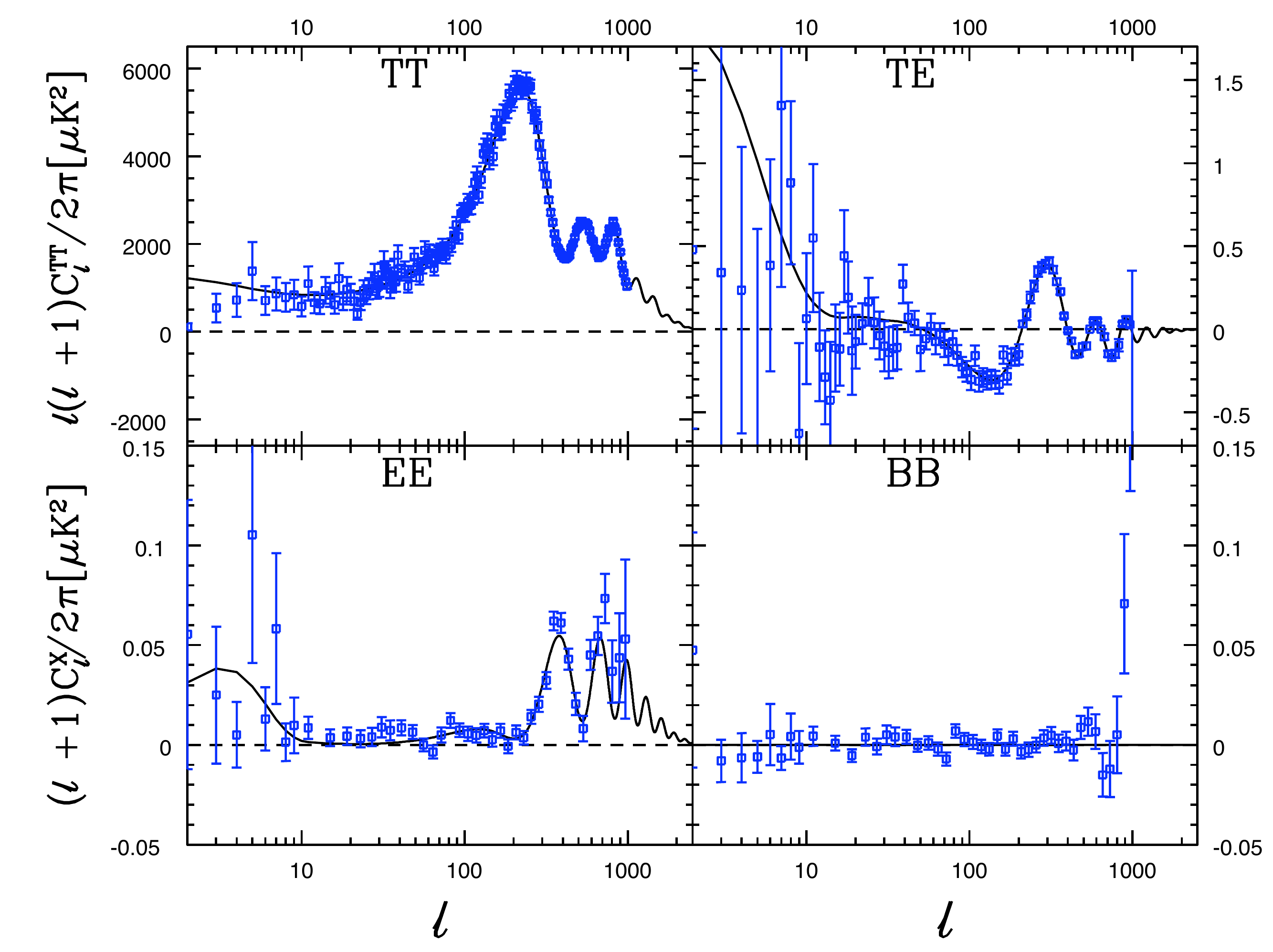}
}}
\caption{ Power spectrum estimated with {\tt XFaster} and $1 \sigma$ error bars. Top row---Phase~1a (left hand side) and Phase~1b (right hand side) for map generated with a quadruplet of detectors. Bottom row---Phase~2a symmetric beam (left hand side) and  Phase~2b asymmetric beam (right hand side) for map generated with all twelve detectors,.
The plot displays the estimated power spectrum (blue) of the observed map, overplotted with the $C_{\ell}$ fiducial model used as input in Phase~2 signal simulations, first year WMAP+CBI+ACBAR best fit model (black).}
\label{pse-all}
\end{center}
\end{figure}

\begin{figure}
\begin{center}
\hbox{
\includegraphics[width=6cm]{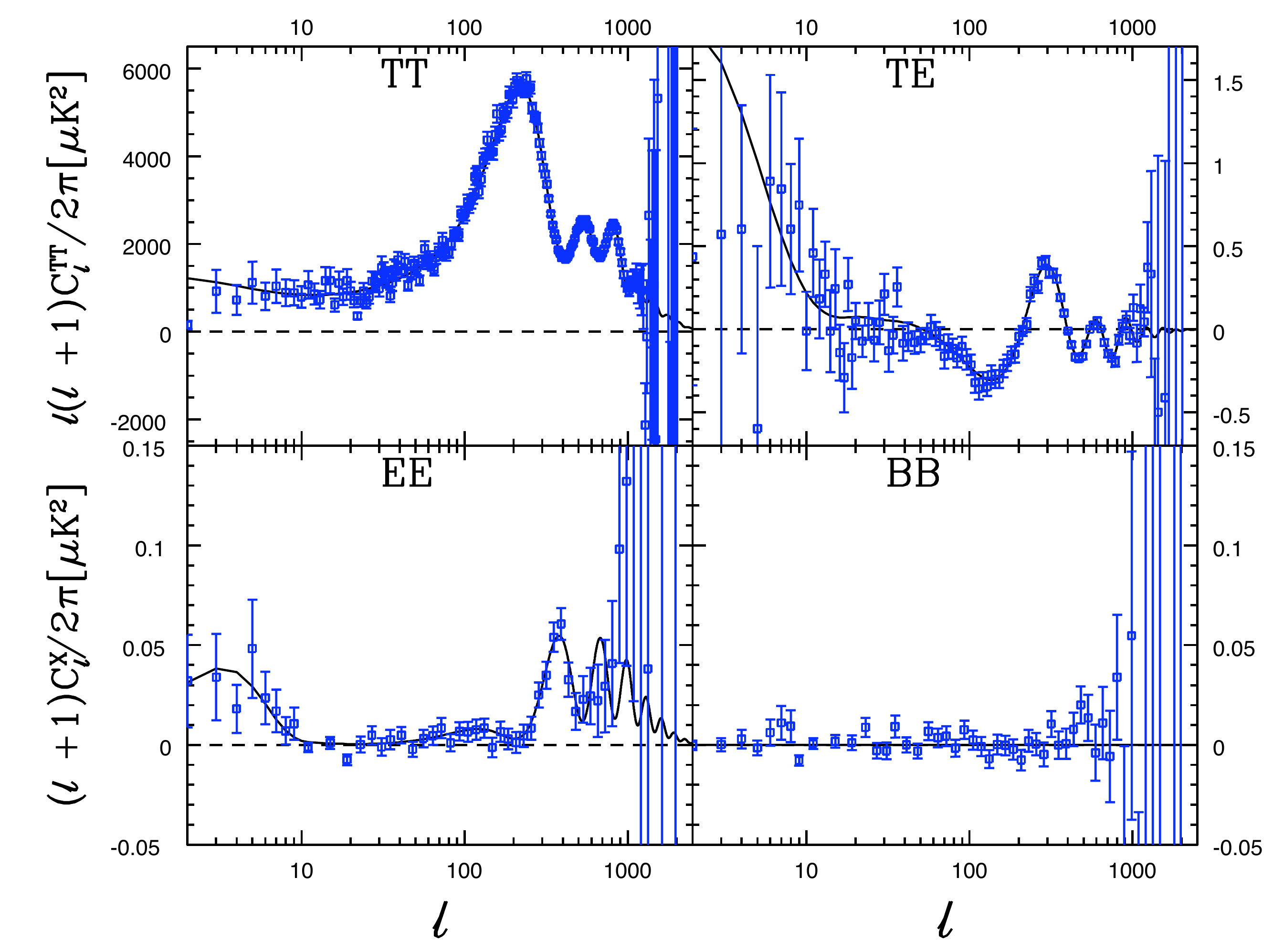}
\includegraphics[width=6cm]{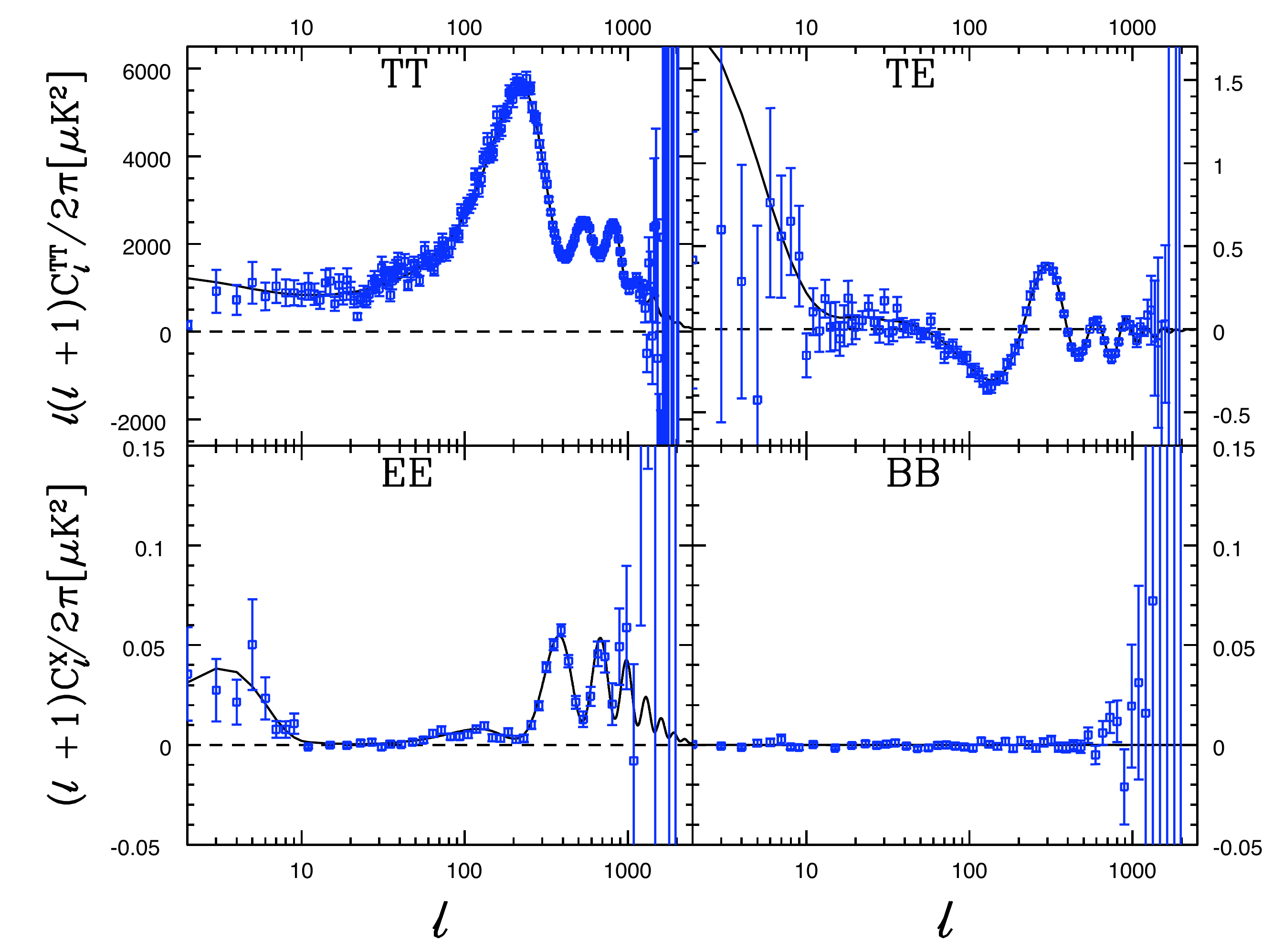}
\includegraphics[width=6cm]{figures/Xfaster_phase2_symm_all_pse_c_A}
}
\caption{ Power spectrum estimated with {\tt XFaster} and $1 \sigma$ error bars, considering transfer function=1 for Phase~1a, for map generated with a quadruplet of detectors (left hand side), with all twelve detectors (middle) and for Phase~2a, map generated with all twelve detectors and convolved with a symmetric beam (right hand side). This plot displays the estimated power spectrum (blue) of the observed map, overplotted with the $C_{\ell}$ fiducial model used as input in Phase~2 signal simulations, first year WMAP+CBI+ACBAR best fit model (black).}
\label{pse-phase1a-II}
\end{center}
\end{figure}

Figure~\ref{pse-all} shows the power spectra estimated for the observed map for Phase~1 and Phase~2 and their $1 \sigma$ error bars.
The power spectra recover accurately the input power spectrum in the middle range of multipoles, $30 \leq \ell \leq 1000$.  At high-$\ell$ ( $\ell \geq 1000$), they are  impacted by the  noise.
At low-$\ell$, as the large scale structure of the observed map is a WMAP constrained realization, the estimated power spectrum is not necessarily consistent with, and exhibits a dispersion around,  the best-fit spectrum.
\
Comparing the power spectra for the diverse phases we conclude that the cut-sky anisotropic noise case (Phase~1b) exhibits slightly greater uncertainties and slightly larger multipole to multipole variations at low-$\ell$ than the full-sky, isotropic noise case (Phase 1a).
This is expected, as the cut-sky will induce correlations at low-$\ell$.  The kernels correct these correlations; however, there is still a small residual dispersion. On the other hand the anisotropic noise will enhance the overall white noise level increasing the power spectrum uncertainty.
For Phase~2 the dispersion of the power spectrum and uncertainties at low-$\ell$ are enhanced due to the residuals of correlated $1/f$ noise.
 As in Figure~\ref{pse-all}, the power spectrum for Phase~1 is estimated from maps generated with a smaller number of detectors (four) than those for Phase~2 (twelve), and is therefore noisier by a factor of the order $\frac{1}{\sqrt{3}}$. Therefore the error bars of the Phase~2 power spectrum are smaller than those of Phase~1. The right thing to do though is to compare the power spectra estimated for the same number of detectors.
In Figure~\ref{pse-phase1a-II} we compare the power spectra for Phase~1a (middle plot) and Phase~2 (right hand side), both estimated on maps generated with all twelve detectors.  As expected,  the uncertainty
for Phase~1a is now smaller than that of Phase~2.

The power spectra for Phase~2 for both the symmetric and asymmetric beams are highly consistent. Therefore we conclude that the beam asymmetry is reasonably well-handled by {\tt XFaster} (see Section~\ref{sym-asym} for more details).

\begin{figure}
\begin{center}
\vbox{
\hbox{
\includegraphics[width=9cm,height=6cm]{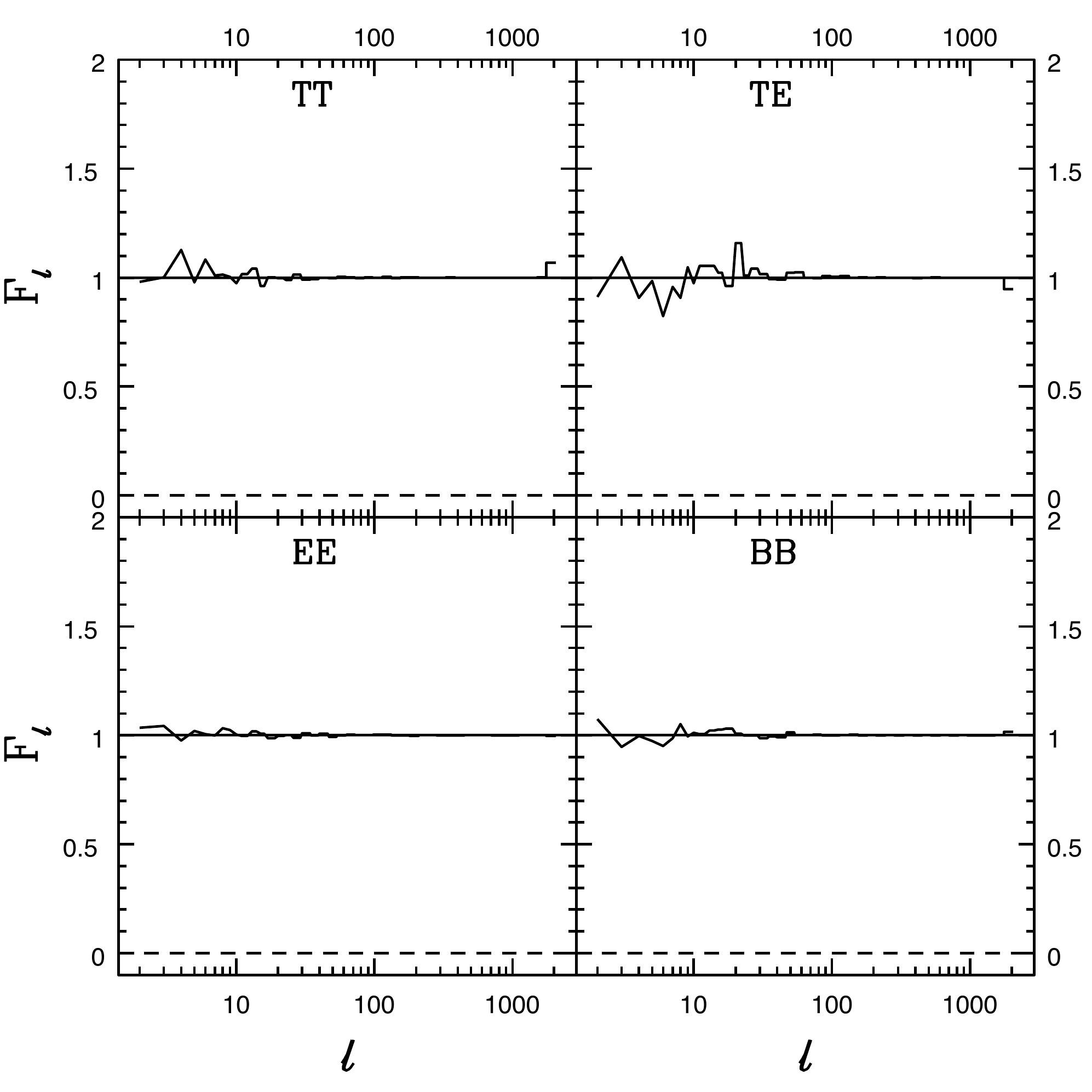}
\includegraphics[width=9cm,height=6cm]{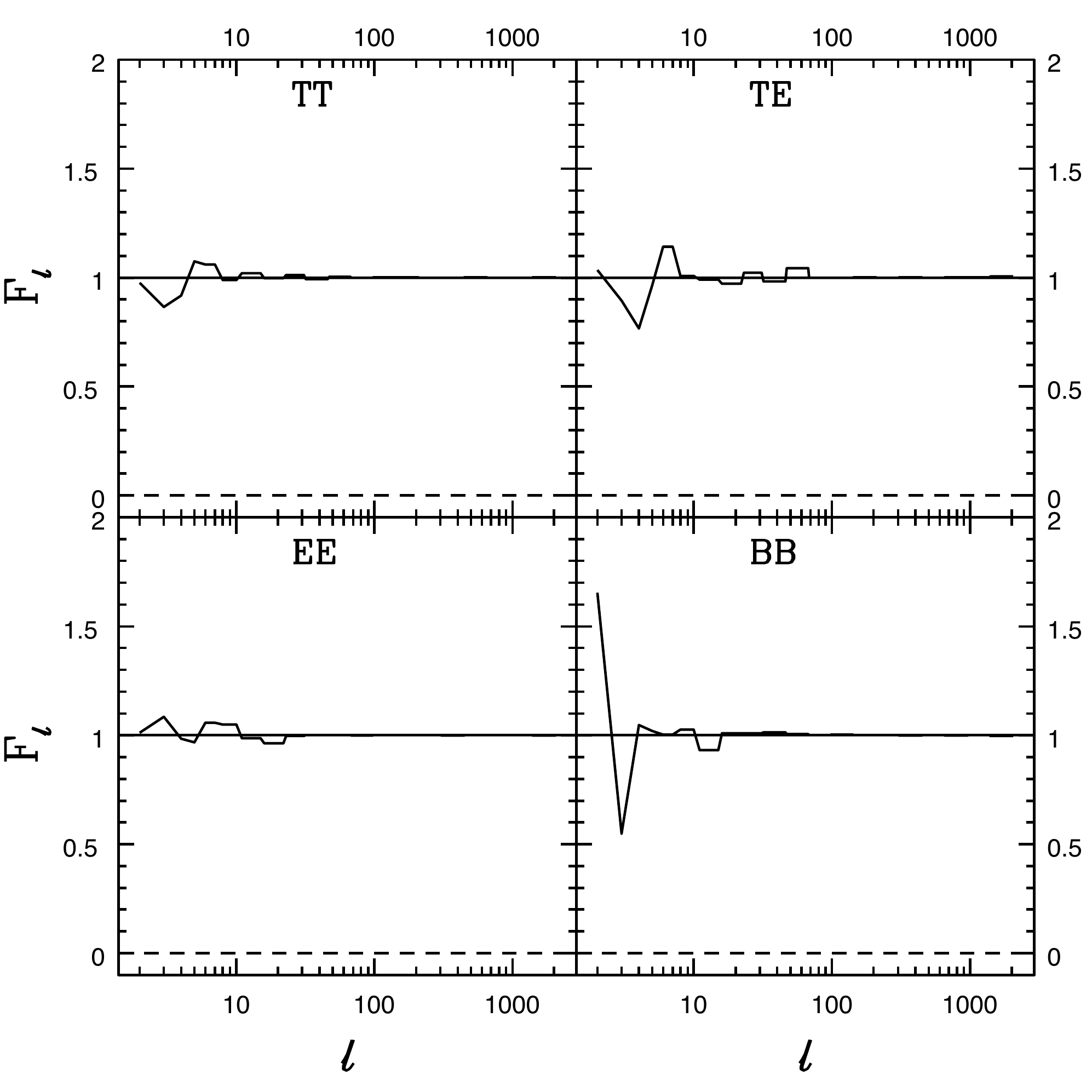}
}
\hbox{
\includegraphics[width=9cm,height=6cm]{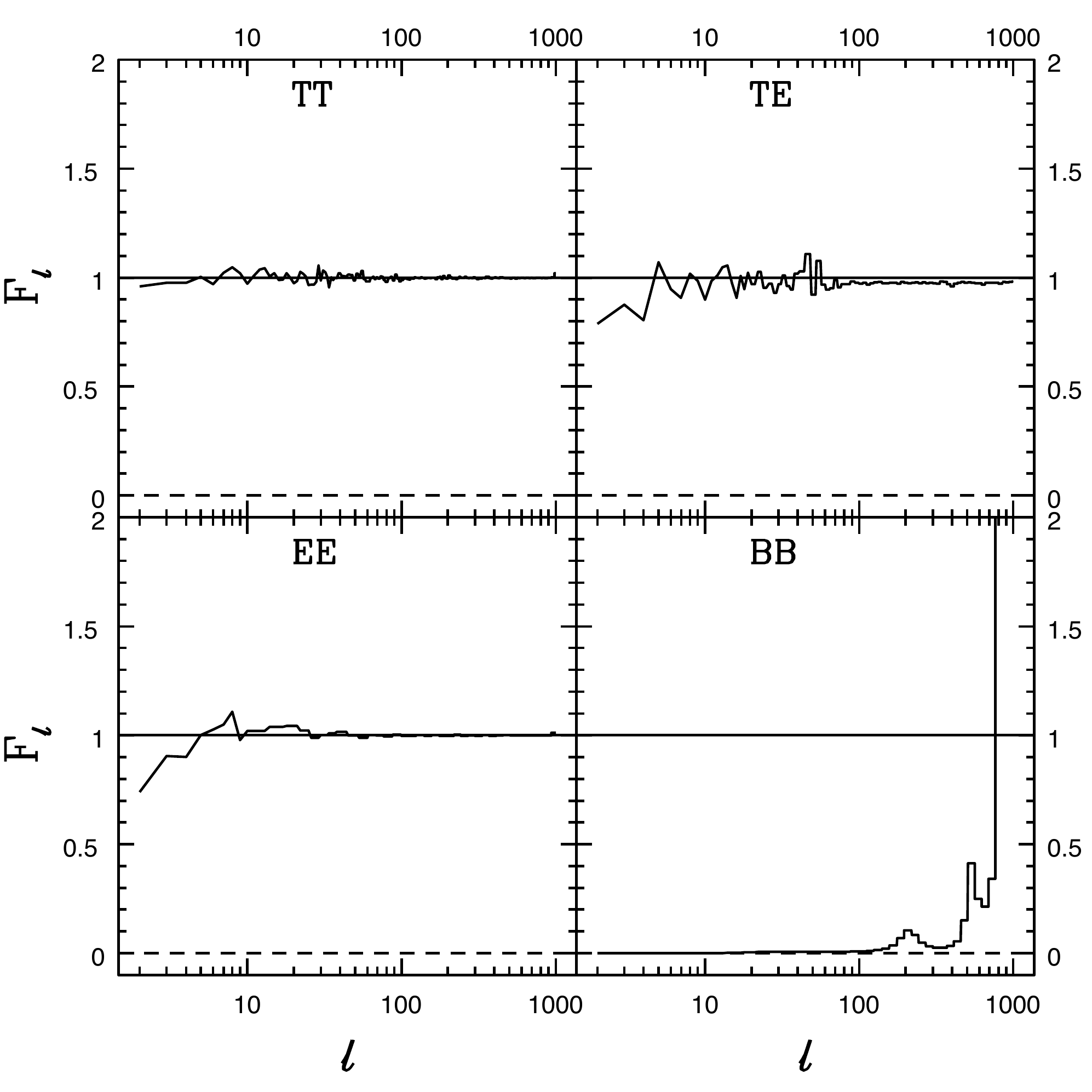}
\includegraphics[width=9cm,height=6cm]{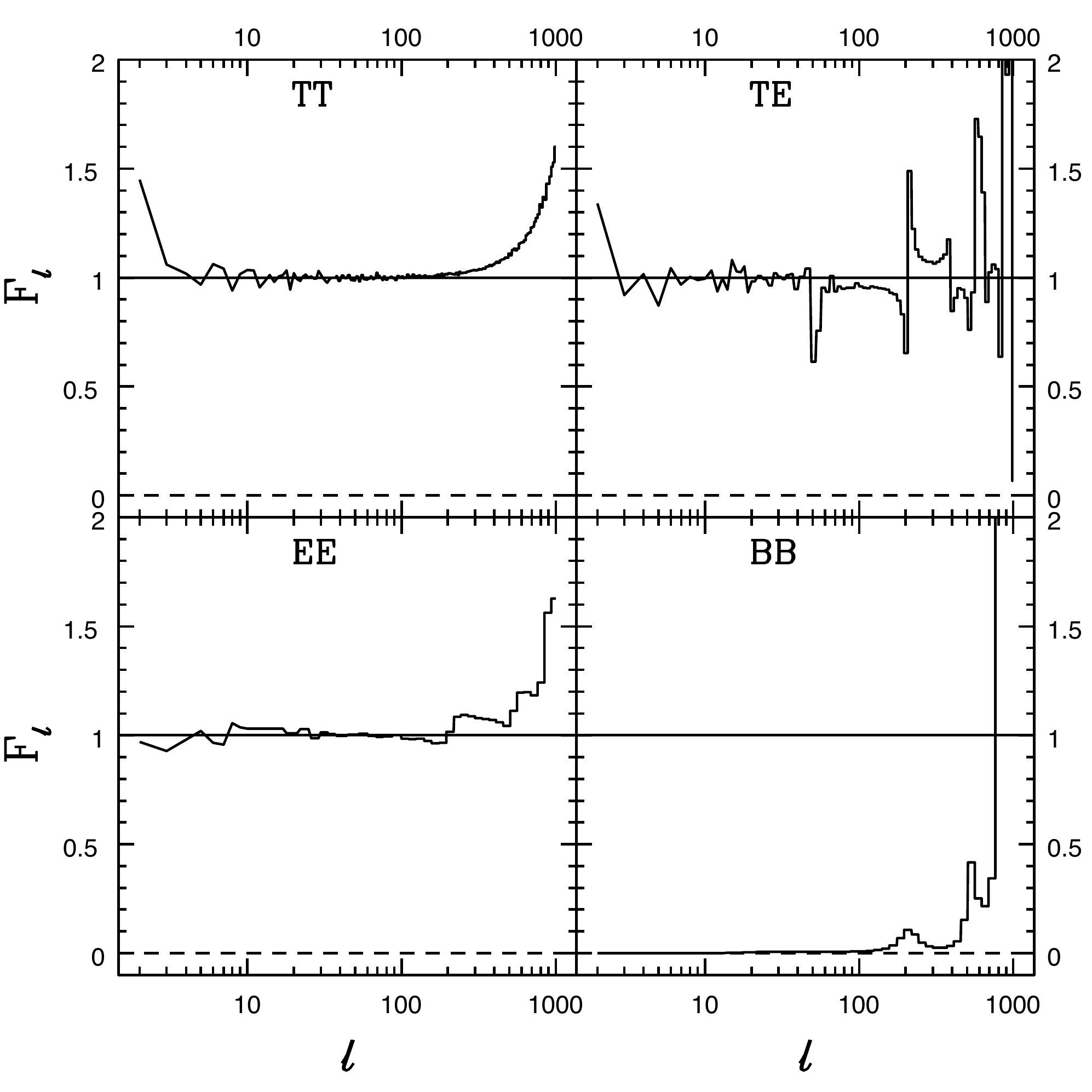}
}}
\caption{Transfer Function: Top row - for Phase~1a (left hand side) and Phase~1b (right hand side); Bottom row - for Phase~2a symmetric beam (left hand side) and Phase~2b asymmetric beam (right hand side).}
\label{tr-all}
\end{center}
\end{figure}

Figure~\ref{tr-all} shows the transfer functions for Phase~1 and Phase~2.
Except for Phase~2b, all are close to $1$. This is expected, as we do not pre-filter the TOD and the only effect at low-${\ell}$ is that due to the mapmaking step
and to the limited number of Monte Carlos of the signal maps available. 
This means there is remaining sample variance on the large-scales although not very significant. We included the transfer function estimates because they are an integral part of the method and in real life they will not be equal to 1.
However to investigate and show that  the significance of these small  departures from 1 are not significant we estimated the power spectrum with transfer function=1 as plotted in Figure~\ref{pse-phase1a-II}.
 We also highlighted the fact that the transfer functions are estimated consistently between polarization types (i.e. taking into account cross-correlation between polarization modes in any realization).
Note that in this work we did not take into account the remaining MC error in the transfer function in the final estimate of the power spectrum since any production run used in the real case will include many more realizations than used in this work (they could be easily included by adding to the final Fisher matrix if needed).
However, for the asymmetric beam case the transfer function exhibits an upturn at high-$\ell$. This upturn tries to correct the mismatch between the 'real' asymmetric beam and our assumed symmetric beam (as discussed in Section~\ref{sym-asym}).
As the input BB power spectrum model of the signal Monte Carlo simulations for Phase~2 is set to zero, we cannot constrain the BB transfer function. The transfer function obtained reflects the inadequacy of the input model and hence is close to zero. When estimating the power spectrum we replace the BB transfer function by the EE transfer function.

\begin{figure}
\begin{center}
\vbox{
\hbox{
\includegraphics[width=9cm]{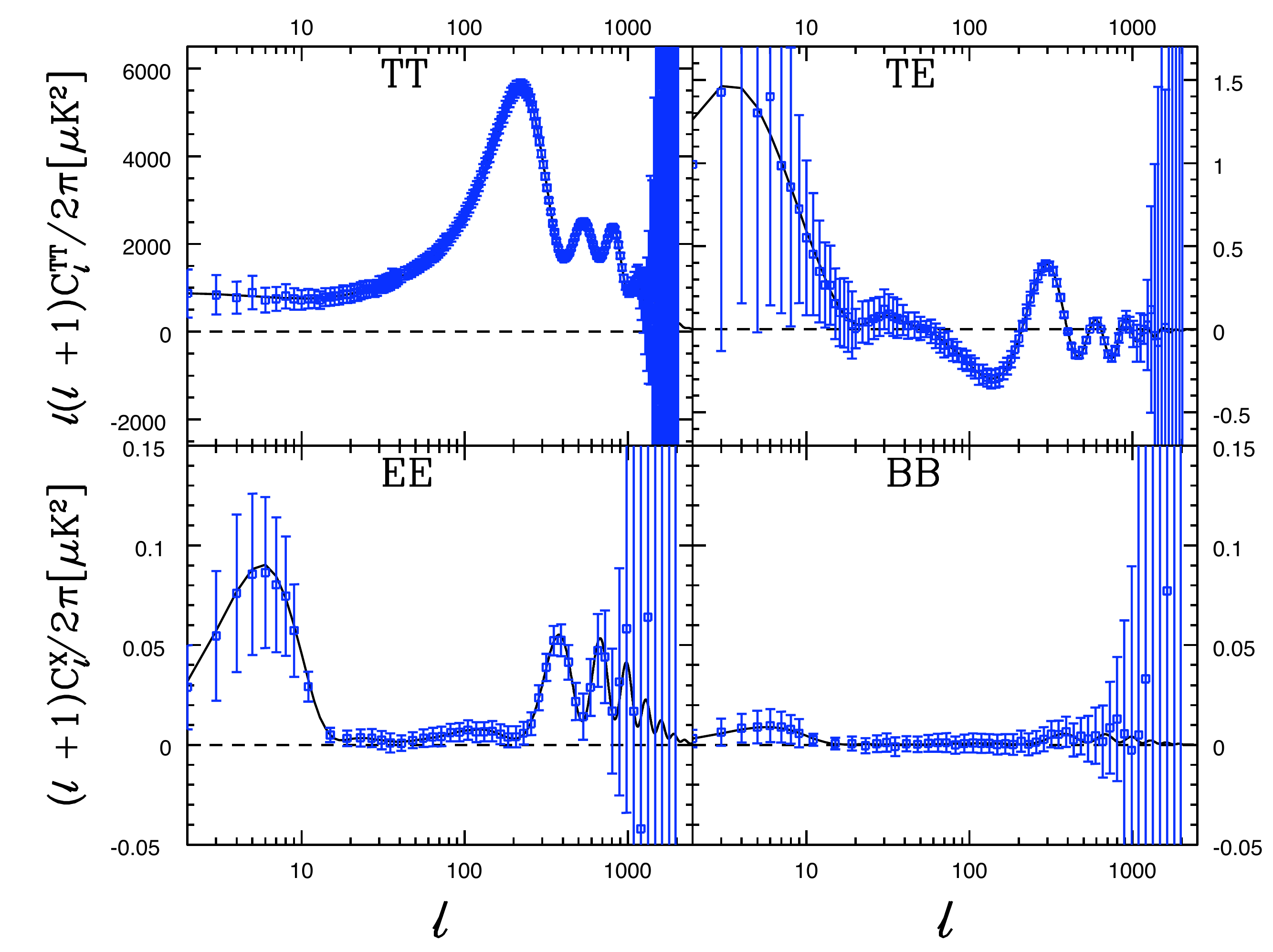}
\includegraphics[width=9cm]{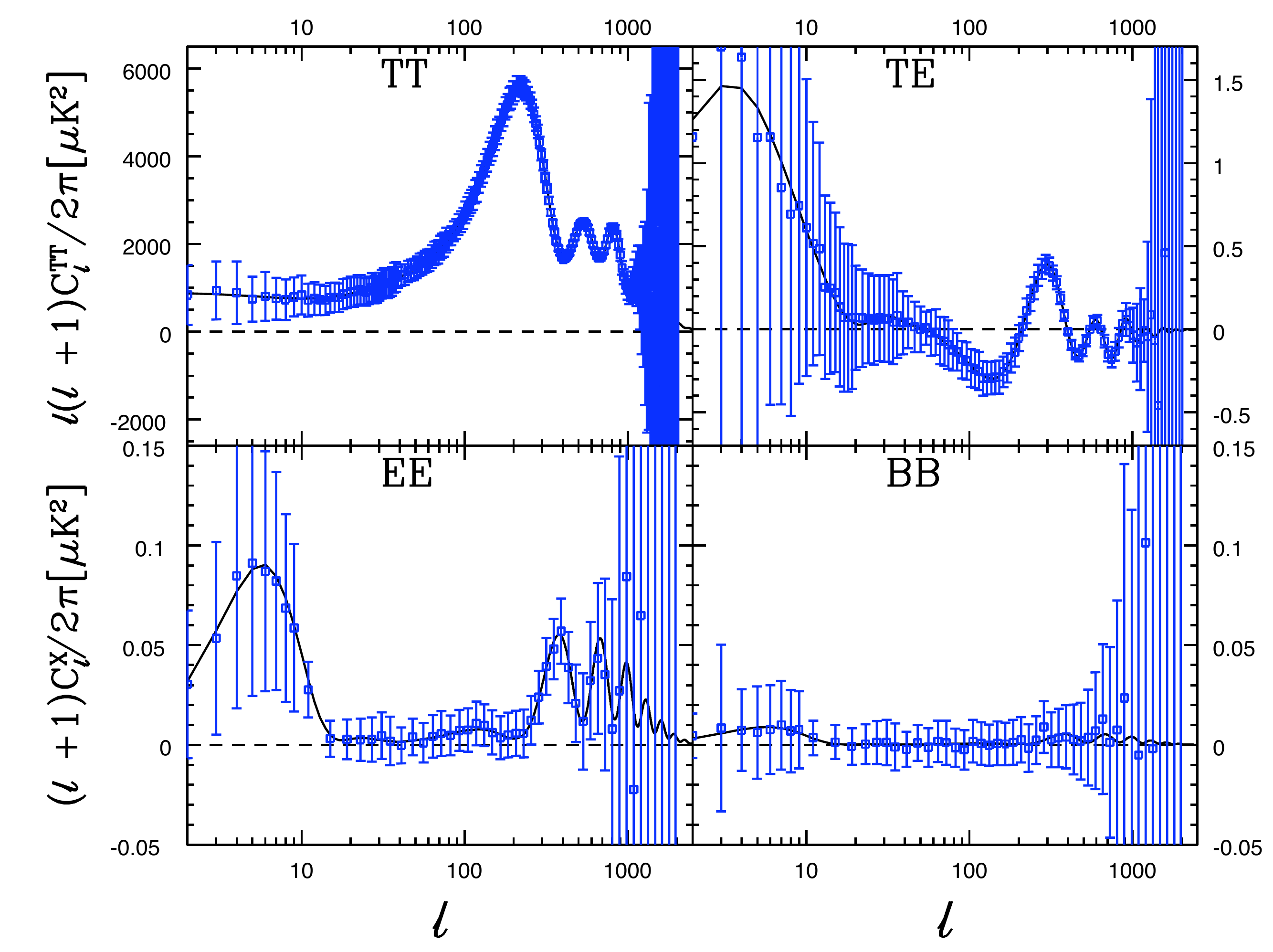}
}
\hbox{
\includegraphics[width=9cm]{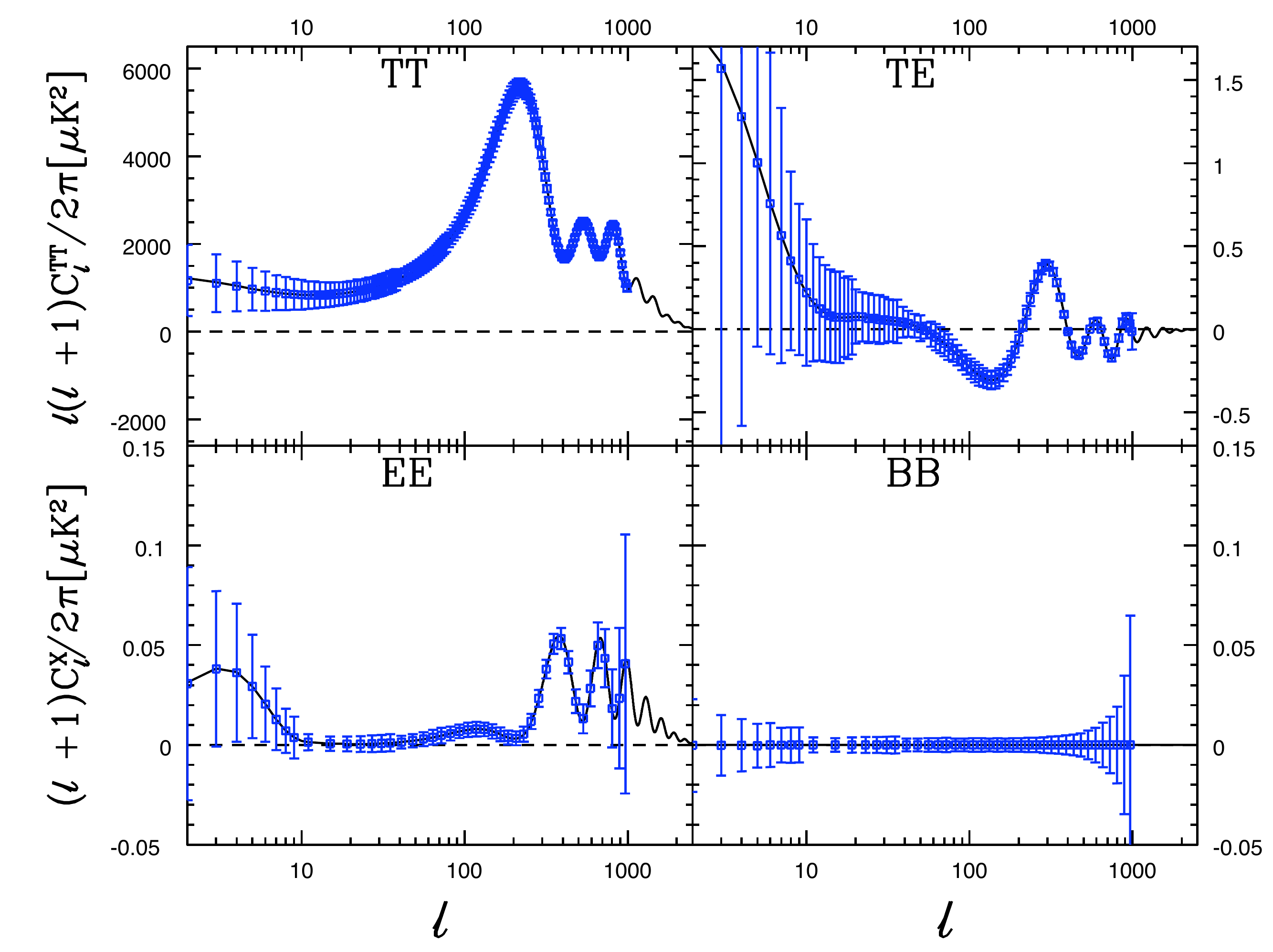}
\includegraphics[width=9cm]{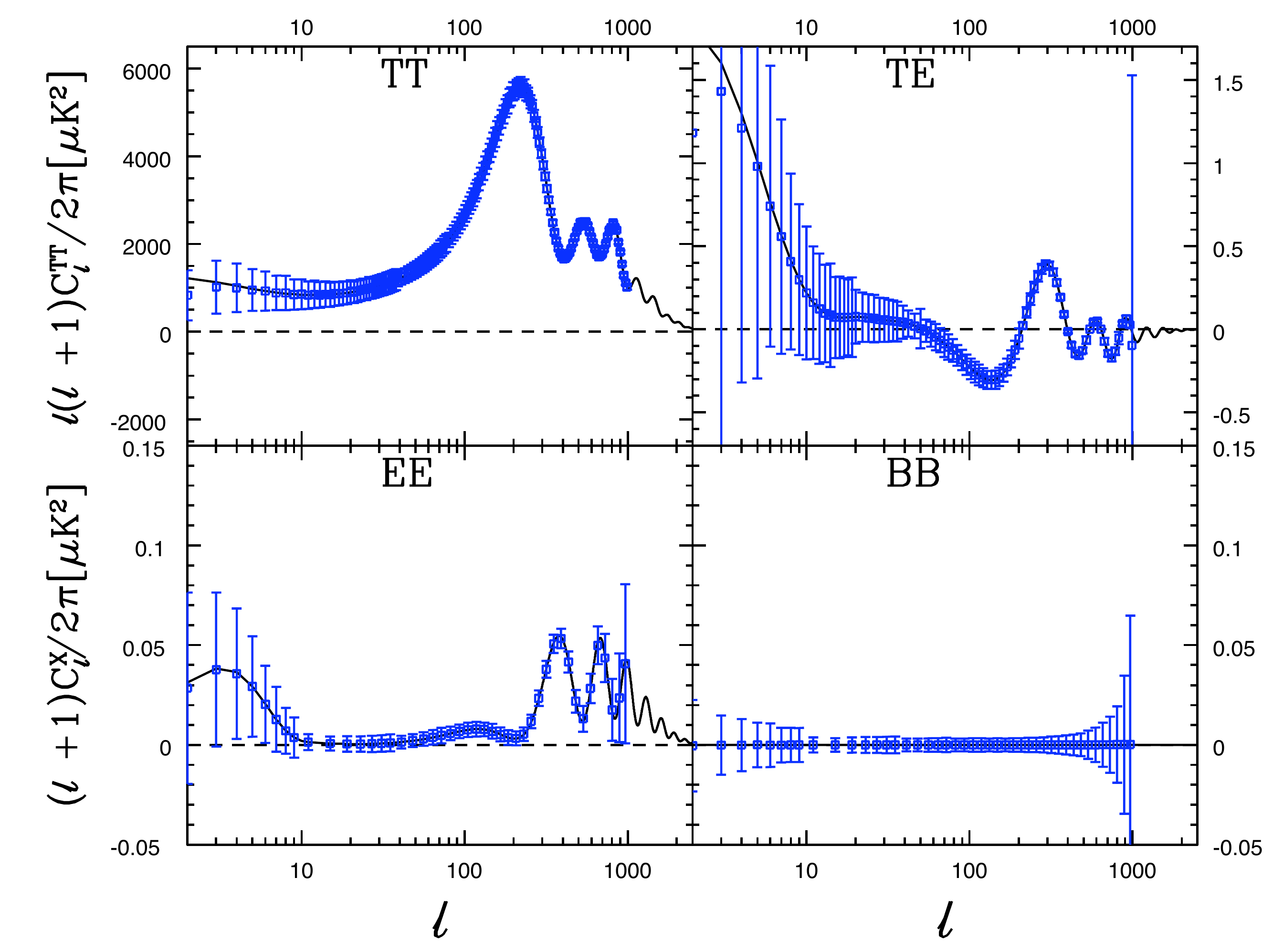}
}}
\caption{Power spectrum obtained with {\tt XFaster} in average mode, meaning that we replaced the observed map by the average of the signal+noise simulated maps (blue). Top row---Phase~1a (left hand side) and Phase~1b (right hand side) for map generated with a quadruplet of detectors. Bottom row---Phase~2a symmetric beam (left hand side) and Phase~2b asymmetric beam (right hand side) for map generated with all twelve detectors, overplotted with the $C_{\ell}$ fiducial model used as input in our Phase~1 signal simulations, first year WMAP best fit model (black) for Phase~1 and first year WMAP+CBI+ACBAR best fit model (black) for Phase~2.  It serves the purpose of checking for possible biases of the power spectrum estimator---in principle the power spectrum estimated in average mode should follow closely the input signal $C_{\ell}$ model used to generate the signal simulations (black). }
\label{avg-all}
\end{center}
\end{figure}

\begin{figure}
\begin{center}
\includegraphics[width=9.5cm]{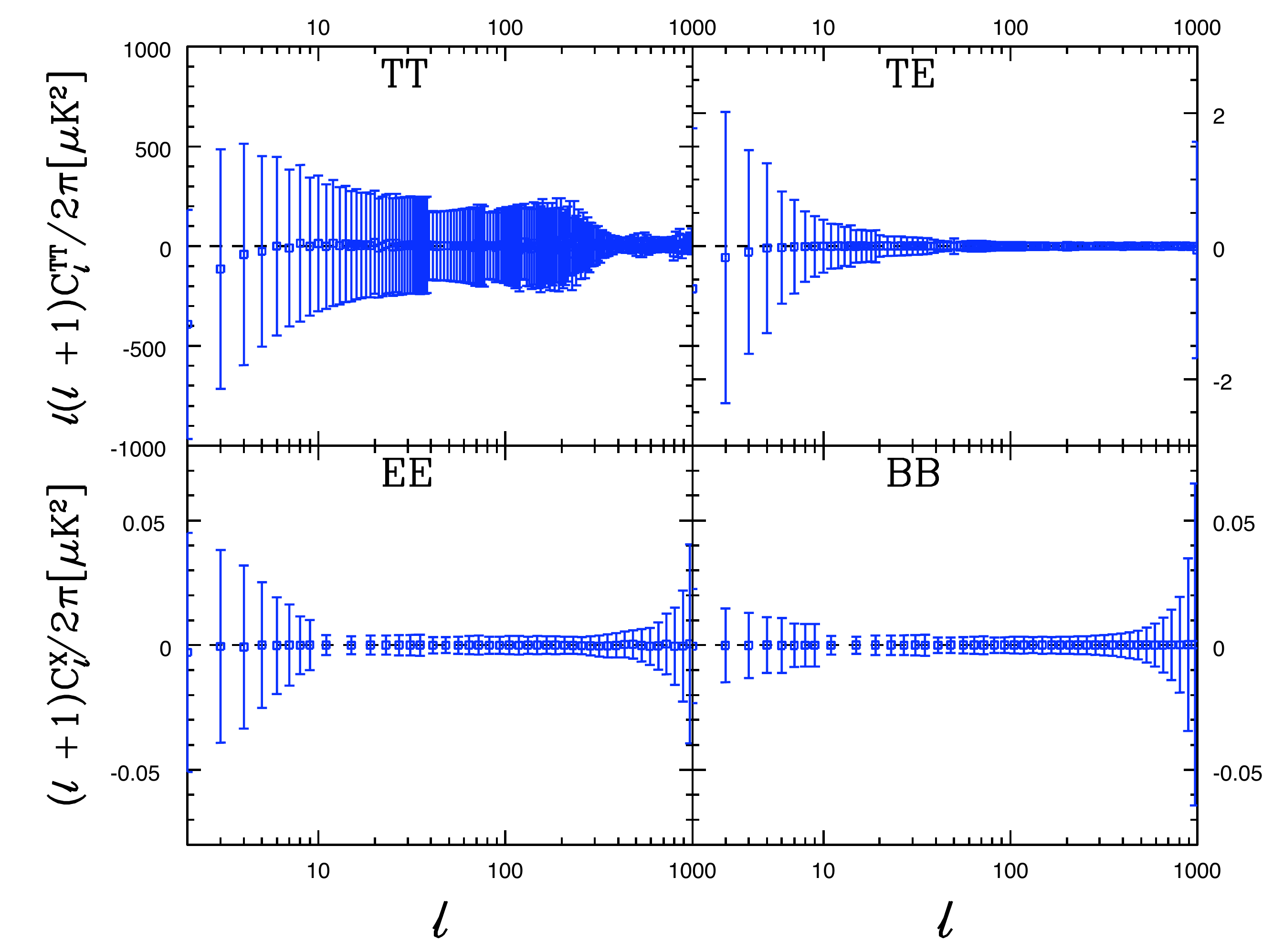}
\caption{Difference of the power spectrum obtained with {\tt XFaster} in average mode, (meaning that we replaced the observed map by the average of the signal+noise simulated maps) and the $C_{\ell}$ fiducial model used as input (first year WMAP+CBI+ACBAR best fit model), for phase2, asymmetric beam case.
The stepwise decreases of the amplitude of the error bars are caused by the changes of the $C_{\ell}$ bins size.}
\label{avg-all-diff}
\end{center}
\end{figure}

Figure~\ref{avg-all} shows the power spectrum estimated for the average mode run. 
Whereas Figure~\ref{avg-all-diff} shows the difference between the power spectrum obtained for the average mode run (with error bars) and the fiducial model, $C_{\ell}$,  used as input, for phase2, asymmetric beam case. The stepwise decreases of the amplitude of the error bars are caused by the changes of the $C_{\ell}$'s bins size.
These plots show that the power spectrum follows the input signal $C_{\ell}$, confirming that {\tt XFaster} is an unbiased estimator. 
These results were obtained using 100 Monte Carlo simulations. Considering 500 simulations for phase1a the small departures of the transfer function from 1 reduces slightly.
Due to computational constraints this was not feasible for phase2. However increases in computational capability over time make it now possible to generate thousands of  simulations. Results for 143GHz channel will be presented in our upcoming papers \cite{GRhfiXFaster10,varenna10}.

\begin{figure}
\begin{center}
\vbox{
\hbox{
\includegraphics[width=9cm]{figures/Xfaster_phase2_symm_all_pse_c_A}
\includegraphics[width=9cm]{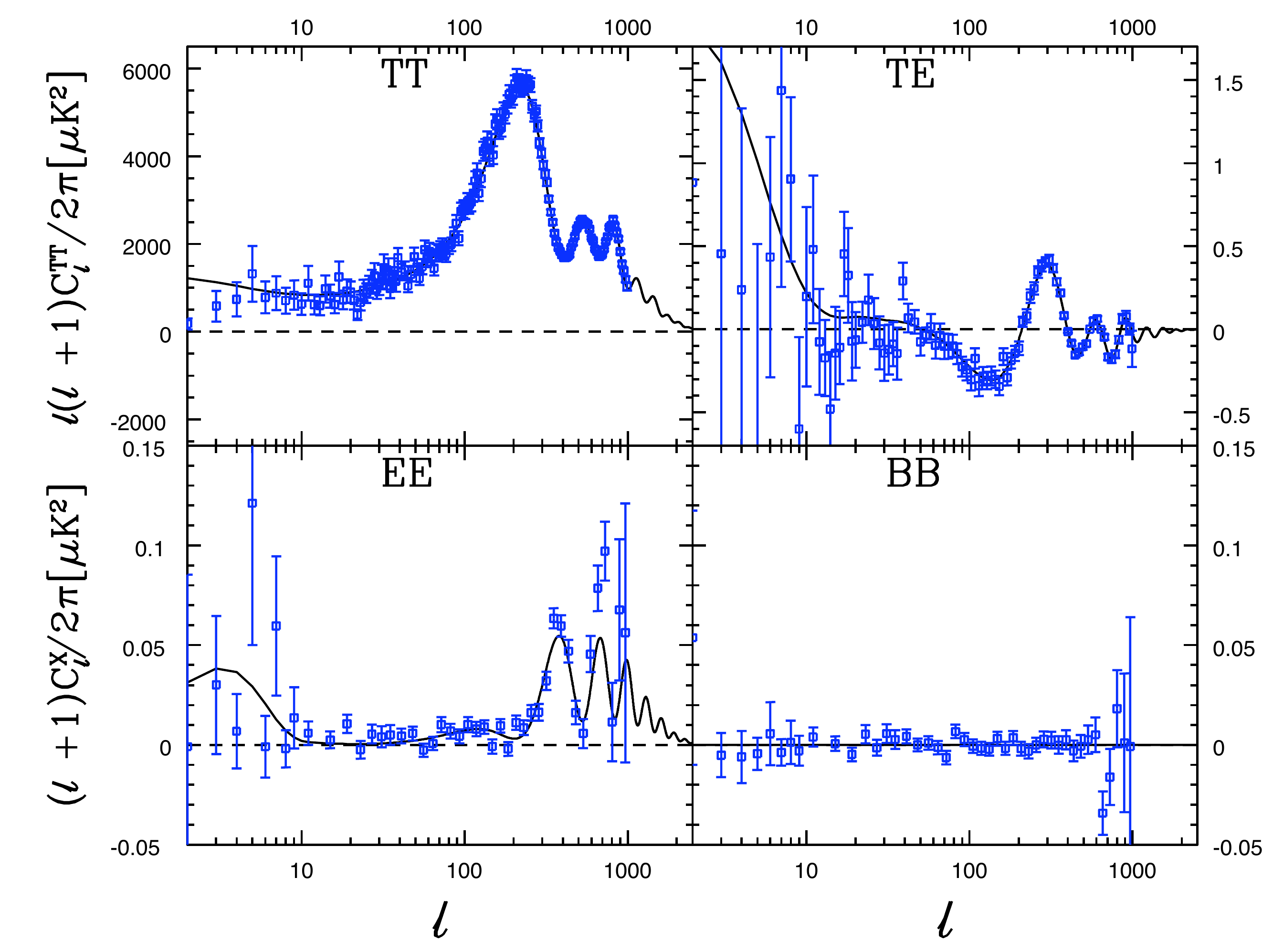}
}
\hbox{
\includegraphics[width=9cm]{figures/Xfaster_phase2_asymm_all_pse_c_A}
\includegraphics[width=9cm]{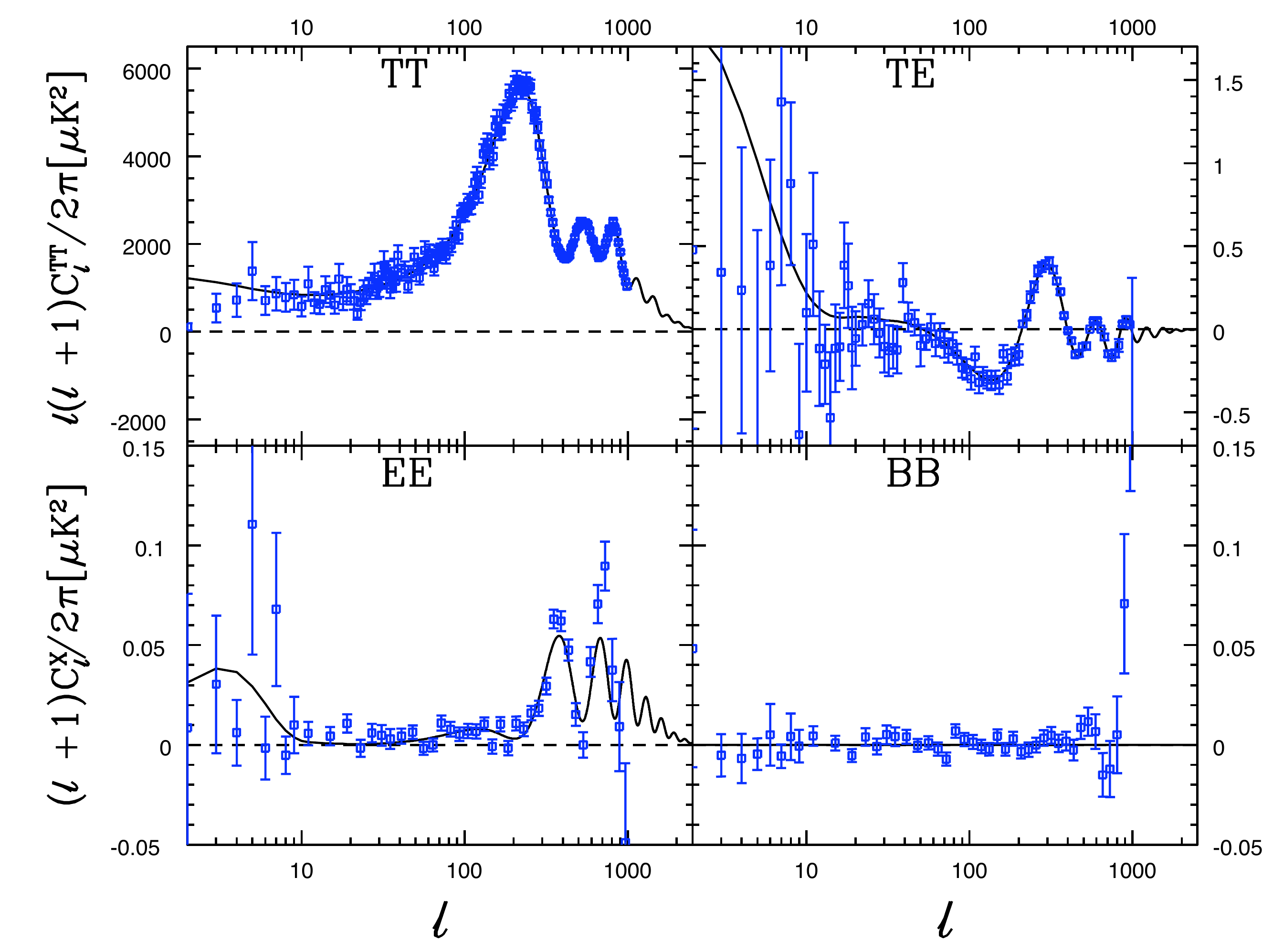}
}}
\caption{Power spectrum estimated with {\tt XFaster} and $1 \sigma$ error bars for Phase~2, for map generated with all twelve detectors, for symmetric (top row) and asymmetric (bottom row) beams. Left hand side plot displays the estimated power spectrum (blue) if the observed map, overplotted is the $C_{\ell}$ fiducial model used as input in our signal simulations, first year WMAP+CBI+ACBAR best fit model (black); Right hand side plot displays the estimated power spectrum with noise Monte Carlo simulations replaced by the BB power spectrum.}
\label{pse-phase2-bbnoise}
\end{center}
\end{figure}

Figure~\ref{pse-phase2-bbnoise} shows a variation of the above plots in which the BB power spectrum replaces  the noise Monte Carlo simulations. These plots show that whenever the noise Monte Carlo simulations exhibit issues in the sense that they do not reflect accurately the noise characteristics of the observations, replacing them  by the BB power spectrum is an adequate procedure, as for Planck at 70\,GHz the BB power spectrum is mostly dominated by noise.

The power spectrum estimated for the observed map has been compared with those from several other methods (\cite{varenna10}).
To further assess the power spectrum estimator we propagated this analysis to cosmological parameter estimation using the new {\tt XFaster}  likelihood as described in Section~\ref{par-results}


\subsubsection{Symmetric and asymmetric beams}
\label{sym-asym}

To study the impact of beam asymmetries on power spectrum and cosmological parameter estimation, we took a minimal, non-informative approach. When computing the power spectrum for the observed maps convolved with  symmetric and asymmetric beams we assumed a FWHM = 14\arcm\ symmetric beam for both cases.

\begin{figure}
\begin{center}
\includegraphics[width=9cm,height=6cm]{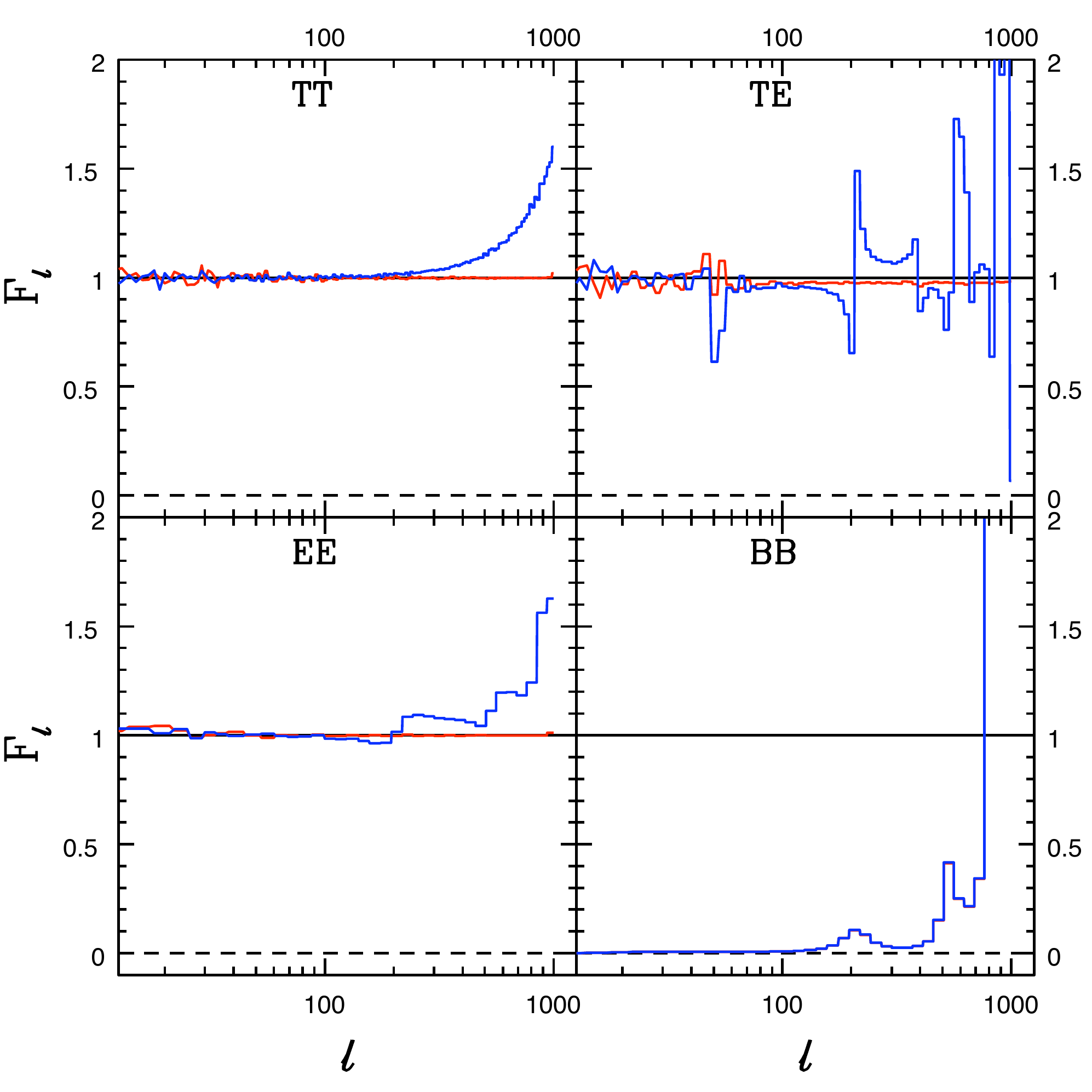}
\caption{Comparison of transfer (filter) functions for Phase~2, for map generated with all twelve detectors, for symmetric beam (red solid line) and for asymmetric beam case (blue solid line).}
\label{tr-sym-asym}
\end{center}
\end{figure}

We started by investigating the effect of this assumption on the shape of the transfer function, $F_{\ell}$, in Equations~\ref{Cl-cut} and ~\ref{Clcut}, or $F_{b} $ in Equations~\ref{Fb}, \ref{qbFb-t}, and \ref{qbFb-p}. In effect we compute not only $F_{\ell}$, but also $F_{\ell} \delta{b_{\ell}}$, where $\delta{b_\ell}$ is the correction to the beam transfer function $B_{\ell}$. This correction arises from assuming a symmetric beam when estimating the power spectrum of an observed map that in fact has been convolved with an asymmetric beam. Call this function the generalized transfer function, $(BF)_{\ell}$.  As mentioned in Section~\ref{pse-results}, since we do not pre-filter the TOD the only effect at low-${\ell}$ is that due to the mapmaking step. Therefore the transfer function should be very close to unity, particularly for the symmetric beam.
Figure~\ref{tr-sym-asym} shows $(BF)_{\ell}$ for both cases.  For the symmetric case this function is very close to $1$ as expected, with tiny oscillations around $1$ at low-${\ell}$ consistent with our expectations. Apart from the transfer function for the BB mode for phase2 which is close to zero. Since the input BB power spectrum model of the signal Monte Carlo simulations for phase2 is set to zero we cannot constrain the transfer function for the BB mode. We resort to using the transfer function for the EE mode to estimate the BB power spectrum instead.

\begin{figure}
\begin{center}
\hbox{
\includegraphics[width=9cm]{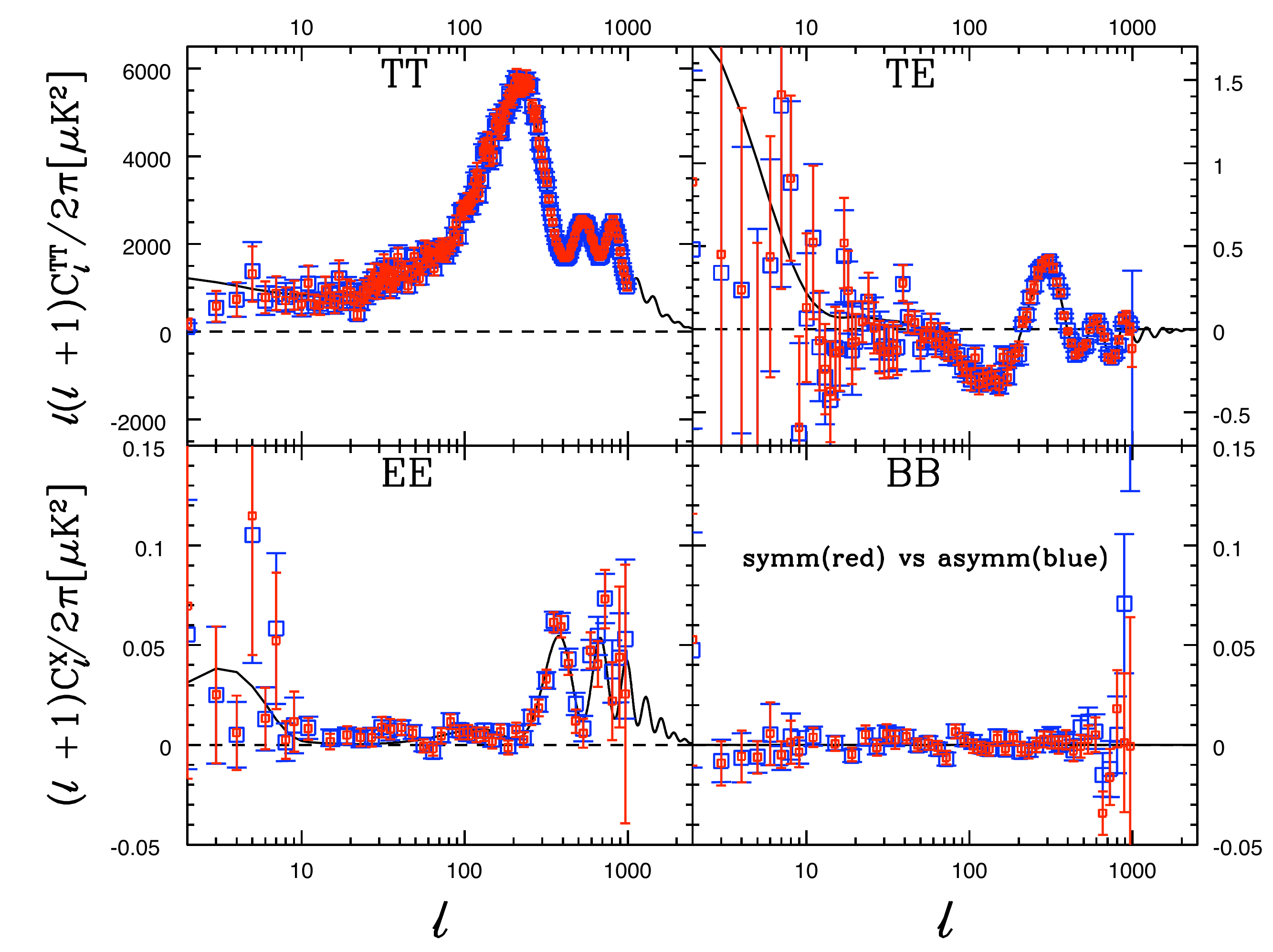}
\includegraphics[width=9cm]{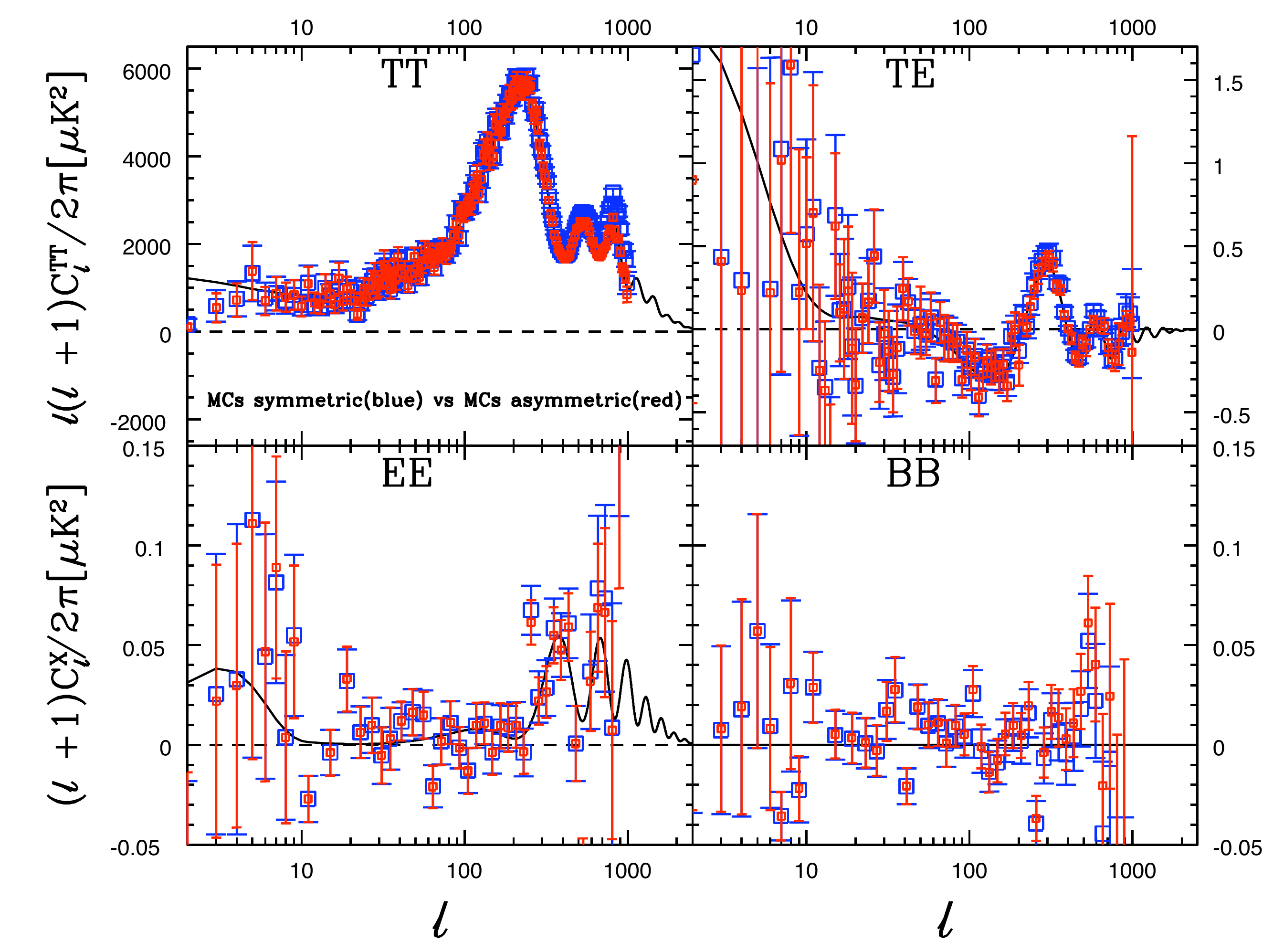}
}
\caption{Power spectrum estimated with {\tt XFaster} and $1 \sigma$ error bars for Phase~2. The lefthand plot displays the power spectrum for a map generated with all twelve detectors, for both symmetric (red) and asymmetric (blue) beam cases. For both runs a symmetric beam with FWHM = 14\arcm\ is assumed.  The resulting power spectra are highly consistent, the compensation is achieved via the generalized transfer function $(BF)_{\ell}$. The righthand plot displays the power spectrum estimated for map generated with a quadruplet of detectors, with two different sets of Monte Carlo simulations.  In one case we use the Monte Carlo simulations convolved with the symmetric beam (blue); in the other we use the correct Monte Carlo simulations convolved with the asymmetric beam (red). Making use of the Monte Carlo simulations for the symmetric case gives rise to a bias high at high-$\ell$.}
\label{pse-phase 2-asym-sym}
\end{center}
\end{figure}

The symmetric and asymmetric $(BF)_{\ell}$ differ. In the asymmetric case, the function exhibits an upturn at high $\ell$. This upturn tries to correct the mismatch between the 'real' asymmetric beam and our assumption. Hence the resulting power spectrum is pretty consistent for both cases, as shown in Figure~\ref{pse-phase 2-asym-sym}. Considering the input symmetric beam with  FWHM = 14\arcm\  and the estimated FWHM$ \simeq 13\arcm$ for the asymmetric beam,
this effect should be approximately $(14/13)^2$, although anisotropic pixel filtering will add an extra component of aliasing in the maps at high-${\ell}$.

\begin{figure}
\begin{center}
\hspace*{-1cm}
\vbox{
\hbox{
\includegraphics[width=10cm]{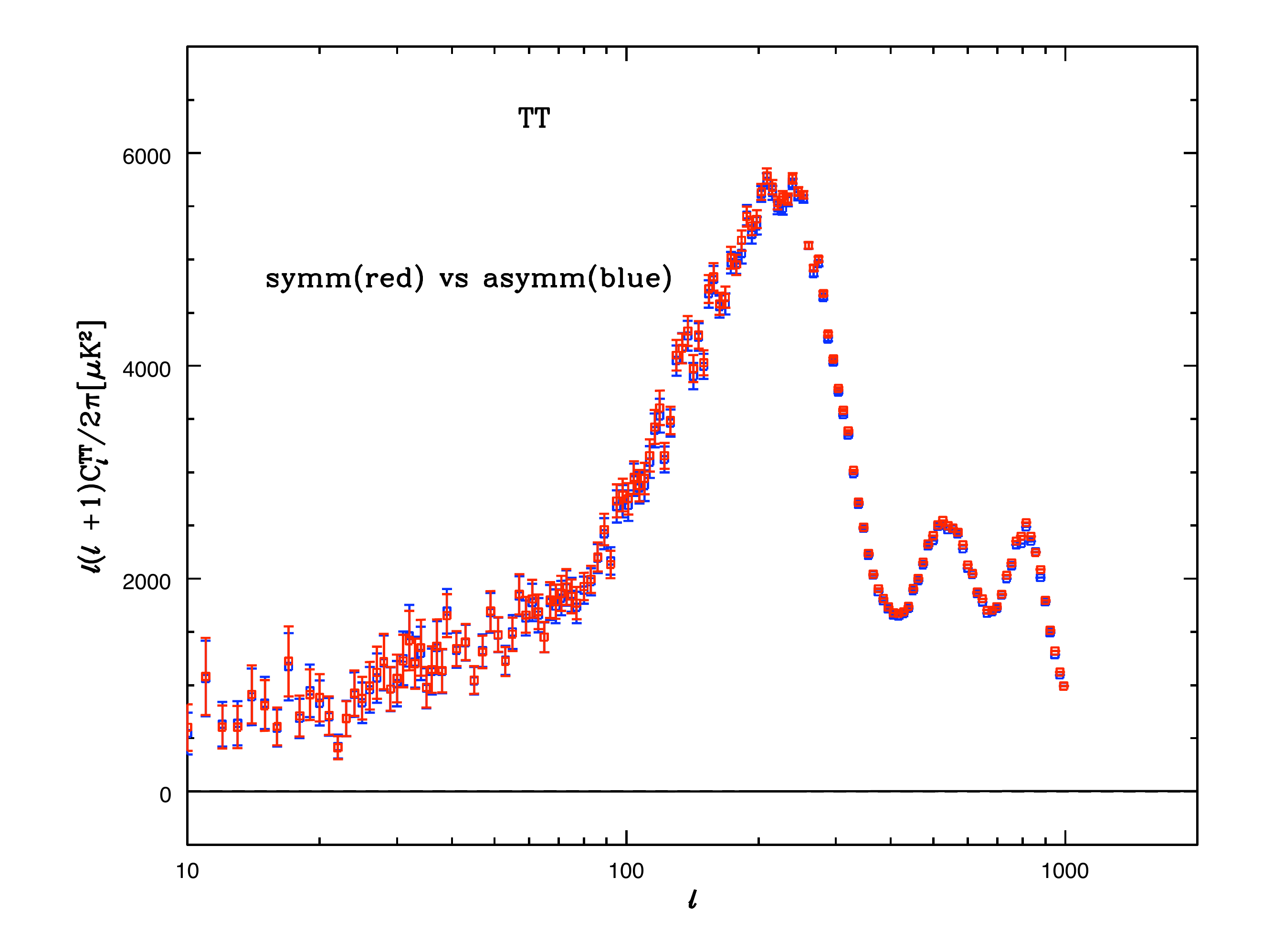}
\includegraphics[width=10cm]{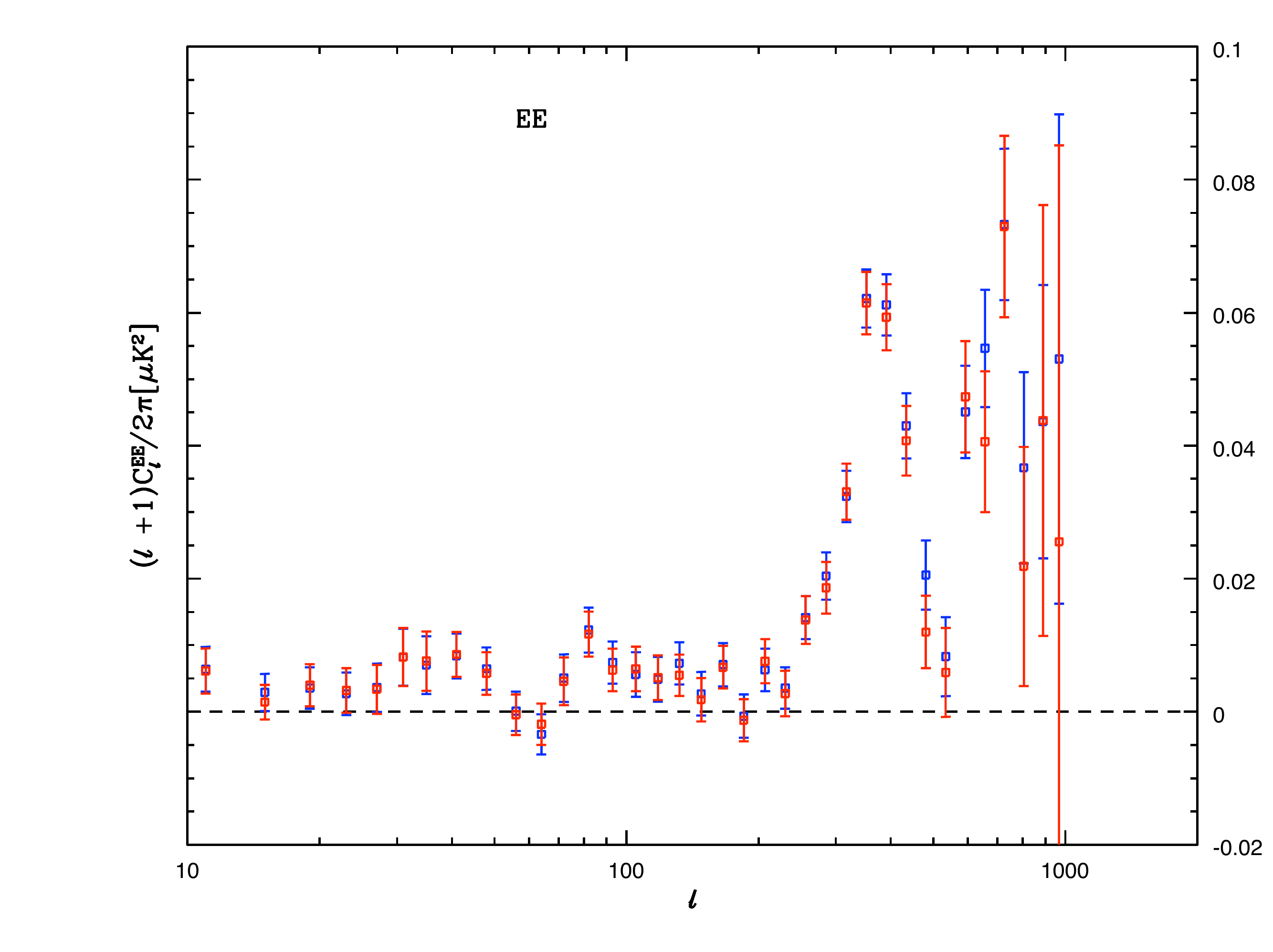}
}}
\includegraphics[width=10cm]{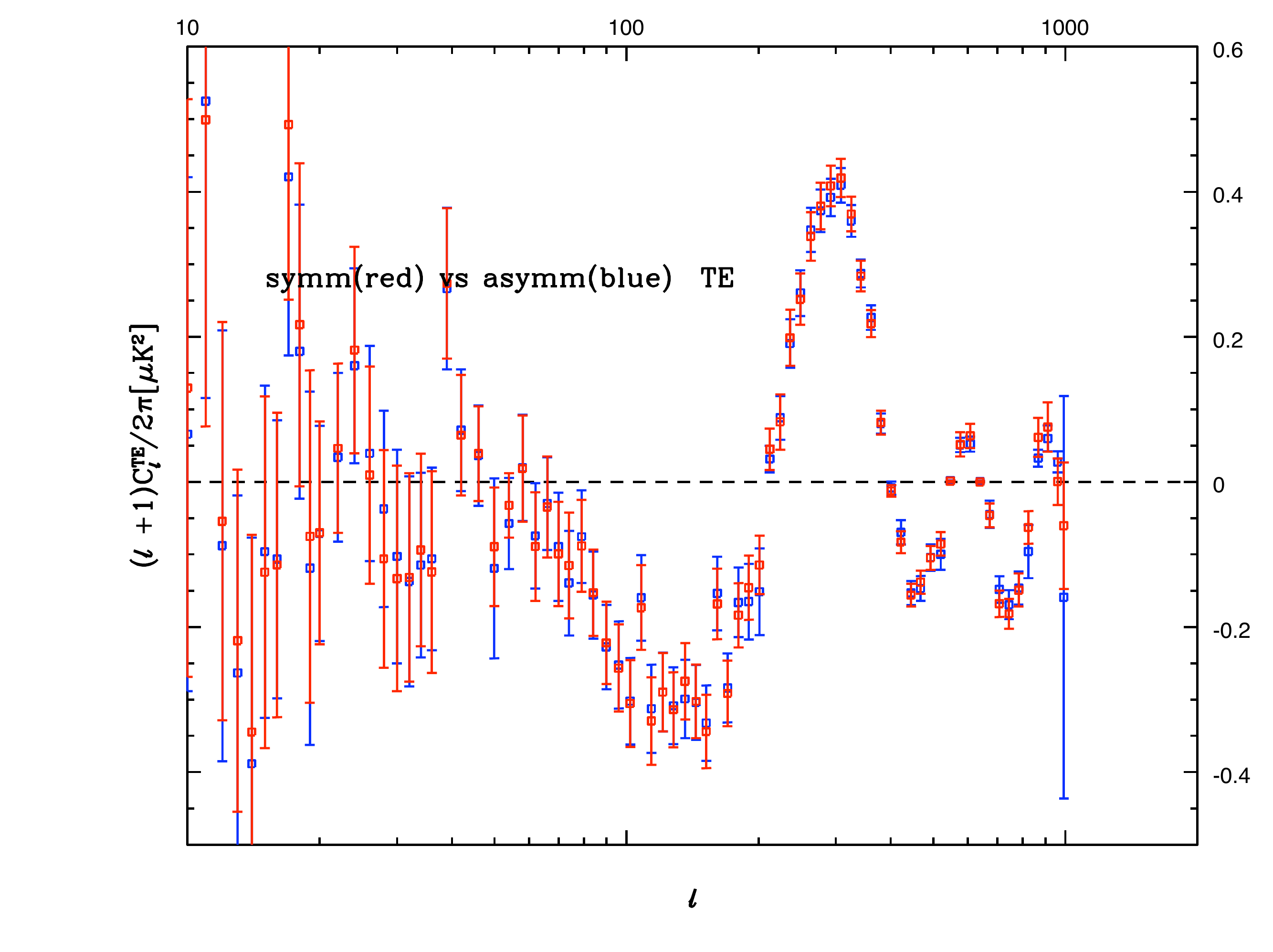}
\caption{Power spectrum estimated with {\tt XFaster} and $1 \sigma$ error bars for Phase~2, for map generated with all twelve detectors, for both symmetric (red) and asymmetric (blue) beam cases. For both runs a symmetric beam with FWHM = 14\arcm\ is assumed, the resulting power spectrum is consistent with each other, the mismatch is corrected via the generalized transfer function, $(BF)_{\ell}$. There is still a slight bias for the asymmetric case in agreement with the small differences of the estimated cosmological parameters (see section~\ref{par-results}).}
\label{pse-phase 2-asym-sym-2}
\end{center}
\end{figure}

Although the power spectra look consistent, the parameter estimation shows departures of the order of $\sigma/2$ for some of the parameters, as shown in Section~\ref{par-results}.  To investigate this further we enhanced the previous plot into Figure~\ref{pse-phase 2-asym-sym-2}.  Though there is very good agreement between the two power spectra, there is still a slight bias for the asymmetric beam case. This bias is consistent with the small differences in the estimated parameters, in particular for $n_{s}$, $\sigma_{8}$ and $\log[10^{10} A_s]$ (see Section~\ref{par-results}).

Next we study the impact of using two different sets of Monte Carlo simulations on the estimation of the power spectrum for the asymmetric beam case, Phase~2b. One set is the Monte Carlo simulations for the symmetric beam (Phase~2a) and the other the correct Monte Carlo simulations for the asymmetric beam (Phase~2b) study. On the right hand side of Figure~\ref{pse-phase 2-asym-sym} we plot the power spectrum estimated using both sets of simulations. The power spectrum of the observed map of Phase~2b (convolved with an asymmetric beam), estimated using the Monte Carlo simulations for Phase~2a (convolved with the symmetric beam), is biased high at high-$\ell$, as expected.

\subsection{Results: Likelihood \label{like-results}}
\label{like-results}

\begin{figure}
\begin{center}
\includegraphics[width=15cm]{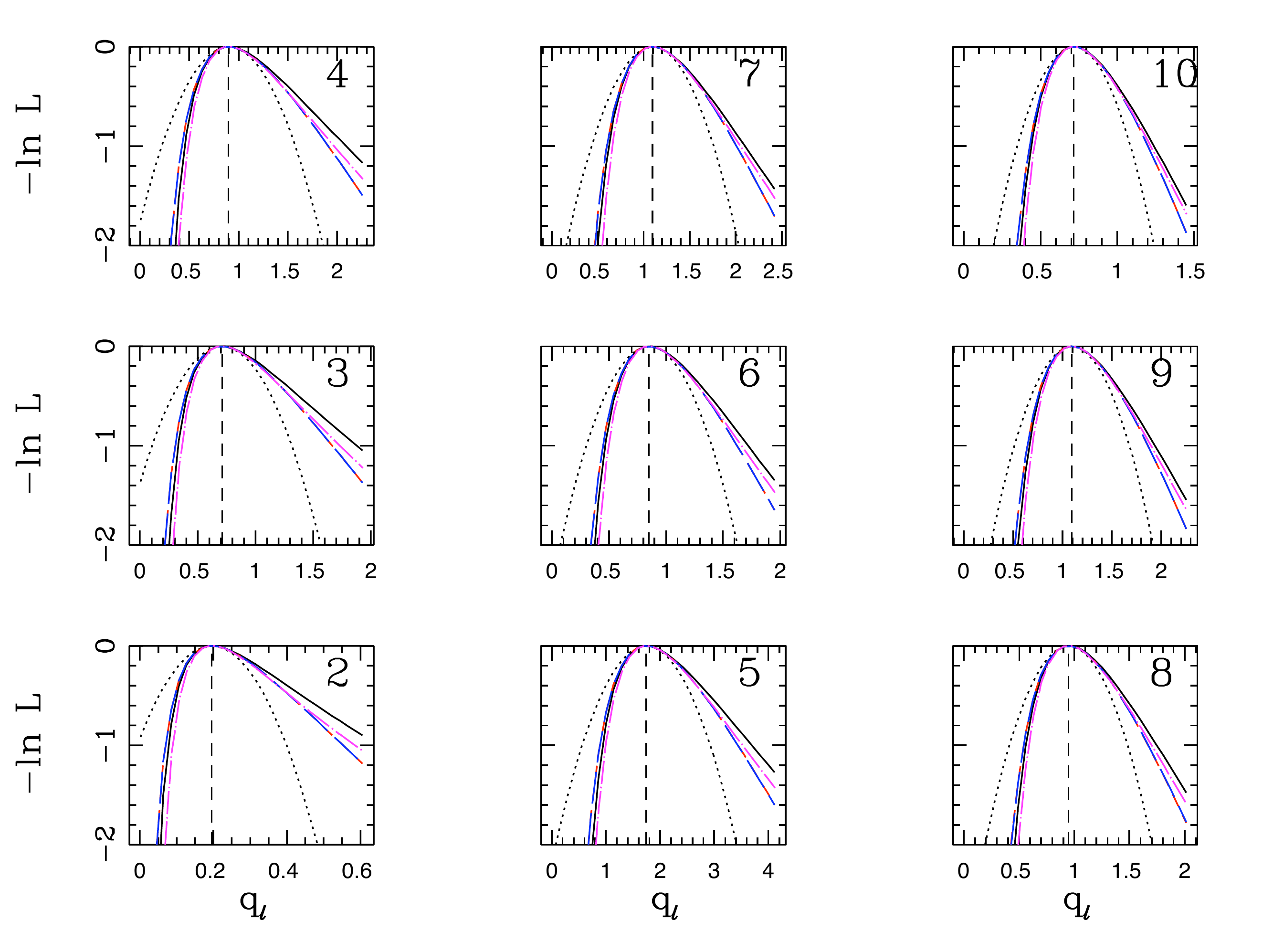}
\caption{ Likelihoods for Phase~2b (12-detector map convolved with an asymmetric beam): Likelihood functions are sampled in each bandpower direction while fixing the other bands at the maximum likelihood values. The black (dotted) curve is the Gaussian approximation given by the Fisher matrix. The blue (dashed) curve is the offset lognormal approximation using the noise $q_{b}^{N}$. The magenta (dash-dotted) curve is the equal variance approximation.  The red (dashed) curve is lognormal distribution.The black (solid) curve is the {\tt XFaster} likelihood estimated for temperature and polarization. The numbers in the right upper corner indicate the multipole $\ell$ or $\ell_{\rm effective}$ of the binned multipoles. As for this set of multipoles $\Delta \ell = 1$ these numbers are the single multipole $\ell$}.
\label{like-XFaster-TP}
\end{center}
\end{figure}

\begin{figure}
\begin{center}
\includegraphics[width=15cm]{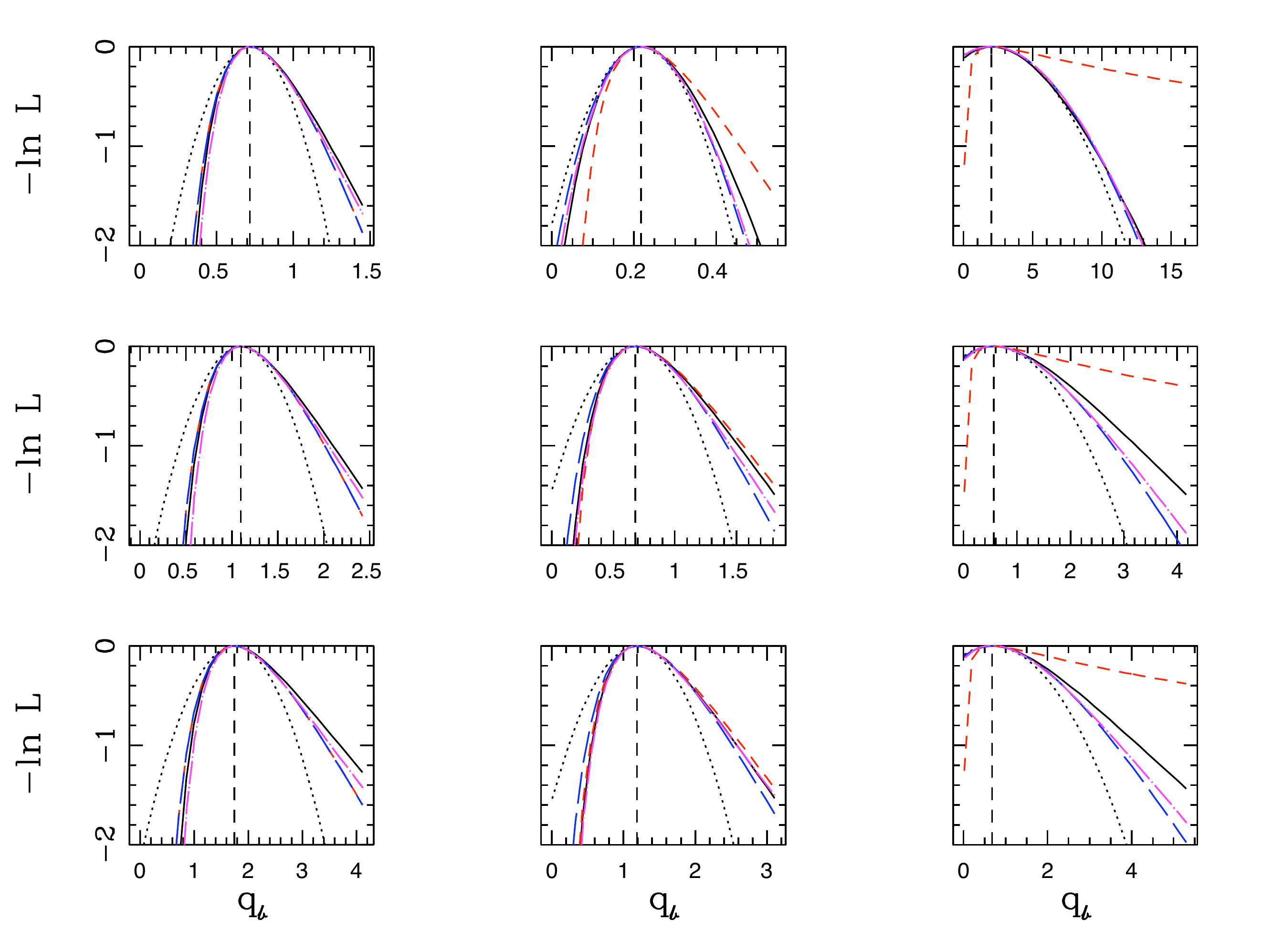}
\caption{ Likelihoods for Phase~2b (12-detector map convolved with an asymmetric beam): Likelihood functions are sampled in each bandpower direction while fixing the other bands at the maximum likelihood values. The black (dotted) curve is the Gaussian approximation given by the Fisher matrix. The blue (dashed) curve is the offset lognormal approximation using the noise $q_{b}^{N}$. The magenta (dash-dotted) curve is the equal variance approximation.  The red (dashed) curve is lognormal distribution.The black (solid) curve is the {\tt XFaster} likelihood estimated for temperature and polarization, for TT  (first column), EE (second column) and TE (third column); for $\ell=5$ (first row), $\ell=7$ (second row), $\ell=10$ for TT and TE and for bin with $\ell$ in [10,13] for EE mode (third row).}
\label{like-XFaster-TP2}
\end{center}
\end{figure}

\begin{figure}
\begin{center}
\includegraphics[width=9cm]{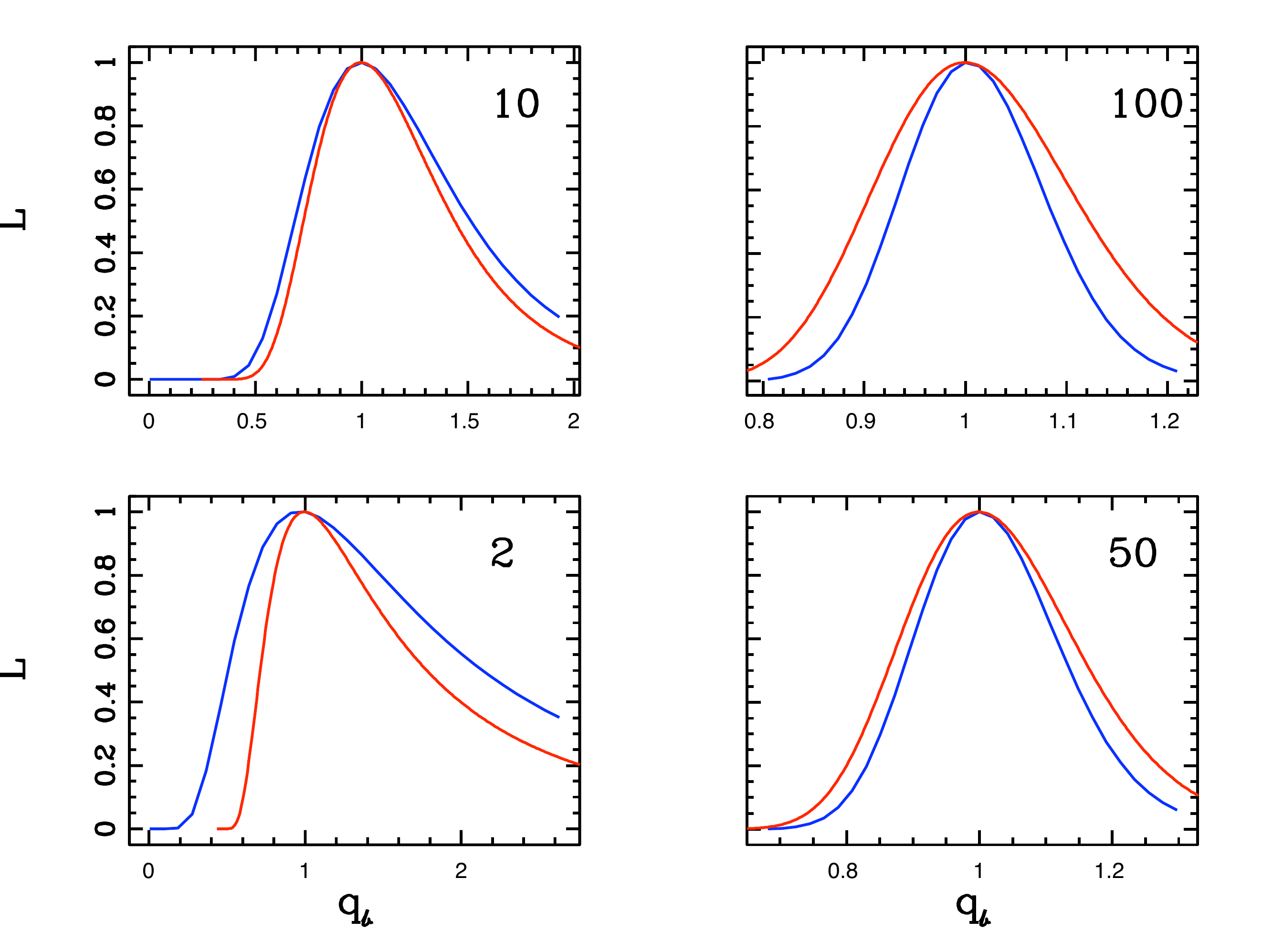}
\caption{ Likelihoods for Phase~2a (12-detector map convolved with a symmetric beam) for TT mode. {\tt XFaster} likelihood (Phase~2 - cut sky) (blue solid line) vs Exact Full Sky likelihood (Phase~1a - full sky) (red solid line). At low-$\ell$ the {\tt XFaster} likelihood is wider due to the correlations induced by the cut-sky while at high-$\ell$ {\tt XFaster} likelihood is narrower due to the binning effect; The numbers in the right upper corner indicate the multipole $\ell$ or the bin number. Up to 10 the bin number is the single multipole $\ell$ as $\Delta \ell=1$, whereas ${\rm bin}=50$ corresponds to $\ell$ in [61,62]  and ${\rm bin}=100$ corresponds to $\ell$ in [257,263].}
\label{like-XFaster-exact}
\end{center}
\end{figure}

\begin{figure}
\begin{center}
\hbox{
\includegraphics[width=9cm]{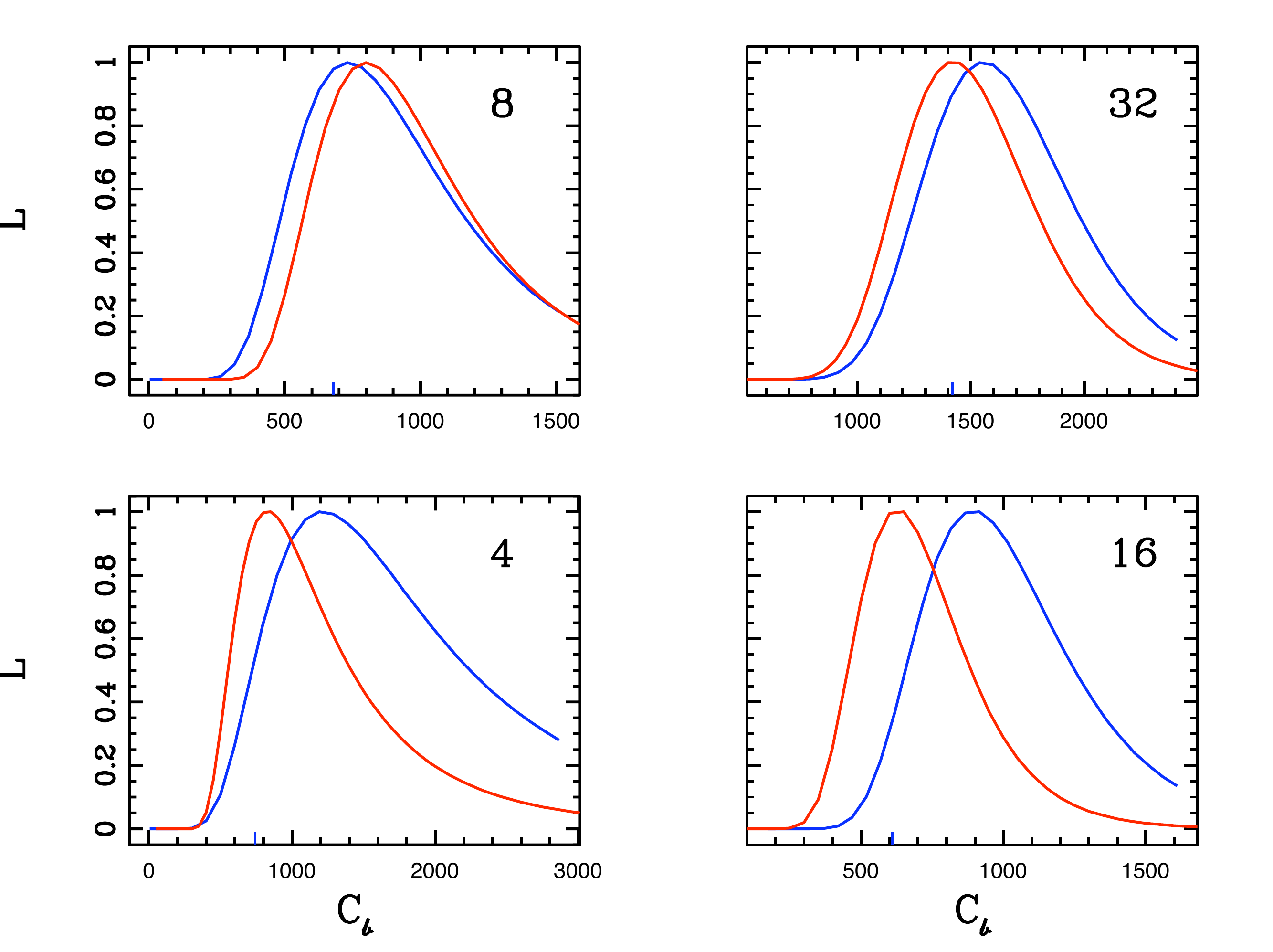}
\includegraphics[width=9cm]{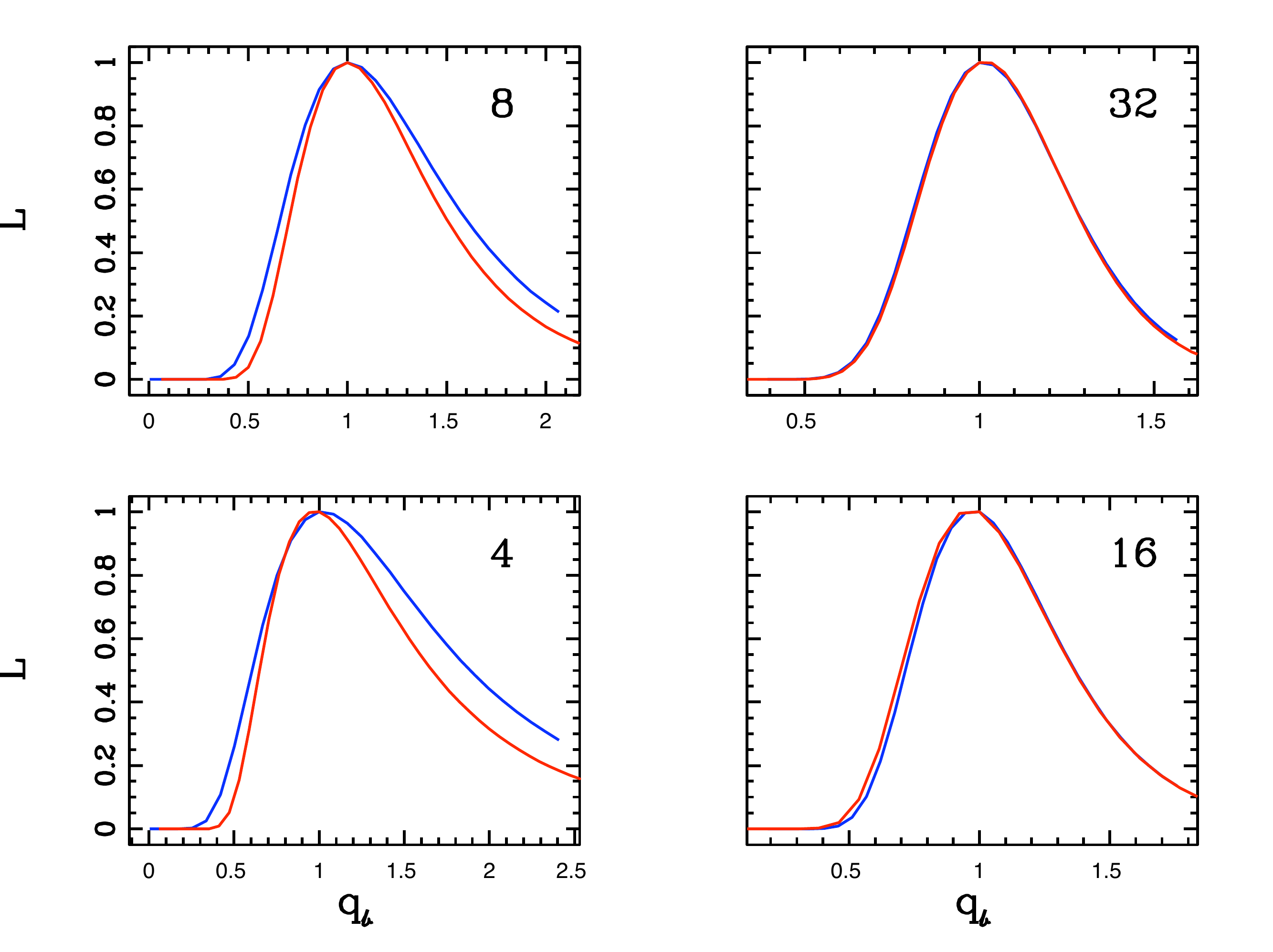}
}
\caption{ Likelihoods for Phase~2a (12-detector map convolved with a symmetric beam) for TT mode. 
 Left hand side plot:  {\tt XFaster} likelihood slices (blue solid line) versus {\tt BFLike}, Pixel based likelihood slices (red solid line), both curves are estimated on the cut-sky map of Phase~2.
This plot can be misleading as the width of both distributions depend on their peaks locations.
Right hand side plot: {\tt XFaster} likelihood slices (blue solid line) versus {\tt BFLike}, Pixel based likelihood slices (red solid line), both curves are estimated on the cut-sky map of Phase~2, assuming they both peak at the same value, ie after dividing both distributions by their peaks values. The agreement is already apparent at $\ell$ as low as $\ell=16,32$. The numbers in the right upper corner indicate the multipole $\ell$ or $\ell_{\rm effective}$ of the binned multipoles. As for this set of multipoles $\Delta \ell = 1$ these numbers are the single multipole $\ell$}.
\label{like-XFaster-pbl}
\end{center}
\end{figure}

Following Section~\ref{like} we use one dimensional slices as an approximation to investigate the Non-Gaussianity of the likelihood, sampling in each $q_{b}$ direction around the maximum likelihood solution $q_{b}^{*}$.  This approximation is adequate if the band powers are not heavily correlated. Note that the likelihood slices are estimated along the $q_{b}$ band power deviations and not along the $q_{\ell}$ power deviations for each multipole $\ell$, and hence are affected by the binning procedure.  These slices are plotted in Figures~\ref{like-XFaster-TP}, \ref{like-XFaster-TP2}, \ref{like-XFaster-exact}, \ref{like-XFaster-pbl}.

Figures~\ref{like-XFaster-TP} and \ref{like-XFaster-TP2} compare the {\tt XFaster} likelihood to four other likelihood approximations, Gaussian, Lognormal, Offset Lognormal, and Equal Variance (\cite{DABJK00,GRLike09}).
A thorough account of these likelihood approximations is given elsewhere (see for instance \cite{GRLike09}), here we give a brief account of their definitions. In what follows $\hat{C}$ means the measured or observed quantity $C$.\\
The Gaussian approximation, (\citet{DABJK00}), is a likelihood that is 
Gaussian in the $\hat{C}_\ell$ i.e.
\begin{equation}
P(\hat{\bmath{C}}| \bmath{C}) \propto
\exp \left\{
-\frac{1}{2} (\hat{\bmath{C}} - \bmath{C})^T
\bmath{S}^{-1} (\hat{\bmath{C}} - \bmath{C})
\right\}
\end{equation}
where $\bmath{C}$ is a vector of $C_\ell$ values (and similarly
$\bmath{\hat{C}}$)
and $\bmath{S}^{-1}$ is the inverse signal covariance matrix.\\
The Offset Lognormal likelihood, (\citet{DABJK00}), is given by:
\begin{equation}
P_{LN}(\hat{\bmath{C}}| \bmath{C}) \propto
\exp \left\{
-\frac{1}{2} (\hat{\bmath{z}} - \bmath{z})^T
\bmath{M}(\hat{\bmath{z}} - \bmath{z})
\right\},
\end{equation}
where $z_\ell = \ln(C_\ell + x_\ell)$ and
the matrix $\bmath{M}$ is related to the inverse
covariance matrix by
\begin{equation}
M_{\ell \ell'} = 
(C_\ell + x_\ell) S^{-1}_{\ell \ell'} (C_\ell' + x_\ell')
\end{equation}
(The offset factors $x_\ell$ are simply a function of the
noise and beam of the experiment.)\\
The Equal Variance likelihood, (\citet{DABJK00}) is given by:
\begin{equation}
\ln L = -\frac{1}{2}G \left[  e^{-(z-\hat{z} ) }- \left(1-(z-\hat{z}) \right)    \right]
\end{equation}
with
\begin{equation}
z= \ln \left(q_{b} + q_{b}^{N} \right)   
\end{equation}
and
\begin{eqnarray}
G = \left [e^{-\sigma_{z}} - (1-\sigma_{z})  \right]^{-1}  & {\rm with} & \sigma_{z} =\frac{\sqrt{{\cal{F}}_{bb'}^{-1}}}{(q_{b} + q_{b}^{N})}       
\end{eqnarray}
The noise offset $q_{b}^{N}$ is estimated using the equation of the maximum likelihood solution for the $q_{b}$ replacing the observed map with the average of the noise Monte Carlo simulation power spectra $ \left<  \tilde{N_{\ell}} \right > $.

We use the values of the power spectrum and Fisher errors estimated with {\tt XFaster}  for Phase~2b (map generated with all twelve detectors and convolved with the asymmetric beam).
As we are primarily interested in the shape of the likelihoods not on the actual value of the peaks, the comparison of the slices is done assuming they all peak at the same value, ie we use the band power specta estimated with XFaster and the functional shape of the other high-l likelihoods. This is the same to say that we first compute the band power spectra and apply the functional forms whereas XFaster likelihood slices are computed when estimating the band-power spectra.
Figure~\ref{like-XFaster-TP} shows temperature slices of the {\tt XFaster} joint  temperature and polarization likelihood.  Figure~\ref{like-XFaster-TP2} shows slices for TT (first column), EE (second column), and TE (third column) (more precisely $-\ln L$).
At the lowest multipoles the approximations differ, but as we move towards higher $\ell$ all but the Gaussian likelihood converge to the same functional form. For EE, however, the approximations differ noticeably for the Gaussian and Lognormal likelihood approximations (at $\ell \simeq 10$, for instance).

We further compare the {\tt XFaster} likelihood to the exact likelihoods at low multipoles. The purpose is twofold.  On one hand we want to validate the {\tt XFaster} approximation, on the other hand we want to determine the $\ell$ range at which the approximations used for the high-$\ell$ power spectrum estimator breaks down.

Figure~\ref{like-XFaster-exact} shows {\tt XFaster} likelihood slices for the Phase~2 binned power spectra and the exact full-sky likelihood estimated for Phase~1a. At low-$\ell$ the correlations induced by the cut-sky  widen the {\tt XFaster} likelihood, while the binning effect at high-$\ell$ results in a narrower  distribution (when compared to the full-sky exact likelihood). Both effects are given by

 \be
 \sigma = \sqrt{\frac{2}{(2\ell+1) f_{sky} \Delta \ell}} C_{\ell},
 \ee
where $f_{sky}$ is the fraction of the sky observed and $\Delta \ell$ is the width of the multipole bins.

To make a direct comparison of the {\tt XFaster} likelihood with the exact likelihood at low-${\ell}$ estimated in the same cut-sky map (Phase~2), which accounts for the correlations induced by the sky cut, Figure~\ref{like-XFaster-pbl} shows the {\tt XFaster} likelihood versus the pixel-based likelihood slices for temperature alone (Phase~2b, asymmetric beam case).
The pixel based likelihood, {\tt BFLike}, is a brute force likelihood evaluation of the multivariate Gaussian in pixel domain for a low-resolution map. The low-$\ell$ dataset  of the CTP Phase~2 simulations was generated directly at $N_{\rm side} =16$. In computing the slices, we conditioned on the remaining $TT$ multipoles, $C_{\ell'}^{TT}$ with $\ell' \ne \ell$, and on all multipoles of the $TE$ and $EE$ spectra (for details see \cite{GRLike09}).
 As for this case the {\tt BFLike} estimates its own peak by computing a brute force pixel based likelihood on a downgraded map, we plot in Figure~\ref{like-XFaster-pbl} both likelihoods  with its own peaks locations (left hand side) and assuming they peak at the same value (right hand side), ie after dividing both distributions by their peaks values. The plot on the left hand side might be misleading as the width of both distributions depend on their peaks locations. As we are mostly interested in comparing the shape of the likelihoods we will pay particular attention to the plot in the right hand side of Figure~\ref{like-XFaster-pbl}.

At $\ell=32$ the agreement of both likelihoods is already quite remarkable, suggesting that a transition between high-$\ell$ and low-$\ell$ estimators around $\ell_{\rm trans} \simeq 30$--40 may be appropriate for this dataset.
Hence a Planck hybrid likelihood built out of these two likelihoods (namely  {\tt PiXFaster}, see eg \cite{GRLike09}), with a transition range around $\ell \simeq 30$--40 should be a viable hybridization scheme.

\subsection{Results: Cosmological parameters \label{par-results}}
\label{par-results}

To compare the theoretical power spectrum with the observed power spectrum and estimate parameters, we need an operator to extract theoretical bandpowers
from model power spectra $C_{\ell}^{T}$.
In \cite{GRLike09} we considered two types of windows, a top hat window per bin, and the appropriate Fisher-weighted window or {\tt XFaster} band power window function.
These window functions have been used in association with the Offset Lognormal Bandpower likelihood (\cite{GRLike09}). The {\tt XFaster} likelihood is estimated multipole by multipole ie for each $\ell$, hence no window function is required.
In this mode {\tt XFaster} can go straight from the map (via its raw pseudo-$C_{\ell}$) to parameter estimation, bypassing the band power spectrum estimation step.

We implemented the {\tt XFaster} likelihood in a (suitably-modified) version of the publicly-available software {\tt CosmoMC} \footnote {http://cosmologist.info/cosmomc/} (\citet{Lewis:2002ah})) for cosmological parameter Markov Chain Monte Carlo
estimations. The {\tt XFaster} likelihood code computes the likelihood of a model passed to it by \texttt{CosmoMC}. There is no need for window functions or the band power spectrum itself.  The inputs are the raw pseudo-$C_{\ell}$ of the observations plus the kernel and transfer function required by {\tt XFaster} to relate the cut-sky pseudo-$C_{\ell}$ to the full-sky $C_{\ell}$.  

\begin{figure}
\begin{center}
\includegraphics[width=15cm]{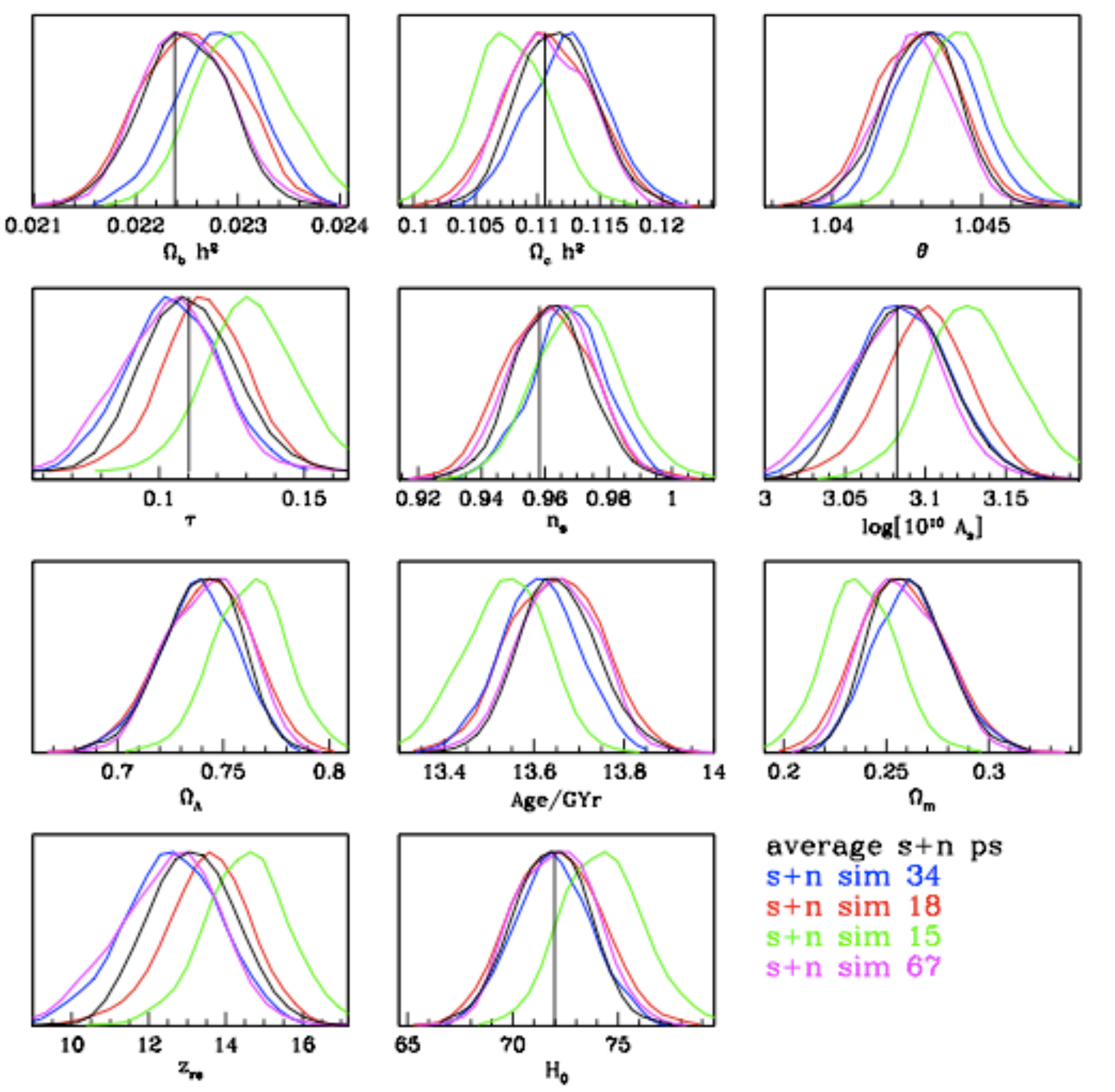}
\caption {Parameter constraints from Phase~2a (12-detector map convolved with a symmetric beam). The 1-dimensional marginalized posteriors are from {\tt XFaster} likelihood (with inclusion of all modes, TT, EE, BB, TE): for ensemble average of signal+noise Monte Carlo simulations, i.e., average run (solid black lines), for several single signal+noise Monte Carlo simulations, and for values of  the fiducial best fit input parameters (black vertical lines). The parameters for the average power spectrum recover the true input parameters. Furthermore this plot shows that there is no systematic bias for each Monte Carlo simulation. This is to be expected as the Monte Carlo simulations are realizations of the WMAP best fit model.}
\label{par-XFaster-mc-avg}
\end{center}
\end{figure}

\begin{figure}
\begin{center}
\includegraphics[width=15cm]{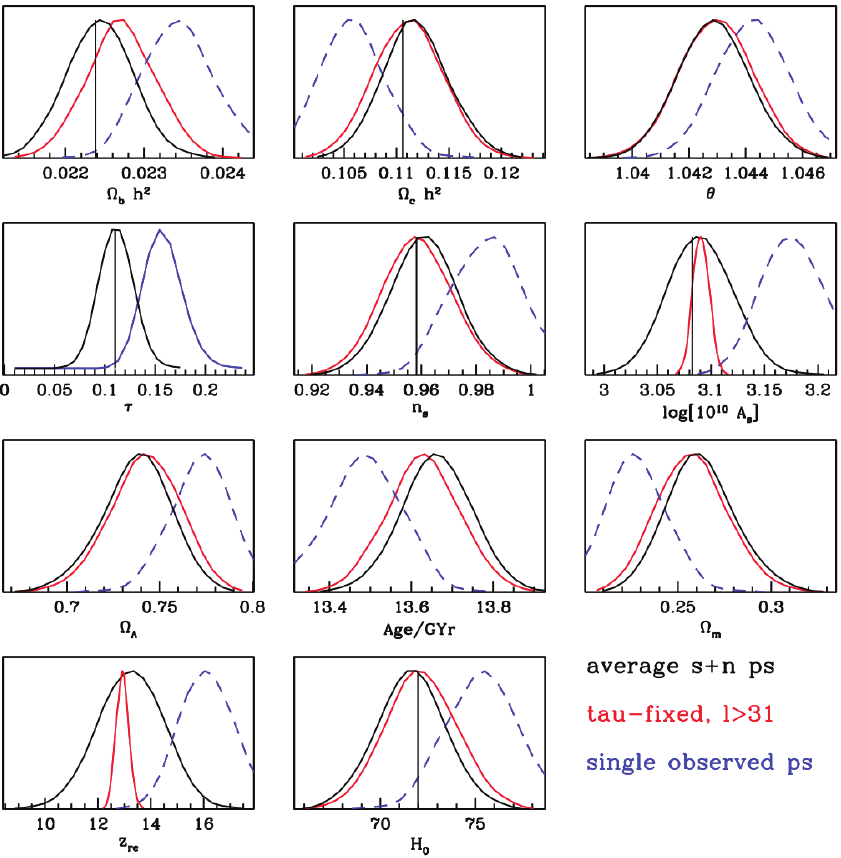}
\caption{Parameter constraints from Phase~2a (12-detector map convolved with a symmetric beam). The 1-dimensional marginalized posteriors are from {\tt XFaster} likelihood (with inclusion of all modes, TT, EE, BB, TE): for ensemble average of signal+noise Monte Carlo simulations, i.e., average run (solid black lines), for single observed data (dashed blue lines) and for the run with $\tau = \tau_{fiducial}$ using {\tt XFaster} likelihood for $\ell > 30$ (solid red lines), with overplotted values of  the fiducial best fit input parameters (black vertical lines). The parameters for the average power spectrum recover the true input parameters. However the parameters for the observed map shift from the input parameters, particularly so for the parameter Amplitude $A_s$.  As the large scale structure of the observed map is a WMAP constrained realization we do not expect the estimated parameters to agree with WMAP best fit parameters. Fixing $\tau$ to the input value regularizes the amplitude in the likelihood runs, now the estimated parameters of the observed map shift towards the input parameter values recovering those estimated for the ensemble average of the Monte Carlo simulations.}
\label{par-XFaster-mc-obs-fixtau}
\end{center}
\end{figure}

\begin{figure}
\begin{center}
\hspace*{-1cm}
\hbox{
\includegraphics[width=10cm]{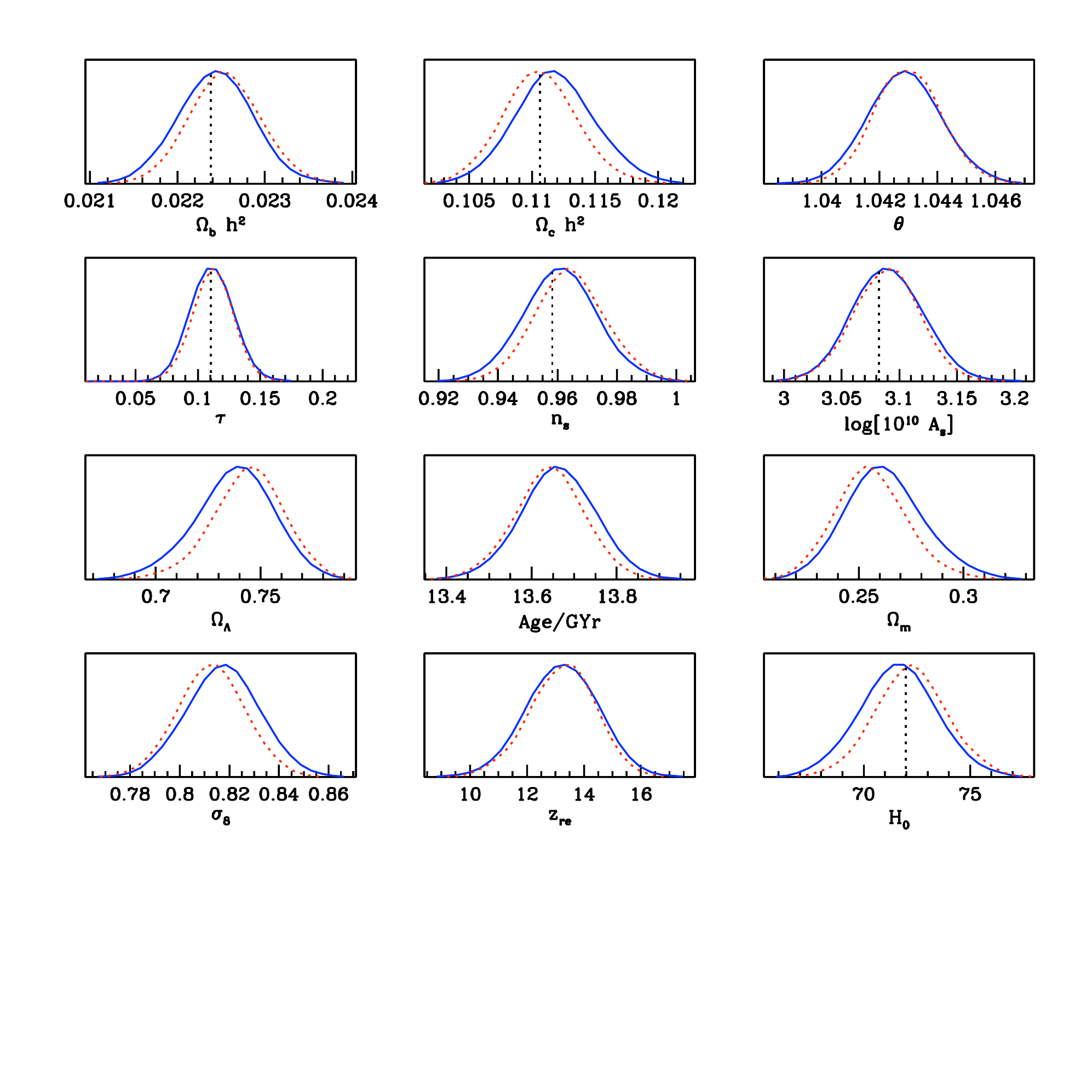}
\includegraphics[width=10cm]{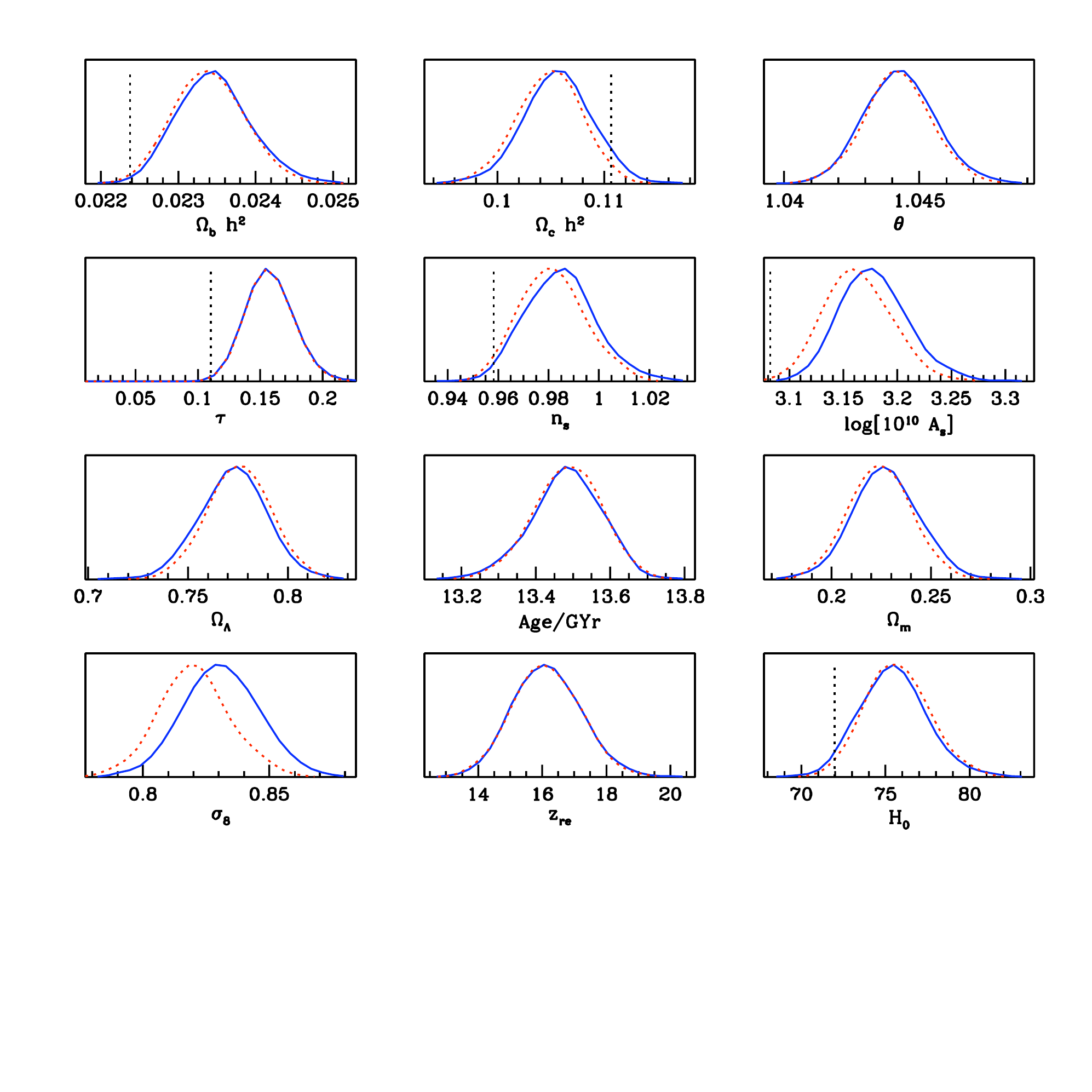}
}
\caption{Parameter constraints from Phase~2a (12-detector map). The 1-dimensional marginalized posteriors are from {\tt XFaster} likelihood (with inclusion of all modes, TT, EE, BB, TE):
for the averaged spectra (left hand side) and for single realization spectra (observed map) without fixing $\tau$ (right hand side),  asymmetric beam (solid blue lines) and symmetric beam (dashed red lines) cases. }
\label{par-XFaster-symm-asymm-avg-obs}
\end{center}
\end{figure}

Figures~\ref{par-XFaster-mc-avg} through \ref{par-XFaster-symm-asymm-avg-obs} show results for a simulation using Phase~2a (symmetric beams) and Phase~2b (asymmetric beams) data.  The parameters considered are the baryon, cold dark matter and cosmological constant densities,
$\omega_b=\Omega_bh^2$ and $\omega_c=\Omega_ch^2$ and $\omega_{\Lambda}=\Omega_{\Lambda} h^2$ respectively, the ratio of the sound horizon to the angular diameter distance at decoupling, $\theta_s$, the scalar spectral index $n_s$, the overall normalization of the spectrum $\log[10^{10} A]$ at $k=0.05$ Mpc$^{-1}$ ($A_s$), the optical depth to reionization $\tau$, the age of the universe, the Hubble constant $H_{0}$, and the reionization redshift $z_{re}$.

Figure~\ref{par-XFaster-mc-obs-fixtau} shows parameters estimated for the average power spectrum of the signal+noise Monte Carlo simulations and for the observed power spectrum (i.e., the power spectrum estimated  for the observed map). The parameters for the average simulated data recover the true input parameters, while those for the observed map shift from the input values, particularly for $A_s$.
As mentioned before, the observed map is a WMAP-constrained realization, i.e., it uses the $a_{lm}$ with phases measured by WMAP up to $\ell=70$ to reproduce the large-scale structure observed by WMAP, and a best-fit model to the WMAP observations for $\ell>70$.
The WMAP best-fit parameters are obtained with considerable marginalization of the low-$\ell$ points by foregrounds.  They are therefore unaffected by the low-$\ell$ anomalies. This means that unless we do such analysis too, we would not expect our  observed realization to agree with the WMAP best fit model. This is clearly shown in Figure~\ref{par-XFaster-mc-obs-fixtau}.
On the other hand, as Monte Carlo simulations are realizations of the WMAP best-fit model, one should expect no systematic bias from the ensemble of simulations, as confirmed in Figure~\ref{par-XFaster-mc-avg}.

As discussed in section~3, {\tt XFaster} assumes that the noise is white (uncorrelated), i.e., that the noise covariance matrix is diagonal.  Also, the {\tt XFaster} likelihood is estimated multipole by multipole, hence to estimate the transfer function properly requires a larger number of Monte Carlo simulations to beat down the correlations between multipoles introduced by, e.g., sky cuts required for foreground removal.  These simulations include both correlated noise and a sky cut.  To assess whether the low-$\ell$ inadequacy of the likelihood is indeed the cause of the parameter offsets seen in Figure~\ref{par-XFaster-mc-obs-fixtau}, we recalculated parameters, this time fixing $\tau$ to the input model value in the simulations. Since $\tau$ and only $\tau$ is constrained almost entirely by the $\ell<30$ data, by fixing $\tau$ we mimic the effect of using a likelihood evaluator that takes full cognizance of correlations in the noise and between multipoles at low $\ell$.   The results are shown in Figure~\ref{par-XFaster-mc-obs-fixtau}.  The estimated parameters shift towards the input parameter values, recovering those estimated for the average case in agreement with our postulated hypothesis.

Table~\ref{parXFaster} compares parameter constraints for the symmetric beam case for the ensemble average power spectrum of the Monte Carlo simulations run (avg) and the observed power spectrum run (pse) without $\tau$ fixed. The parameter constraints tabulated are from the marginalised distributions.
 The parameter constraints for the observed power spectrum run (pse) when $\tau$ is fixed to the fiducial input value are indistinguishable from those derived from the average run, hence very close to the input parameter values. Therefore we do not include them in Table~\ref{parXFaster}.


Figure~\ref{par-XFaster-symm-asymm-avg-obs} shows constraints from symmetric and asymmetric beam case for the average power spectrum of the Monte Carlo simulations and the actual observed estimated power spectrum without fixing $\tau$.
Most of the parameters for both cases are consistent with each other.
Investigating the plot for the average mode, we see deviations of the order of $\sigma/2$ for $\Omega_{c} h^{2}$, $\sigma_{8}$, $n_{s}$ and $H_{0}$. There is an obvious degeneracy between $\sigma_{8}$ and $n_{s}$.  For the observed case these deviations are noticeable mostly in $A_s$ and $\sigma_{8}$. Once again these parameters are degenerate.

The overall agreement  in the parameter constraints from both symmetric and asymmetric beam cases is quite impressive. This reflects the adequacy of our procedure when dealing with beam asymmetries.

\begin{table*}
\caption{\label{parXFaster} Parameter constraints from Planck 70\,GHz data, Phase~2a, for map generated with all twelve detectors and convolved with a symmetric beam, using {\tt XFaster} power spectrum and likelihood estimator for the average of the signal+noise Monte Carlo simulations power spectrum (avg) and the actual observed power spectrum (pse) without $\tau$ fixed. As mentioned in the text, the parameter constraints for the actual power spectrum run (pse) when $\tau$ is fixed to the fiducial input value are indistinguishable from those derived from the average run, hence very close to the input parameter values. Parameter constraints displayed here were obtained from the marginalised distributions.}
\begin{tabular}{cccc} \\
&& \\ \hline\hline
param   &    bestfit (avg) $\simeq$ bestfit (pse with fixed $\tau$) & bestfit (pse with varying $\tau$)       & input    \\ \hline
& & \\
 $        \omega_b$ & $  0.0225^{+0.00042}_{-0.00042 } $      & $  0.0231^{+0.00046}_{-0.00046 } $ & 0.02238 \\
 & & \\
$        \omega_c$ & $  0.1115^{+0.00309}_{-0.00305 } $   & $  0.1077^{+0.00307}_{-0.00314 } $    & 0.11061  \\
& & \\
$          \theta$ & $  1.0430^{+0.00120}_{-0.00122 } $      & $  1.0445^{+0.00130}_{-0.00137 } $          \\
& & \\
$            \tau$ & $  0.1105^{+0.00643}_{-0.00771 } $    & $  0.1573^{+0.00833}_{-0.00950 } $  & 0.1103         \\
& & \\
$n_s$ & $  0.9621^{+0.01130}_{-0.01170 } $    & $  0.9757^{+0.01341}_{-0.01353 } $          &  0.95820             \\
& & \\
$     log[10^{10}A_{s}]$ & $  3.0874^{+0.02690}_{-0.02748 } $    & $  3.1727^{+0.03270}_{-0.03324 } $ & 3.0824  \\
& & \\
$  \Omega_\Lambda$ & $  0.7394^{+0.01750}_{-0.01843 } $    & $  0.7633^{+0.01710}_{-0.01695 } $   &    \\
& & \\
$             Age$ & $ 13.7^{+ 0.1}_{- 0.1 } $     & $ 13.5^{+ 0.1}_{- 0.1 } $  &  \\
& & \\
$        \Omega_m$ & $  0.2606^{+0.01843}_{-0.01749 } $   & $  0.2367^{+0.01698}_{-0.01710 } $  &   \\
& & \\
$          z_{re}$ & $ 13.1^{+ 1.1}_{- 1.1 } $       & $ 16.3^{+ 1.1}_{- 1.1 } $ &   \\
& & \\
$             H_0$ & $ 71.82^{+ 1.74}_{- 1.81 } $   & $ 74.46^{+ 1.95}_{- 1.94 } $   & 71.992   \\  \hline
& &
\end{tabular}
\end{table*}
%

\section{Conclusions \label{conc}}
\label{conc}

The {\tt XFaster} power spectrum estimator is fully adequate to estimate the power spectrum of Planck data in the high-$\ell$ regime.  It also performs well at moderately low multipoles, as long as the low-$\ell$ polarization and temperature
power is properly accounted for, {\it e.g.}, by adding an adequate
low-$\ell$ likelihood ingredient. Our minimal non-informative approach enables us to recover most input parameters regardless of the asymmetry of the beam.

\section*{Ackowledgments}

  The work reported in this paper was partially done within the CTP Working Group of
  the {\sc Planck} Consortia. {\sc Planck} is a mission of the
  European Space Agency.
   GR would like to say a special thank you to Charles Lawrence for his insightful comments and dedicated help during the writing up of this manuscript.
  GR is also grateful to Jeff Jewell for useful discussions.
  This research used resources of the National Energy Research
  Scientific Computing Center, which is supported by the Office of
  Science of the U.S. Department of Energy under Contract
  No. DE-AC03-76SF00098.
  This work has made use of the
  {\tt HEALPix} package (\cite{healpix}); and of the {\sc Planck} satellite simulation
  package, {\tt LevelS}, (\cite{levels}) which is assembled by the Max Planck Institute
  for Astrophysics {\sc Planck} Analysis Centre (MPAC).
   GR {\sc Planck} Project is supported by the NASA
  Science Mission Directorate.
  The research described in this paper was partially carried out at the Jet propulsion Laboratory, California Institute of Technology, under a contract with NASA.
Copyright 2009. All rights reserved.



\label{lastpage}

\end{document}